\documentstyle[prb,preprint,aps]{revtex}

\def\prb{ Phys. Rev. B }

\def\beq{\begin{equation}}
\def\eeq{\end{equation}}
\def\bear{\begin{eqnarray}}
\def\eear{\end{eqnarray}}

\begin{document}
\draft
\date{\today}

\title{Theory of strongly correlated electron systems.\\
I. Exact Hamiltonian, Hubbard-Anderson models and 
perturbation theory  near atomic limit within non-orthogonal basis set}

\author{Igor Sandalov$^{1,2}$, B\"{o}rje Johansson$^1$, Olle
Eriksson$^{1}$ } 

\address{
$^{1}$Condensed Matter Theory Group, Institute of Physics,
Uppsala University, Box 530, SE-751 21 Uppsala, Sweden, \\
$^{2}$Dept.\ of Physics, Link\"oping University,
SE-581 83 Link\"oping, Sweden}

\maketitle

\begin{abstract}
The theory of correlated electron systems is formulated in a form which
allows to use as a reference point an {\em ab initio} band structure
theory (AIBST). The theory is constructed in two steps. As a
first step the total Hamiltonian is transformed into a correlated form. In
order to elucidate the microscopical origin of the parameters of the
periodical Hubbard-Anderson model (PHAM) the terms of the full Hamiltonian which
have the operator structure of PHAM are separated. It is found that the
matrix element of mixing interaction includes ion-configuration and
number-of-particles dependent contributions from the Coulomb interaction. In a
second step the diagram technique (DT) is developed by means of 
generalization of the Baym-Kadanoff method for correlated systems.
The advantages of the method are that i) a non-orthogonal basis can be used,
particularly, the one which is generated by AIBST; ii) the  equations for 
Green's functions (GFs) for the Fermi- and Bose-type of quasi-particles
can be formulated in the form of {\em closed} system of functional 
equations. The latter allows to  resolve
some difficulties of previous versions of the DT. Although the expressions for
all interactions
depend on the overlap matrix, it is shown that the theory is formally equivalent
to one with orthogonal states with redefined interactions. When 
the PHAM is treated from the atomic-limit side the vertexes
are generated by kinematic interactions. The latter arise due to non-trivial
commutation relations between
 $X$-operators and come from the mixing, hopping and
the overlap of states. The equations for GFs are derived in
different approximations for vertexes, corresponding to the Hubbard-I, mean
field, and random phase approximations, respectively. The technique 
is also extended to the case of intersite
 Coulomb interaction. The self-consistent equation 
for "Hubbard $U$-s" is derived. Both interactions, kinematic mixing and 
hopping, and Coulomb interaction of $f$- and non-$f$-electrons contribute 
to the screening of  bare $U$. 
 The diagram techniques for pure Hubbard,
Anderson models, $s-f$-model and for iso- and anisotropic spin Hamiltonians
can be obtained from the developed approach as special cases.
\end{abstract}



\section{Introduction}

The calculations of properties of real materials in the language of
many-body wave functions are incomparably more effort- and time- consuming
than the ones based on the density functional theory (DFT).  However, 
DFT is designed for
calculation of {\em ground-state} properties and, by construction, is
expected to have limited use if one is interested in {\em dynamic}
properties. The only available regular way to calculate quasiparticle 
properties is to
make use of some diagram technique. The latter usually 
involves expressions for
secondary quantization. Therefore, anyhow one
needs some single-electron equation which would generate a set of wave
functions with proper boundary conditions. 
 The DFT-based {\em ab initio}
band structure calculations (AIBST) seems to be a reasonable starting point
at least for the conduction electrons. Therefore, the
corresponding 
Schr\"{o}dinger equation can be used as the one which generates a
convenient basis set of wave functions for formulation of secondary 
quantized Hamiltonian. On the other hand it is far from obvious how DFT can be used for correlated electrons. Materials which
usually are classified as systems with strong electron correlations (SEC)
show wide spread properties such as magnets with localized and partly
localized moments, Mott insulators, Kondo lattices, heavy fermion systems,
and high-$T_c$ superconductors. These materials contain elements with an
open $d-$ or $f-$shell which often are main source of difficulties in a
theoretical description both within the local spin density approximation\cite
{dft} (LSDA) and model approaches. The classical example of a ground state
property which the LDA approach is not able to describe is the Mott
insulating state where materials with odd number of electrons per primitive
cell are insulators\cite{brandow}. The special features which arise in these
systems due to SEC are configuration-dependent mixing and hopping (see ref.%
\cite{sand_fil}) and the hybridization-dependent spectral weight transfer%
\cite{meinders}. Even much simpler properties as
 the atomic-like Hund's rules are not reproduced
within DFT. The common way to correct the situation in AIBST is to use
phenomenological methods like the self-interaction correction\cite
{temmerman,perdew_zunger}(SIC), and the so-called LDA+U method of Brandow%
\cite{brandow,boring,AZA,olof}. These methods provide a picture close to the
discussed one by Brandow\cite{brandow} in the Hartree-Fock approximation:
the SIC corrected
 orbitals (selected at the onset of the calculation)
 are pushed down by the 
SIC potential, usually to
8-11 eV below Fermi energy, $\epsilon _F$, while all other $f(d)$-orbitals
remain above $\epsilon _F$ and are
empty. From the point of view of
phenomenological theories of SCES, like the
periodical Anderson model (PAM), the
picture generated by SIC is not satisfactory since usually mixing
interaction disappears when $f(d)$-level is below the bottom of conduction
band\cite{olof}. This makes it
impossible to check the PAM mechanisms within
AIBST. The LDA+U method gives a
much smoother picture, since the values of the
Hubbard $U$ are taken to be 2-6 eV. The models used are also
different and there is no clear arguments why in one case should us
the Hubbard model (i.e. mixing is equal to zero) whereas
another case the PAM (i.e. hopping is neglected). This raises the question
whether these arguments can be derived from strict microscopical approach
or not as well as
whether some minimal model can be derived from the 
exact full Hamiltonian or
not. Below we will show that when the system contains a subsystem 
with (quasi)localized electrons, the
full Hamiltonian $H$ always contains a part, which corresponds to 
the periodical Hubbard-Anderson model (PHAM). Then,  
  representing $H$ in the form $H=H_{PHAM}+H'$ we will
 construct a diagram technique which will make possible both calculation 
of renormalization of $H_{PHAM}$ by remaining Coulomb terms from $H'$,  
 and to understand to what extent picture obtained within the PHAM 
is accurate.

Another problem, which is difficult to tackle within the AIBST, is 
to calculate the
spectrum of quasi-particles.
Although there are no grounds to consider the eigenvalues of Kohn-Sham
equations as energies of quasi-particles, this is quite common practice due
to its relative simplicity. For some cases a simple formula which connects
these energies can be written\cite{maksimov}. Part of the
dispersion and 
life-time of quasi-particles, nevertheless, seem to be lost.
 However, the DFT potential treated as a
self-energy\cite{sham} by
construction can neither provide the energy dependence of the
real part of the self-energy, nor reflect the scattering of quasi-particles, since it does not
have imaginary part. This is especially important for the case of SEC, where 
$\Sigma _{k\lambda }(E)$ sometimes can give three solutions of the equation
for the quasiparticle energy $E=\epsilon _{k\lambda }+Re\Sigma _{k\lambda
}(E)$ (e.g., satellites in the photoelectron spectra of $Ni$) while the DFT
potential, $v_{DFT},$ due to absence of a dependence on energy can provide
only one solution to this equation. Thus, in order to obtain information 
about these properties of the system one needs a many-body theory. 
 In the present paper we intend
to develop an approach for the systems with $f(d)$-electrons in the
strong-coupling regime, therefore, we choose perturbation theory from the
atomic limit. As follows from the above argument it is desirable to start
with self-consistent AIBST where the localized electrons are treated as core
electrons. Here we will treat these localized (or almost localized), electrons
with the help of Hubbard operators\cite{hubb1,hubb2} $X^{pq}\equiv |p\rangle
\langle q|$. The latter describe transitions from the state $p$ of the {\em ion}
to the state $q$. The first step then is a formulation of the Hamiltonian of
the system in terms of these variables. In the case of rare earth metals
(Russel-Saunders coupling) and a model of homogeneous gas for conduction
electrons the work by Irkhin and Irkhin\cite{irkhiny} can be of great help.
Particularly, a way to construct many-electron secondary-quantized
operators, which describe creation and destruction of the $f$-ion states
with the help of the standard technique used in atomic spectroscopy, is
thoroughly described therein. This technique has been also successfully used
for derivation of the form of $f$-density for use in DFT-based AIBST\cite
{brooks}. The case of $j-j$-coupling is developed much less since an
accurate account of effects of non-local exchange interaction should 
be performed within a
fully relativistic field-theory 
(the case of strong 
spin-orbit coupling); the cases when a relativistic
effects can be taken into account via perturbative Hamiltonian, can be
treated within our approach too. There is one more important
motivation for a 
microscopical derivation of the correlated form of
Hamiltonians and an effective Hamiltonians. An extension of AIBST to
thermodynamics is based usually on mapping of the constrained AIBST to some
model; the parameters of the model are calculated from fitting
the model's energy surface to the DFT one for some selected configurations
of the system\cite{rosengaard}. Obviously, the choice of the model is not
unique, {\em i.e.}, for sufficiently large number of parameters many
different models can reproduce the DFT-derived dependence of energy on the
 parameters involved. Besides, being within framework of density functional 
formalism it is very difficult to avoid classical approximations,
that can introduce uncontrolled errors. Below we
shall show that such derivation can be performed for arbitrary localized set
of orbitals. This means that the parameters of the model, as well as 
correcting terms, are representation-dependent. 
Probably, the best starting step is 
 a DFT-AIBST. As will be seen, our approach for derivation of the
correlated form of Hamiltonian in the particular case of 
Russel-Saunders coupling is not fully identical to the one, 
suggested by Irkhin\cite{irkhiny},
but their main results can easily be
used for our purposes. The 
 formalism will be developed in general notations, so that different 
 cases can be considered. Compared to the approach developed in 
Ref.\cite{irkhiny} the present paper the
main emphasis 
 will be put on the next logic step of development of the theory,
namely, on the derivation of a {\em regular} diagram technique 
from atomic limit for the
evaluation of Green's functions within the 
basis chosen.

The fact, that SEC have \emph{local, site-centered } nature dictates
a requirement of the approach: the strong on-site Coulomb interactions
should be taken into account first. This can be done much easier within the
representation of some site-centered wave functions. Then the question
arises immediately: should one
first orthogonalize the wave functions
centered on different atoms and then take into account correlations or, 
\emph{vice versa}, first to take into account local interactions and then
treat somehow non-orthogonality? The answer, of course, depends on the ratio
of the bandwidth, generated not only by  non-orthogonality, but by all
possible mechanisms (due to self-consistency) and the largest matrix element
of local Coulomb interaction. It is clear that a big enough Coulomb
interaction can a
cause strong shift of the levels of orbitals, that, in turn,
can strongly change the
overlap with neighboring atoms. This picture has been
many times confirmed by the calculations, which use LDA+U-, SIC
(self-interaction correction), or Hartree-Fock approximation: the orbital,
which was delocalized before application of, say, SIC, becomes completely
localized after applying SIC, loosing simultaneously all mixing and overlap
with neighbors. Thus, for  such materials the answer seems to be clear and
conceptually important: first should be taken into account SEC and only
after that the overlap, mixing and hopping should be treated. Strictly
speaking, the final band structure should not depend on our choice of
perturbation theory: from atomic or itinerant limit we develop perturbation
theory. In the latter case a necessity arises to generalize the three-body
Faddeev equations\cite{faddeev} to the $n$-body ones and to work with
many-electron GFs. Although this is not done yet, one can expect that an
analysis of the contributions from the closed channels of scattering 
should raise the same problems, which we have deal with 
when start from the atomic limit. Returning to our approach, 
we immediately find that the non-orthogonality of orbitals
causes a problem:
a regular diagrammatic approach for the
description of correlations of electrons
in open many-electron inraatomic shells, for the case when the system is near the
atomic limit and the basis used is non-orthogonal, has not been developed yet. 
We have
not found
 a method that
allows to treat the 
non-orthogonality within the versions of the DT,
developed by Kuramoto\cite{kuramoto} and Grewe\cite{grewedt}, as well as for
other version of the DT, that has been developed by Westwansky and
Pawlikowski\cite{westwansky} and by Zaitsev\cite{zaitsev} (WPZ) for the $s$%
-band Hubbard model (the group $SU(4)$).
The problem is that in the case of a non-orthogonal basis the picture
of interaction cannot be used since we are not able to calculate the time
dependence of the conduction-electron operators $c_{{\bf k}\lambda }(t)$ and
of the $X_n(t)$-operators even for the zero Hamiltonian (18): $c(t)$ and $%
X(t)$ are coupled by the overlap matrix. In the WPZ-technique one more open
question exists. Namely, as has been shown\cite{nikolaev}, compared to the
conventional DT for fermions and bosons, the set of diagrams in the DT for
the s-band Hubbard model is not unique, being dependent on the order in
which different $X$-operators are picked up from the chronologically ordered
product $\langle T\{X_1(\tau _1)...X_m(\tau )...X_n(\tau _n)\}\rangle $ for
Wick's decoupling. In order to avoid under- or double counting, this way of
the formulation of the DT must be complemented by a system of priority, or
hierarchy, for the $X$-operators. The recipe for the $s$-band Hubbard model
is developed. However, it is difficult to use this recipe in practical
calculations for the $d$- and $f$-systems due to very large number of the
electron transitions involved (this number quickly grows with the
 number of
electrons in the shell). Another important moment is connected with 
existence of {\em closed} system of equations for GFs. The latter 
give us an opportunity to introduce the accurate definitions for self-energy 
and vertexes. We shall show below 
that an absence of such definitions in earlier 
theories can lead to erroneous results.
As will be seen, the mentioned difficulties can be resolved 
by making use of a properly generalized method of Baym and Kadanoff\cite
{baym_kadanoff,rucksmr2} where the diagram technique can be constructed via
equations of motion. Within this technique the hierarchy is automatically
established by the choice of the variables, Grassman or boson ones, which
are used in the Hamiltonian of external fields. 

The aims of the present paper can be formulated as follows:\\
1) To derive an
approximate Hamiltonian in a local many-electron representation, which
allows to use as a reference point the LDA Hamiltonian, $H_{LDA}$, for a
description of delocalized electrons, but the many-electron description of
the localized or, almost localized, electrons. To separate from
 the total 
Hamiltonian  the one corresponding to the
generalized Hubbard-Anderson model in order to elucidate 
the microscopical origin of the parameters of this model.
2) To
derive a {\em closed} system of equations for the Green's functions and to
develop a diagram technique, that allows one
to work with a non-orthogonal
basis set which is free from the problem of a hierarchy between Hubbard
operators. \\
3) To derive a few ''common'' approximations for the case of a
generalized Hubbard-Anderson model within a non-orthogonal basis. 
Particularly, to derive the self-consistent equations for 
the set of "Hubbard $U$-s. \\
 4) To
inspect the role of remaining terms of the Coulomb Hamiltonian in formation
of the Hubbard-Anderson model and to 
make an attempt to formulate ''minimal'' model, which
describes formation of the low-energy physics in the SEC system. \\ 
5) We will finally compare our
equations with some of available results of other authors.

It is worth noting that i) although our derivation of perturbation theory is
based on the ideas similar to ones
 used by Ruckenstein and Schmitt-Rink\cite{rucksmr2}
(Schwinger's source theory\cite{schwinger}) the resulting Green's function
technique is closer to the one used by Zaitsev\cite{zaitsev} and Izyumov et
al.\cite{izyumov1} (within "boson" hierarchy) since we are using local
Green's functions as a zero-order limit instead of the itinerant limit used
in ref.24\cite{rucksmr2}; 
ii) we suggest to use the self-consistent LDA potential for generating the set of 
wave functions and eigenvalues as a convenient first step, taking in a proper way 
symmetry and detailed charge distribution of the material of interest, but {\em not} 
the conceptional framework of the density functional theory (DFT). We emphasize 
that we will use a
more or less standard many-body approach, which is
{\em different} from DFT ideology: in our case the expressions for the ground state 
energy, free energy and excitations should be found directly from the 
calculated GFs.

The content of the next sections is the following. In Sec.IIA,  assuming
that the self-consistent solution of the band problem within LDA or 
some other convenient potential with the $ f(d)$-electrons 
treated as core is found, we use a local set of
site-centered non-orthogonal functions for a transformation of the total
Hamiltonian to a form that is
 convenient for our further purposes with
 representation of
non-orthogonal field operators. Then we 
show that the Green's function, defined on 
these non-orthogonal operators, reproduces the LDA problem. In
section IIB we introduce the many-electron representation and 
we derive the generalized Hubbard-Anderson model corresponding to the
secular equation of the LDA problem plus those terms of Coulomb
interaction which are responsible for formation of the $f(d)$/shell 
of the ion.

Sec.III is devoted to the diagram technique itself. We start with a
derivation of commutation relations between many-electron operators of the $%
f $-shell and delocalized electrons. Then with the help of these relations
the equations of motion for the operators of quasi-fermion and quasi-boson
intra-atomic transitions are derived. The latter equations allow us to find
out what type of Schwinger's sources should be introduced into the theory
and to derive a closed system of equations for Green's functions (GFs) in
terms of functional derivatives of generating functional.

In Sec.IV we rewrite the theory in a more compact form, introducing the
self-energies and vertexes, and derive a few approximations. The equation
for the self-energy is written here in terms of functional derivatives,
which is convenient for iterations. After that the equations for GFs within
the "{\em Hubbard-I}" approximation are formulated. 
This approximation corresponds to the vertex equal to zero. 
Next approximations require knowledge of the vertex. Therefore,
we formulate the
equation for the self-energy in terms of the
vertex. This allows
 us to derive a 
{\em mean field} theory, beyond the Hubbard-I approximation, in which the
energies of $f$-transitions are self-consistently renormalized by
{\em kinematic} interactions caused by mixing
interaction and hopping. 
The abovementioned self-consistent equation
for Hubbard $U$-s is composed of these equations. At last, by
making use of transformation to a description in
terms of an effective field, we formulate the equations for the "{\em random
phase approximation}" (RPA), which describe the self-dressing of kinematic
interactions.

In Sec. V we consider some of remaining terms of Coulomb interaction.
 The purpose of this section is
two-fold: i) to generalize the equations and method 
for the case of Coulomb interaction; 
ii) to discuss the correction
 from Coulomb interaction  of $f$-electrons and conduction electrons 
to the self-consistent equation for
''Hubbard $U$-s''. The paper is completed by Sec.VI 
where discussions and conclusions are given. Applications of the
presented technique to concrete problems will be given in forthcoming 
papers~\cite{lundin}.

\section{Hamiltonian  and The Hubbard-Anderson Model}
\subsection{Non-orthogonal representation}
In order to have a possibility to correct the band structure and the ground
state properties, obtained by the LDA calculation in a regular way
we first introduce
the {\em exact} Hamiltonian of the system in second quantization
using some complete set of wave functions, for example, plane waves.
One can expect that the many-body corrections to the spectrum and wave 
functions of conduction electrons will be small\cite{zein1,zein2}, 
which motivates to
start evaluations with the LDA Hamiltonian. We note, however, 
that the construction described below allows to start from {\em any} convenient 
single-electron potential (e.g. a self-interaction corrected one).
Let us  
separate the LDA part of full Hamiltonian and
develop a many-particle perturbation theory for the
Green's functions over the deviation between the exact and LDA Hamiltonians.
 
The corrections 
which we are going
to discuss come mainly from on-site-localized Coulomb repulsion in
the open $d$- or $f$-shell. From a technical point of view 
a serious disadvantage of local representation is the
non-orthogonality of wave functions, centered on different sites. The
corresponding diagram technique should necessarily include overlap
matrices.
The same reason, however, turns out to be an advantage since it allows 
to easily separate
the on-site interactions from the contributions coming from other sites.
In the present paper we restrict ourselves with a non-relativistic
consideration, i.e.,  the total Hamiltonian of the
system we shall start with is: 
\begin{eqnarray}
H &=&(T+H_{en})+H_{ee}+H_{nn} \nonumber \\
 &=&\sum_{\sigma} \int dr\psi _\sigma ^{\dagger }(r)\left(
\frac{p^2}{2m}-\sum_j\frac{%
Z_je^2}{|r-R_j|}\right) \psi _\sigma (r)  \nonumber \\
&+&\frac 12 \sum_{\sigma} \sum_{\sigma '}
\int dr\int dr^{\prime }\psi _{\sigma ^{\prime }}^{\dagger
}(r)\psi _{\sigma }^{\dagger }(r^{\prime })\frac{e^2}{|r-r^{\prime
}|}\psi _{\sigma ^{\prime }}(r^{\prime })\psi _\sigma (r)+\frac
12\sum_{j\neq j^{\prime }}\frac{Z_jZ_{j^{\prime }}e^2}{|R_j-R_{j^{\prime
}}|}%
,
\end{eqnarray}
where the first term includes the kinetic energy and the 
electron-nuclei interaction, the second term the electron-electron 
interaction and the last term the nucleus-nucleus interaction.
It should be emphasized that the electron operators 
$\psi _\sigma (r)$ are constructed from some
complete set of functions and, therefore, the field operator $\psi $ 
and, correspondently, Hamiltonian $H$, are exact.
Let us now introduce  the set of wave functions,
$\phi _{k\lambda }(r)$,
which
are the solution to the LDA band problem:
\begin{equation}
\lbrack \frac{p^2}{2m}+v_{LDA}(r)]\phi _{k\lambda }(r)=\epsilon _{k\lambda
}\phi _{k\lambda }(r),
\end{equation}
where $v_{LDA}(r)$ is the LDA effective potential, $ \epsilon
_{k\lambda}$ the Kohn-Sham eigenvalue, and $\lambda $ is a band index. 
We will use below the solutions of this equation only as an almost 
complete set, on the one hand, 
in order to reformulate our full problem into convenient 
for our aims representation , and in order to express the 
energies of conduction-electron quasi-particles
 in terms of the Kohn-Sham eigenvalues, on the other hand.  
Since the set $\phi _{k\lambda }(r)$ is orthonormalized,
corresponding annihilation operators can 
be obtained simply by projection from the full field operator,
$ \hat{a}_{k\lambda \sigma} = 
\int dr\phi _{k\lambda }^{*}(r)\hat{\psi} _{\sigma}(r)$.
Next, we separate the full field operator into two
parts,
\begin{equation}
\psi _\sigma (r)=\psi _\sigma ^{(LDA)}(r)+(\psi _\sigma (r)-\psi _\sigma
^{(LDA)}(r))=\psi _{1\sigma }(r)+\psi _{2\sigma }(r),
\end{equation}
where
\begin{equation}
\psi _{1\sigma }(r)=\psi _\sigma ^{(LDA)}(r)=\sum_{k\lambda }\phi
_{k\lambda
}(r)\hat{a}_{k\lambda }.
\end{equation}
Let us now insert $\psi =\psi _1+\psi _2$ into the exact Hamiltonian (1), and 
of 
all terms which arise we take into account only the ones
which contain only $\psi_1$. We stress that the
approximation made here should be kept in mind, since for a
description of some experiments, for instance the ones 
involving large-energy perturbation, 
(e.g. photoelectron spectroscopies) this set may be insufficient. 
For these cases a
consideration within this reduced Hamiltonian would miss essential physics
of the phenomena (essential additional terms in the Hamiltonian).

The obvious, strict way to construct the
standard weak-coupling perturbation theory which takes
into account the effects of Hubbard $U$ and starts from 
 $H_{LDA}$ as the zero Hamiltonian is to represent $H$ in the form:
\begin{equation}
H=(T+H_{en}+V_{LDA})+H_{nn}+(H_{ee}-V_{LDA})=H_0+H_{int}
=H_0+H_{int}.
\end{equation}
It worth noting that although we use $V_{LDA}$ one can start with 
any convenient potential. This opportunity may be especially important for 
consideration of problems with constrained boundary conditions 
(like, for example. quantum dots).
In consideration below we will work, however, with the approximate 
Hamiltonian, having neglected all the $\psi_2$-dependent terms:
\begin{eqnarray}
H &=&\int dr\psi _{1\sigma }^{\dagger }(r)\left( \frac{p^2}{2m}%
+v_{LDA}(r)\right) \psi _{1\sigma }(r)  \nonumber \\
&+&\left( H_{nn}+H_{ne}+H_{ee}-\int dr\psi _{1\sigma }^{\dagger
}(r)v_{LDA}(r)\psi _{1\sigma }(r)\right),
\end{eqnarray}
with conventional notations.
 
The further steps are the following.
 The first one is to construct expansion of field operators in 
the formalism of secondary quantization
in such a way that the equation of motion for the Matsubara (or,
Baym-Kadanoff) GF in the $\chi _{jL}$-representation reproduced Eq.10 
{\em automatically}. The latter means that we have to define the field 
operators
in a proper way, then, using this definition, to calculate the commutation
relations between them and rewrite the full secondary quantized Hamiltonian
in this representation. Then we will use these fermion operators for
constructing the many-electron operators, which describe SEC, while the
elementary commutation relations for these fermions will be used for
calculation of the commutation relations between different complex
combinations of them, which arise in the theory. In the second step we will
separate the Hubbard-Anderson model from the full Hamiltonian. This should
help to clarify the microscopical origin of the parameters of these models,
and the nature of remaining Coulomb interactions. In third step we will
generalize the Baym-Kadanoff formalism to the case of these complex
non-Fermi and non-Bose operators which are generated by the equations of
motion.

In cellular methods of AIBST the  Bloch functions $\varphi _{k\lambda }(r)$ of the
band $\lambda $ are often represented in the form of expansion over site-centered
functions $\chi _{jL}(r)\equiv \chi _L(r-R_j)$. Here $L$ is complex index: $%
L=\{l,m_{l,}\sigma \}$. The secular equation in this representation has the
form: 
\begin{equation}
(EO_{jL,j_1L_1}-h_{jL,j_1L_1})\chi _{j_1L_1}(r)=0;
\end{equation}
here $O_{jL,j_1L_1}\equiv \langle \chi _{jL}|\chi _{j_1L_1}\rangle $ is
overlap matrix. The functions $\varphi $ and $\chi $ are connected by
relation 
\begin{equation}
\varphi _{k\lambda }(r)=\sum_{\alpha \;jL}\tilde{u}_\alpha ^{*\lambda
}(k)Z_L^\alpha (k)e^{ikR_j}\chi _L(r-R_j)\equiv \sum_{\;jL}u_{k\lambda
}^{jL}\chi _{jL}.
\end{equation}
The coefficients $Z$ here come from the Choleski's factorization of the
overlap matrix, $O=\bar{Z}Z$, while the Hamiltonian $\tilde{h}$ of
corresponding secular problem, 
\begin{equation}
\tilde{h}_{\alpha \alpha ^{\prime }}(k)\tilde{\phi}_{k\alpha ^{\prime
}}(r)=E_\alpha (k)\tilde{\phi}_{k\alpha }(r)
\end{equation}
is connected with the matrix element $h$ as follows: 
\begin{equation}
\tilde{h}_{\alpha \alpha ^{\prime }}(k)=\sum_{LL^{\prime }}[\bar{Z}%
^{-1}(k)]_L^\alpha h_{LL^{\prime }}(k)[Z^{-1}(k)]_{L^{\prime }}^{\alpha
^{\prime }}.
\end{equation}
At last, the equation for corresponding to the Eq.7 fermion GF $F$ has the
form: 
\begin{equation}
(EO-h)_{jL,j^{\prime \prime }L^{\prime \prime }}F_{j^{\prime \prime
}L^{\prime \prime },j^{\prime }L^{\prime }}=\delta _{jj^{\prime }}\delta
_{LL^{\prime }}.
\end{equation}
	Having known the expression for $\varphi $ it is easy to find definition for 
$a_{jL}$. Let us insert Eq.8 into expansion for the field operator (4):
\begin{eqnarray}
\psi _{1\sigma }(r) &=&\sum_{k\lambda }[\sum_{\alpha \;jL}\tilde{u}_\alpha
^{*\lambda }(k)Z_L^\alpha (k)e^{ikR_j}\chi _L(r-R_j)]a_{k\lambda }= 
\nonumber \\
&&\sum_{\;jL}\chi _{jL}[\sum_{k\lambda }[\tilde{u}_\alpha ^{*\lambda
}(k)Z_L^\alpha (k)e^{ikR_j}a_{k\lambda }]\equiv \sum_{\;jL}\chi _{jL}a_{jL},
\end{eqnarray}
\emph{i.e.}, the operator of destruction of electron on the orbital $(jL)$
should be defined as the following combination of normal  band operators:
\begin{equation}
a_{jL}=\sum_{k\lambda }u_{k\lambda }^{jL}a_{k\lambda }.
\end{equation}
It follows from this definition that 
the anticommutator of these operators is equal to the inverse
of the overlap matrix:
\begin{equation}
\{a_{jL},a_{j'L'}^{\dagger }\}=(O^{-1})_{jL,j'L'},
\end{equation}
\[
O_{jL,j^{\prime }L^{\prime }}=\langle \chi
_{jL}|\chi _{j^{\prime }L^{\prime }}\rangle .
\]
Since in practical calculations only finite number of the functions $\chi_{jL}$
can be used,  it is clear that the accuracy of this 
approximation is determined by the possibility to represent $\delta $-function 
in terms of the functions $\chi$, namely, 
\[
\delta (x-x'){\approx}
\sum_{jl,j'L'}\chi_{jL}(x)O^{-1}_{jL,j'L'}\chi_{j'L'}^{*}(x'),
\]
The terms containing $\psi_2$ can alway be taken into account additionally.
From a technical point
of view the case of $f$-electrons differs from the case of $d$-electrons
only by the number of correlated orbitals involved, at least in the initial
steps. Let us now discuss the case of $f$-electrons and use
for the operator $a_{jL}$ with $L=\{l=3,m_l,\sigma \}$ the notation $%
f_{jm_l\sigma }$ , whereas all the non-$f$'s are called $c_{jL}$,
 $a_{jL}=\delta _{L,(3,m_l\sigma )}f_{m_l\sigma
}+(1-\delta _{L,(3,m_l\sigma )})c_{jL})$  . Then the LDA part of
the Hamiltonian can be written in the form:
\begin{eqnarray}
H_{LMTO} &=&\sum_{k\lambda }\epsilon_{k\lambda \sigma }
a_{k\lambda \sigma }^{\dagger }a_{k\lambda
\sigma }=\sum_{j,L,j^{\prime },L^{\prime }}H_{jL,j^{\prime }L^{\prime
}}c_{jL}^{\dagger }c_{j^{\prime }L^{\prime }}+\sum_{j,m_l,\sigma }\epsilon
_f^0f_{jm_l\sigma }^{\dagger }f_{jm_l\sigma }  \nonumber \\
&+&\sum_{jm_l\sigma j^{\prime }m_l^{\prime }}t_{jm_l,j^{\prime }m_l^{\prime
}}^\sigma f_{jm_l\sigma }^{\dagger }f_{j^{\prime }m_l^{\prime }\sigma }
\nonumber \\
&+&\sum_{j,L,j^{\prime }m_l,\sigma }[H_{jL;j^{\prime }\sigma
}c_{jL}^{\dagger }f_{j^{\prime }m_l\sigma }+H_{jm_l\sigma ;j^{\prime
}L}^{*}f_{jm_l\sigma }^{\dagger }c_{j^{\prime }L}].
\end{eqnarray}
Here the hopping matrix element and the position of the center of the $f$-band are
given by\cite{zein2}
\begin{eqnarray}
t_{j,m_l,j^{\prime },m_l^{\prime }}^\sigma  &=&\sum_ke^{ik(R_j-R_{j^{\prime
}})}[H_{m_l,\sigma ,m_l^{\prime },\sigma }(k)-\delta _{m_l,m_l^{\prime
}}\epsilon _0^f]  \nonumber \\
\epsilon _0^f &=&\sum_kH_{m_l\sigma ,m_l\sigma }(k).
\end{eqnarray}
In the case when in {\em ab initio} band structure calculations (AIBST) 
the $f$-electrons have been treated as core electrons, $\epsilon_0^f$ is just 
single-electron level in the "LDA"-atom, since in this case hopping and mixing 
are set equal to zero.
Clearly, we are not allowed to diagonalize this Hamiltonian directly, since we 
deal with a non-orthogonal basis set. 

Let us now write down 
the equation of motion for the Matsubara's Green's function (GF)
\beq
G_{jL,j^{\prime }L^{\prime }}(i\omega )=-i\langle
T[a_{jL}(t)a_{jL}^{\dagger }(t^{\prime })]\rangle _{\omega _n},
\eeq
 in this representation. We find that in the zero order approximation 
this equation has the desired  form
\begin{equation}
\lbrack i\omega _nO_{jL,j^{\prime }L^{\prime }}-H_{jL,j^{\prime }L^{\prime
}}]G_{jL,j^{\prime }L^{\prime }}(i\omega _n)=\delta _{j,j^{\prime }}\delta
_{L,L^{\prime }},
\end{equation}
since the expression in the square brackets exactly 
reproduces the matrix of the LDA secular problem.  

In order to avoid confusion we remind that, as follows from
Eqs. (11?) and (14?), a straightforward diagonalization
of the matrix $H_{jL,j^{\prime }L^{\prime }}$ 
 would lead to
eigenvalues which do not coincide with the band-structure ones since the
functions $\chi _{jL}$ are not orthogonal to each other, $\langle
jL|j^{\prime }L^{\prime }\rangle =O_{jL,j^{\prime }L^{\prime }}\neq \delta
_{jj^{\prime }}\delta _{LL^{\prime }}$ (!), and the overlap matrix $O$ is
hidden in the commutation relations (14).

The other warning is the following:
 although the poles of the GF in Eq.11 coincide 
with the eigenvalues of the Kohn-Sham secular problem (what we 
want to have here), one cannot obtain 
the total energy $E_{tot}$ from the standard formula via GF and its 
self-energy. Therefore, the 
LDA potential in the equation for GF cannot be considered as self-energy.  

The density of 
charge $\rho_{\sigma}(x)$ which enters the $H_{LDA}$, can be found, 
as usual, from GF (10), 
\[
\rho_{\sigma}(x) = \langle \psi_{\sigma}^{\dagger}(x)\psi_{\sigma}(x)
\rangle \approx 
\langle \psi_{1,\sigma}^{\dagger}(x)\psi_{1,\sigma}(x)\rangle = 
\lim_{t'\rightarrow t}
\sum_{jL,j'L'} \chi^{*}_{jL}(x)\chi_{j'L'}(x) 
G_{jL,j'L'}(t,t').
\]
As seen from (5-6) the perturbation theory from weak-coupling limit 
can be constructed with respect to deviations $(H-H_{LDA})$. In the case of 
expansion near atomic limit this is not possible.
 
\subsection{Many-electron representation and periodical Anderson model. }
A guiding idea on
 how to introduce this representation can be obtained from the 
known considerations of the $s$-band Hubbard model.
Namely, the decomposition of the fermion
operator $c_{\uparrow }=c_{\uparrow }\cdot
(1-\hat{n}_{\downarrow })+c_{\uparrow
}\cdot \hat{n}_{\downarrow }$, is 
produced by the Hubbard repulsion which makes
energies of the transitions $c_{\uparrow }(1-\hat{n}_{\downarrow })$ and $%
c_{\uparrow }\hat{n}_{\downarrow }$ different. The physical meaning of this
decomposition is that every electron should ''know'' about populations of
other orbitals at the same site. It is clear from this speculation that, as
an example, for
an $f$-electron in the orbital $m_l=-3,\sigma =\uparrow $ we have to
take into account in the corresponding decomposition as sum of {\em all}
possible products:
\begin{eqnarray}
f_{-3,\uparrow }
&=& f_{-3,\uparrow } \cdot (1 -\hat{n}_{-2,\uparrow})
 \cdot (1-\hat{n}_{-1,\uparrow })\cdot ... \cdot
 (1 -\hat{n}_{+3,\downarrow }) \nonumber \\
&+& f_{-3,\uparrow } \cdot \hat{n}_{-2,\uparrow}
 \cdot (1-\hat{n}_{-1,\uparrow })\cdot ... \cdot
(1 -\hat{n}_{+3,\downarrow }) \nonumber \\
&+& . . . . . . . . . . . . . . . . .  . . . . . . . . . . . . . .
. . . . . . . . . . . . . . .  \nonumber \\
&+& f_{-3,\uparrow } \cdot  \hat{n}_{-2,\uparrow}
 \cdot \hat{n}_{-1,\uparrow }\cdot ... \cdot \hat{n}_{+3,\downarrow }
\end{eqnarray}
This transformation is, obviously, identical.
In the next step, inserting this into the total Hamiltonian we take into
account correlations by making use of the properties of the Fermi operators:
$f^2 = (f^{\dagger})^2 = 0$, $\hat{n}^2 = \hat{n} $. The sums of the
diagonal terms in the Hamiltonian,
which have the same operator structure, give energies of corresponding
many-electron states in the zero approximation.
This step is still
exact. However after this transformation 
returning to the form (6) and (15) of the Hamiltonian
becomes somewhat difficult.  
The fermion GF in the atomic limit now acquires the form\cite{hubb1}:
\begin{equation}
{\cal F}_{\nu}^{(at)} (i\omega )\equiv 
\langle T f_{\nu} f_{\nu}^{\dagger} \rangle |_{i\omega } =
\sum _{\{ \Gamma_n \}} \frac{\langle |\Gamma_n |f_{\nu }|\Gamma_{n+1}\rangle
 |^2
(N_{\Gamma_n}+N_{\Gamma_{n+1}})}{i\omega -(E_{\Gamma_{n+1}}-E_{\Gamma_{n}})
},
\end{equation}
where the spectral weights, say, for $d$-electrons,
 contain non-decoupled correlators, like
\begin{equation}
N_{\Gamma_{n=1,\nu}}\equiv
\langle (1-\hat{n}_1)(1-\hat{n}_2)...(1-\hat{n}_{\nu-1}) \hat{n}_{\nu}
(1-\hat{n}_{\nu +1})...(1-\hat{n}_{10}) \rangle
\end{equation}
and so on. Here $E_{\Gamma_n}$  is energy of $n$-electron 
configuration $\Gamma_n$ of $d(f)$-ion.
The transitions, described by different terms
 in this expansion, have {\it different energies}, and the energy
 separation between different terms is often so large, that their
 contribution to almost all observables is negligible. These terms
 can be neglected without changing the physics. The new 
reduced Hamiltonian
 can still be treated within a single-electron approximation, {\it
 if we put all
 population numbers} of the $f$-orbitals {\em equal to integer numbers}
 (0 or 1). The
procedure, described above, actually, gives a formal derivation
for the approach, exploited usually in band structure calculations
for materials with well-localized
 $4f$-electrons 
:   $f$-electrons are placed in the core 
 and their occupation number is taken to be equal
to an integer number.  
For $d$-s and $5f$-s, and even for the first part
of $4f$-series
 this procedure is often not sufficiently accurate.
Thus, if either mixing or hopping are allowed, or
 some of the many-electron states have close energies,
 we come to the case of strong electron correlations. The key question is,
 of course, how large is Hubbard $U$ in  a particular
compound with $d$- or $f$-electrons under interest. It should be noted that
 the definition of
this magnitude and the way of calculation of it in different papers 
is different\cite{AZA,zein1,zein2,hubb1,cooper2,gunn,mcmahan}.
Hubbard in his first paper\cite{hubb1}, i.e. just the matrix 
element,
which stands in the Hamiltonian. Note, however, 
that if it is convenient, any 
"preliminary" value can be chosen for $H_0$, since it will be 
taken into account later via $H-H_0$.
 Sometimes the  shift $U \rightarrow U_{model}$ is desirable, 
since this
decreases the strength of perturbation.
Since we started from an exact Hamiltonian, we have an opportunity
to calculate $U_{model}$ in terms of Coulomb interactions, existing 
in the system. We will return to this question after including Coulomb 
interaction into our diagram technique.

As we noted above, an alternative way to transform the
Hamiltonian to the many-electron form, 
which describes the correlated motion of $f$-electrons in the 
case of Russel-Saunders type of coupling
used by Irkhin\cite{irkhiny}. 
After the transformation is done, the terms in the expansion,
which clearly give small contribution
can be neglected (for example, one can hardly
expect that the Ni ion can be found in $5-$ or $6-$valent states). 
Below we will use an equivalent, but more 
convenient way, compared to the above described method, 
to introduce the many-electron representation.

In the following part of this section we shall perform the following three steps: 
1.  Introduce a many-orbital representation. 
2. Rewrite our Hamiltonian in this representation. 
3. Separate out the main terms of the perturbation Hamiltonian.
These steps will allow us to derive microscopically the
multi-orbital periodic Hubbard-Anderson model for this 
particular representation. 

Let us now
construct the orbital representation for atomic states.
 It is convenient to use the following
short notations for the $f$-orbitals. The orbitals with spin ''up'' starting
from $m_l=-3$ to $m_l=3$ we shall label by $\mu =1,2,...,7$, whereas the
ones with spin ''down'' from $m_l=-3$ to $m_l=3$ by $\mu =8,...,14$. For
instance, $%
|m_l=-2,\downarrow \rangle \equiv |\mu =9\rangle $. The states of an $f$-ion
with different number of electrons in the $f$-shell will be described in the
orbital representation as $|\Gamma _0\rangle \equiv |0\rangle $, $|\Gamma
_\mu \rangle \equiv |\mu \rangle =f_\mu ^{\dagger }|0\rangle $, $|\Gamma
_{\mu ,\nu }\rangle \equiv |\mu ,\nu \rangle =(\theta _{\mu \nu }f_\mu
^{\dagger }f_\nu ^{\dagger }|0\rangle +\theta _{\nu \mu }f_\nu ^{\dagger
}f_{mu}^{\dagger }|0\rangle $ and so on. Here $\theta _{\nu \mu }=1$ if $\nu
<\mu $ and $\theta _{\nu \mu }=0$ if $\mu <\nu $. In other words, the
following convention about the order of orbitals in any many-particle state $%
|\Gamma _n\rangle \equiv |\mu _1,\mu _2,...,\mu _n\rangle \equiv f_{\mu
_1}^{\dagger }f_{\mu _2}^{\dagger }...f_{\mu _n}^{\dagger }|0\rangle $ is
taken: $\mu _1<\mu _2<...<\mu _n$ always holds.
  The advantage of the orbital
representation is that it presents the simplest possible way to transform
the Hamiltonian to the atomic representation and gives a very simple form for zero
fermionic Green's functions. Another advantage is that 
almost all rare earths have single-determinant ground state.
On the other hand, the set of wave functions $|\Gamma _n \rangle $ 
is a complete set for any fixed $n$, therefore, more complicated states 
can be easily combined of these wave functions. This representation can be 
good starting point for description of $d$-electrons in $3d$-materials 
or for $5f$-electrons in actinides. For the case of $4f$-electrons the 
Russel-Saunders coupling should be used, {\em i.e.} many-electron states 
have to be constructed according to the scheme (see recent review by
 Irkhin and Irkhin\cite{irkhiny} and references therein):
\begin{equation}
|\Gamma_n \rangle = A^{\dagger}_{\Gamma_n}|0\rangle =
\frac{1}{\sqrt n}\sum_{\gamma, \Gamma_{n-1}}G^{\Gamma_n}_{\Gamma_{n-1}}
C^{\Gamma_n}_{\Gamma_{n-1}}f^{\dagger}_{\mu}A^{\dagger}_{\Gamma_{n-1}}
|0\rangle
\end{equation}
where $G^{\Gamma_n}_{\Gamma_{n-1}}\equiv 
G^{S_n L_n \alpha _n}_{S_{n-1}L_{n-1}\alpha _{n-1}}$
are fractional parentage coefficients ($\alpha$ is Rakah's seniority), and 
 $C^{\Gamma_n}_{\Gamma_{n-1}}\equiv
C^{L_n,M_n}_{L_{n-1}M_{n-1}lm}C^{S_n,\mu_n}_{S_{n-1}\mu_{n-1},1/2}$ are 
Clebsch-Gordan coefficients. For actinides $LS$-coupling should be used (with 
spin-orbit coupling and other relativistic terms added). For $3d$-materials, 
probably, crystal-field combinations are a proper choice. Thus, since
 different materials require different combinations of wave functions 
(or, operators), below we will use general notation for 
expansion of $f$-operators, so that a combination needed for diagonalization 
of $H_{ion}$,
\begin{eqnarray}
H_{ion}&=&\sum_{j\mu}\epsilon^0_{j\mu}a^{\dagger}_{j\mu}a_{j\mu}  \nonumber \\
&+& \frac{1}{2}\sum U_{\mu_1 \mu_2 \mu_3 \mu_4}
a^{\dagger}_{j\mu_1}a^{\dagger}_{j\mu_2}a_{j\mu_3}a_{j\mu_4} -
\sum v^{(LDA)}_{\mu_1\mu_2}a^{\dagger}_{j\mu_1}a_{j\mu_2},
\end{eqnarray}
is assumed to be used. Here
the $\epsilon_{\mu}$ are core and $f$-levels from LDA-calculation. 
Note, that in the case of Russel-Saunders coupling
 the Coulomb part of interaction  can be rewritten in terms 
of Slater integrals and conveniently separated into the terms 
describing spin-spin, orbital-orbital interactions, etc..\cite{irkhiny}.

Let us now introduce a description in terms of the Hubbard's
operators $X^{\Gamma \Gamma^{\prime }}
=|\Gamma \rangle \langle \Gamma ^{\prime }|$. In order
to see explicitly the process involved in the particular term of
perturbation theory, it is convenient to use different notations for
the diagonal and non-diagonal operators, distinguishing also quasi-fermionic and
quasi-bosonic operators:
For diagonal operators we use the notation: $\hat{h}^\Gamma \equiv |\Gamma \rangle
\langle \Gamma |$. For non-diagonal Fermi-like we use the notation: $%
X^{\Gamma _n,\Gamma _{n\pm 1}^{\prime }}\equiv |\Gamma _n\rangle \langle
\Gamma _{n\pm 1}^{\prime }|$. For non-diagonal Bose-like, with no change in the number of
particles, we use $Z$: $Z^{\Gamma
_n,\Gamma _n^{\prime }}\equiv |\Gamma _n\rangle \langle \Gamma _n^{\prime }|$
and for Bose-like operators where the number of particles changes by an even
number  we use $Z^{\Gamma _n,\Gamma
_{n\pm 2}}\equiv |\Gamma _n\rangle \langle \Gamma _{n\pm 2}|$.
Below we shall use also shorter notations for the transitions; for the Fermi-like
transitions we denote
 $X^a$ with $a\equiv [\Gamma_n,\Gamma_{n + 1}^{\prime }]$ ,
 and for Bose-like transitions we use capital as super scripts: $Z^\xi$
 with $\xi \equiv [\Gamma_n,\Gamma_{n}^{\prime}]$ (including
 $h^{\Gamma} \equiv Z^{\Gamma,\Gamma}$ ), or
 $A\equiv [\Gamma_n,\Gamma_{n \pm 2}^{\prime}]$ .
 The transitions in opposite direction will be denoted
 by a bar: $\bar{a}=[\Gamma _{n+1}^{\prime },\Gamma _n]$. We also
introduce special notations for the anticommutator of $X$-operators and its
average:
\begin{equation}
\{X_j^a,X_j^b\}=\hat{Q}_j^{ab}\;\;\mbox{ and}\;\;\;\langle
\{X_j^a,X_j^b\}\rangle =Q_j^{ab}\mbox{ with}\;\;\langle \hat{A}\rangle =%
\frac{TrAe^{-\beta H}}{Tre^{-\beta H}}.
\end{equation}
Here $\beta =1/T$, and $T$ is the temperature in units $k_B=1$. The fermion operator
can be written in this representation as
\begin{equation}
\hat{f}_{i\gamma }=(f_\gamma )^aX_i^a=
\sum_{\gamma_1} f^{\gamma}_{0,\gamma_1} X_i^{0,\gamma_1 }
+\sum f_{\gamma^{\prime },\Gamma }^\gamma X_i^{\gamma ^{\prime },\Gamma }
+\sum f_{\Gamma^{\prime },\Lambda }^\gamma X_i^{\Gamma ^{\prime },\Lambda }
+...,
\end{equation}
where  
$\gamma $ describes single-electron states, $\Gamma $ and $\Lambda $
stand for two- and three-particle states correspondingly. The matrix
elements $f^{\gamma}_{0,\gamma_1} = (0|\hat{f}_{\gamma} |\gamma_1 )$,
 $f_{\gamma ^{\prime }\Gamma }^\gamma =(\gamma ^{\prime }|\hat{f}%
_\gamma |\Gamma )$, $f_{\Gamma ^{\prime }\Lambda }^\gamma =(\Gamma ^{\prime
}|\hat{f}_\gamma |\Lambda )$ of the fermion operator $\hat{f}_\gamma $ (using 
many-electron states $|A\rangle =|\gamma \rangle ,|\Gamma \rangle ,|\Lambda
\rangle $) describe selection rules
and reflect the representation chosen for description of $f(d)$-states:
 $j-j$-, Russel-Saunders coupling 
or the orbital one. In the case when the $f$-(non-$f$) overlap integrals are neglected,
these coefficients acquire the  simplest form  in the
orbital representation. For example, $\langle \gamma ^{\prime }|f_\gamma
|\Gamma \rangle \neq 0$ only if $|\Gamma \rangle =|\gamma ^{\prime }\gamma
\rangle $. Therefore, in the expansion (25) the sum over double-electron
states $\Gamma $ is actually, $\sum_{\gamma ^{\prime }}X^{\gamma ^{\prime
},(\gamma ^{\prime },\gamma )}(1-\delta_{\gamma ^{\prime },\gamma })$. In the
last term in (Eq.25) the summation over three-particle states $\Lambda $
includes only the states with $\Lambda =(\Gamma ^{\prime },\gamma )$.
Thus, the sum over $\Gamma ^{\prime }$ contains only those of states $%
\Gamma ^{\prime }$, which do not enter the second term of Eq.(25). 
For the rare earth ions the $f$-electron coefficients of expansion 
over $X$-operators can be expressed via Clebsch-Gordan and fractional 
parentage coefficients\cite{judd,irkhiny}, 
$(f_\gamma ^{\dagger} )^{a=[\Gamma _{n-1},\Gamma _n]} = 
\sqrt{n}G_{\Gamma _{n-1}}^{\Gamma _{n}}C_{\Gamma _{n-1},\gamma}^{\Gamma _{n}}$.
However, when either $f$-orbitals of different sites, or $f$- and (non-$f$)-orbitals, 
are not orthogonal to each other, these coefficients contain inversed overlap matrices.
The Hubbard's representation, taking into account
correlations, builds into the theory the physical fact that in a system
with SEC around atomic limit, the single electron transition,
 described by a fermion operator, {\it does not exist} as a  
simple one-electron excitation.
Instead it is split into many intra-ion transitions. Formally this
expansion includes the transitions between {\em all} states with $(n+1)$and $%
n$ electrons. However, as was pointed out by Hubbard\cite{hubb1}, the terms
with large deviation of the population number $n_f$ from its equilibrium
value are separated by a very large energy gap from the ground state and
therefore, if they even exist at all, they do not give contribution to
the formation of a crystal (and even liquid). \\ 
Let us change every $f$-operator
in the total Hamiltonian by its expression in terms of $X$-operators
(Eq.25) and separate the
intra atomic part. 
The terms of Hamiltonian which contain more than one Hubbard operator,
belonging to the same ion, should be simplified according to the 
multiplication rule for Hubbard's operators, 
$X^{pq}_jX^{rs}_j=\nu_{qr}X^{ps}_j$, what takes strong interactions 
into account algebraically (here $\nu_{qr}=\langle q|r \rangle$ is sum determinants 
consisting of  matrix elements of corresponding to these states inversed overlap 
matrices; $\nu_{qr} \neq \delta_{qr}$ if $\hat{O}^{-1} \neq \hat{I}$ . 
After that we have to collect the terms which have 
the same operator structure. Performing this transformation with 
the LDA-part of the Hamiltonian (6), and adding the $f-f$-Coulomb single-site
interaction, we come to the first-step" Hubbard-Anderson model:
\begin{equation}
{\cal H}_0=\sum_{i,\Gamma }\tilde{E}_\Gamma ^0h_i^\Gamma +\sum_{jL,j^{\prime
}L^{\prime }}H_{jL,j^{\prime }L^{\prime }}c_{jL}^{\dagger }c_{j^{\prime
}L^{\prime }},
\end{equation}
\begin{equation}
{\cal H}_{int}^{PHAM}=\sum t_{jj^{\prime }}^{\bar{a}b}X_j^{\bar{a}}
X_{j^{\prime }}^{b}+
\sum W_{jL,j^{\prime }a}c_{jL}^{\dagger }X_j^a+H.C..
\end{equation}

 The matrix elements of the Hamiltonian are discussed in Appendix A.\\

We need also the expression for the total number of electrons in this
representation:
\bear
\hat{N}_e &=  O_{j_1L_1,j_2L_2}c^{\dagger}_{j_1L_1}c_{j_2L_2} +
 O_{j_1L_1,j_2M_2}(f_{M_2})^a c^{\dagger}_{j_1L_1}X_{j_2}^a \nonumber \\
 &+  O_{j_1M_1,j_2L_2}(f_{M_1}^{\dagger})^{\hat{a}}
 X^{\hat{a}}_{j_1}c_{j_2L_2} +
 (\hat{n}_M)_{\Gamma,\Gamma}h^{\Gamma}_j \nonumber
 \\ &+ (1-\delta_{j_1j_2})(f_{M_1}^{\dagger})^{\hat{a}}
  O_{j_1M_1,j_2M_2} (f_{M_2})^{b} X^{\hat{a}}_{j_1}X_{j_2}^b.
\eear
As seen in the equations above, due to the  non-orthogonality between states, 
part of the electrons are in mixed states.

Now our Hamiltonian is fully defined, i.e. 
the operator structure is derived 
and a recipe for a calculation of all matrix elements has been given.
This allows us to continue to the next step, {\em i.e.} to the 
construction of the diagram technique.
The reason why we call the Hamiltonian (26-27) 
as the "first-step" one is that remaining 
terms of Coulomb interaction renormalize these parameters. We will return 
to this question later.  First we 
will consider the Hubbard-Anderson Hamiltonian. 
As we said, due to very large number of terms in 
full Hamiltonian we will construct
first the diagram technique for the Hubbard-Anderson Hamiltonian, Eqs.
(26),(27). After that we will return to Coulomb interaction.

\section{ Diagram technique }

\subsection{ Commutation relations}

Since our aim here is to calculate the 
spectra and spectral
weights of the quasi-particles by means of Green's
functions,  the commutation relations between the operators involved 
is a necessary ingredient. 
Let us start from the fermion-like excitations and 
derive the equations of motion $i\partial _tA=[A,H]$ for
the operators $A=c_{jL},X_{j^{\prime }}^a$ for the case when these operators
for different sites neither anticommute with each other nor commute with $%
h_\Gamma $. The correlations here manifest themselves via commutation
relations on the site, to which the $f$-electrons under discussion belong.
The results of 
a commutation $[X_{j^{\prime }}^{a_1},h_{j^{\prime }}^\Gamma ]
$, (which  gives two Fermi-like $X$-operators), and 
an  anticommutation $%
\{X,X^{\dagger}\}$, (which gives Bose-like operators $h,Z$)
 are properties of the
particular algebra used. It is convenient to have this information, written
in terms of the structure constants of the correspondent algebra:
\beq
\{X^{a},X^{\bar{b}}\}=\varepsilon _\Gamma ^{a,\bar{b}}h^\Gamma
+\varepsilon _A^{a,\bar{b}}Z^A+\varepsilon _{\bar{A}}^{a,\bar{b}}Z^{%
\bar{A}} \equiv
\sum_{\xi = \Gamma,A,\bar{A} } \varepsilon _{\xi}^{a,\bar{b}}Z^{\xi},
\eeq
\beq
\lbrack X^a,h^\Gamma ]=\varepsilon _{a_1}^{a,\Gamma }X^{a_1}.
\eeq
When the interaction between $X$- and $c$-subsytems is switched off, in
this particular (orbital) representation
the constants $\varepsilon $ can have values $  0, 1, -1$ and the 
upper indices
show for which operators the (anti)commutator  is calculated, 
while the lower index provides the right choice of all Hubbard operators for the 
r.h.s. of the (anti)commutator. In the case when $X$- and $c$-subsystems 
do interact and corresponding orbitals overlap, the constants $\epsilon$ in (29,30) 
contain sum of products of matrix elements of the inverse of overlap matrix 
(determinants)  and are given by quite cumbersome expressions. Usually, in 
physical situation, only some of $c$-electron bands have non-zero overlap with 
$f$-electron orbitals because of their location in different energy regions. For this 
reason one needs actually to take into account the overlap integrals only with 
those of $f$-orbitals which are located in the vicinity of Fermi energy, {\em i.e.} 
with upper subset of orbitally polarized $f$-shell. In order to give feeling of the 
structure of constants $\epsilon $ we give a simple example in the 
Appendix B. 
These constants can easily be calculated for any given set of many-electron 
states. Below we will assume that $\epsilon ^{ab}_\xi $, etc., are calculated 
and known.  One way to find the commutation relations
is a straightforward calculation by
making use of the facts that we know that
\beq
\{c_{jL},f_{j^{\prime }m_l\sigma }^{\dagger}\}
=(O^{-1})_{jL,j^{\prime }m_l\sigma },
\eeq
and any of the operators $X^a$ or $h^\Gamma $ can be expressed
in terms of $f,f^{\dagger}-$operators. For example,
 using the general expansion 
\beq
\hat{Y}_j=\sum_{\Gamma \Gamma ^{\prime }}\langle \Gamma |\hat{Y}_j|\Gamma
^{\prime }\rangle X^{\Gamma \Gamma ^{\prime }},
\eeq
 and explicit expression for the operator of annihilation of electron on the third 
orbital in the state with second and third orbitals occupied, we see that
\beq
(1-\hat{n}_1)\hat{n}_2\hat{f}_3(1-\hat{n}_4)...(1-\hat{n}_{14})=-X^{(2)%
\left( 23\right) }.
\eeq
This direct way of calculation is quite long. There is, however, a 
way to make it compact.
It is easy to verify that one can work with the 
operator $\sum_{jL}O_{j^{\prime }\mu ,jL}c_{jL}$ 
under an anticommutator sign (as one can with $%
f_{j^{\prime }\mu }$ (here $\mu \equiv (m_l,\sigma )$ ). This observation
immediately leads to the formula
\beq
\{c_{jL},X_{j^{\prime }}^{\bar{a}}\}=\sum_{a_1m_l^{^{\prime }}\sigma}
\sum_{\xi=A,\bar{A},\Gamma }
(O^{-1})_{jL,j^{\prime }m_l^{\prime }\sigma ^{\prime }}(f_{m_l^{\prime
}\sigma ^{\prime }})^{a_1}\varepsilon ^{a_1,\bar{a}}_{\xi}Z_{j'}^{\xi},
\eeq
which reproduces the one we calculate in a
straightforward way from the the expression for the $X$-operator 
in terms of $f,f^{\dagger}$-operators. The
sign of
each term in the sum over $a_1$ is determined by the matrix element $%
(f_{m_l^{\prime }\sigma ^{\prime }})^{a_1}$, which takes into account the
convention about the order of $f$-operators in a $|\Gamma \rangle $-state
automatically. The same argumentation leads to the following 
expression for the commutator
with $h^\Gamma $:
\beq
\lbrack c_{jL},h_{j^{\prime }}^\Gamma ]=\sum_{a_1m_l^{^{\prime }}\sigma
^{\prime }}(O^{-1})_{jL,j^{\prime }m_l^{\prime }\sigma ^{\prime
}}(f_{m_l^{\prime }\sigma ^{\prime }})^{a_1}
\varepsilon ^{a_1,\Gamma}_b X_{j^{\prime}}^{b}.
\eeq
In the non-interacting system ($H_{int}=0$) inverse processes of excitations
are absent since $\{c,f\}=0$. This anticommutator, of course, cannot be changed
by $H_{int}$. However, the partial anticommutators 
$\{c,X^a\}$ may be non-zero, if
they are
 considered separately. As it will be seen later, these anticommutators
may  involve two-particle processes.

Since the majority of rare-earth-based materials have 
a single-determinant ground
state, it is quite obvious that the orbital representation is 
convenient for
the description of ground-state properties. 
The excited states, however, are
described in most cases by many-determinant states (in 
quantum-chemistry language described as ''configurational interaction'')
 and, therefore, the procedure of constructing the 
states $|\Gamma _n\rangle $,
discussed in details by Irkhin and Irkhin\cite{irkhiny} should be used. 
In these cases the structure 
constants $\varepsilon^{ab}_c $ and so on, 
which come from the commutation relations, 
and the matrix
elements $(f_{m_l\sigma })^a\equiv (f_{m_l\sigma })^{\Gamma _n,\Gamma
_{n+1}}=\langle \Gamma _n|f_{m_l\sigma }|\Gamma _{n+1}\rangle $,
 include
also the Clebsch-Gordon and fractional parentage coefficients. 
Here we will not
go here into details of this well-developed theory and refer 
the  reader to ref.18.

\subsection{S-matrix and the exact equations for GFs in 
functional derivatives.}
\subsubsection{Quasi-fermions}

Now we introduce the Matsubara's Green's function (GF) for the fermion-like
transitions. The idea, exploited 
in Refs.\cite{schwinger,baym_kadanoff,rucksmr2} for the
derivation of the diagram technique, is based on the possibility to express
the terms, which come from interaction in the equation of motion for a GF,
in terms of functional derivative of GF over external field, and, in the
next step, on an iteration procedure. 
The equations of motion for Fermi-like and Bose-like operators are given 
in Appendix C. 
 As seen from these equations (Eqs (C5-C16)), in our case
the Hamiltonian generates the terms with the operator $\hat{Q}_j^{ac}$.
Therefore, we have to introduce a type of 
external field ${\cal U}(t)$ which
generates this operator. However, an inspection of the further steps in such
a version of the theory shows that using the field in the form $\int {\cal U}%
_j^{ac}(t)\hat{Q}_j^{ac}(t)dt$ introduces information in a 
too much integrated
form, which does not provide a possibility to calculate the quasi-boson GFs.
Instead we introduce the definition for the 
quasi-fermion GF with different fields for
different quasi-boson transitions;
\begin{equation}
\langle T\eta _{j\lambda }(t)\bar{\eta}_{j^{\prime }\lambda ^{\prime
}}^{}(t^{\prime })\rangle _u=\frac{\langle TS(-i\beta ,0)\eta _{j\lambda }(t)%
\bar{\eta}_{j^{\prime }\lambda ^{\prime }}^{}(t^{\prime })\rangle }{\langle
TS(-i\beta ,0)\rangle }.
\end{equation}
The $S$-matrix is then defined as follows:
\begin{equation}
S(-i\beta ,0)=exp\{-i\int_0^{-i\beta }dt\;[\sum_{j\xi }\;{\cal U}_{j\xi
}^{}(t)Z_j^\xi (t) 
+\sum c_{jL}^{\dagger }(t)F_{jL,j^{\prime }L^{\prime }}(t)
c_{j^{\prime }L^{\prime }}(t)]\}.
\end{equation}
The additional terms which are generated by the external field ${\cal U}$
and which should be added into the  equations for the GFs are seen from the equation:
\begin{equation}
i\partial _t\{TS(-i\beta ,t)\eta _{j\lambda }(t)S(t,0)\}
=\{TS(-i\beta ,t)[i\partial _t\eta _{j\lambda }(t)+\;
A_{j\lambda ,j^{\prime }\lambda ^{\prime }}^{}(t)
\eta _{j^{\prime }\lambda ^{\prime }}(t)]S(t,0)\}.
\end{equation}
The elements of the matrix $A_{j\lambda ,j^{\prime }\lambda ^{\prime }}^{}(t)$ 
are given in the Appendix D.

Second term, $c^{\dagger }Fc$, has the same operator structure as zero
Hamiltonian of conduction electrons, $c^{\dagger }hc$; therefore, we will
take them into account if in the equations of motion for any operator make
exchange $\tilde{h}_{12}\rightarrow h_{12}=\tilde{h}_{12}+F_{12}$. Actually,
we will need this field only for consideration of remaining terms of Coulomb
interaction. On the other hand, it gives an additional convenient freedom to
operate with equations for GFs.

As seen from the equations of motion (see Eqs.C5-C6) for the components 
of the $\eta $-operator, $c$- and $X$-operators, 
the equations for the GFs $ \langle T\eta \bar{\eta}\rangle _u$ 
contain the Greens function $\langle TQ(t)\eta (t)%
\bar{\eta}(t^{\prime })\rangle _u$ and $\langle T\eta (t)Q(t)\bar{\eta}%
(t^{\prime })\rangle _u$ (see Eq.(C10-13)). Both these GFs can be obtained from
the three-time GF $\langle TQ(t^{\prime \prime })\eta (t)\bar{\eta}%
(t^{\prime })\rangle _u$ by taking the corresponding limit, $\lim_{t^{%
\prime \prime }\rightarrow t+0}...$ or $\lim_{t^{\prime \prime
}\rightarrow t-0}...$ . Since $\hat{Q}^{a\bar{b}}=\varepsilon _\xi ^{a\bar{b}%
}Z^\xi ,$ $\delta S/\delta {\cal U}_{j\xi }^{}(t)=-iSZ^\xi ,$ and
\begin{equation}
\;\langle TZ_\xi (t^{\prime \prime })\eta (t)\bar{\eta}(t^{\prime })\rangle
_u\equiv (\langle TZ_\xi (t^{\prime \prime })\rangle _u+i\frac \delta
{\delta {\cal U}_{j\xi }^{}(t^{\prime \prime })})\langle T\eta (t)\bar{\eta}%
(t^{\prime })\rangle _u,
\end{equation}
any of the three-time GFs can be expressed via functional derivative of the
GF $\langle T\eta (t)\bar{\eta}(t^{\prime })\rangle _u$ over corresponding
field ${\cal U}_{j\xi }^{}(t^{\prime \prime })$, another trick exploited 
by Schwinger\cite{schwinger} and 
by Baym and Kadanoff\cite{baym_kadanoff}. After that the equations 
for the GFs 
can be written in closed form straight forwardly from the equations of motion for
$c$- and $X$-operators, given by Eqs.C5 and C6, and the relation in Eqn.38. The diagonal 
expectation values of the diagonal operators, 
$\langle h^{\Gamma }\rangle $ in 
the equilibrium state have physical meaning of population numbers 
of the ion's state $\Gamma $. 
When external time-dependent fields are present, these operators do not commute
with the Hamiltonian even when the interaction is switched off, therefore, 
$\langle h^{\Gamma }\rangle $, strictly speaking, cannot be interpreted as 
population numbers any more.  Although it is not 
completely correct, for briefness we will refer loosely to 
"population numbers" for the expectation values of any Bose-like 
single-site operator.

For briefness we will not write the complete matrix equation 
for $\langle \eta (t)\bar{\eta}\rangle _u$, instead we will specify only the 
left-hand-side $\eta $-operator.
 Then we find that the equation for the set of 
GFs  $\langle Tc_{jL}(t)\bar{\eta}\rangle _u$ has the following form,
\begin{eqnarray}
&&[\delta_{j,j_1}\delta_{L,L_1}i\partial _t-(O^{-1}H)_{jL,j_1L_1}]
\langle Tc_{j_1L_1}(t)\bar{\eta}%
\rangle _u-[A^{(cX)}_{jL,j_1b}(t)+w_{jL,j_1b}]\langle TX_{j_1}^b(t)\bar{\eta}%
\rangle _u  \nonumber \\
&-&A^{(cX^{\dagger})}_{jL,j_1{\bar{b}a}}(t)
\langle TX_{j_1}^{\bar{b}}(t)\bar{\eta} \rangle _u 
=i\langle \{c_{jL},\bar{\eta}\}\rangle _u\delta
(t-t^{\prime })  \nonumber \\
&&+O^{-1}_{jL,j_1M}(f_M)^c[\langle \hat{Q}_{j_1}^{c\bar{a}}(t^{+})\rangle
_u+i\frac \delta {\delta {{\cal U}}_{j_1}^{c\bar{a}%
}(t^{+})}]W_{j_1a,j_2L_2}^{*}\langle Tc_{j_2L_2}(t)\bar{\eta}\rangle _u 
 \nonumber \\
&&+O^{-1}_{jL,j_1M}(f_M)^c[\langle \hat{Q}_{j_1}^{c\bar{a}}(t^{+})\rangle
_u+i\frac \delta {\delta {{\cal U}}_{j_1}^{c\bar{a}}(t^{+})}]t_{j_1j_2}^{%
\bar{a}b}\langle TX_{j_2}^b(t)\bar{\eta}\rangle _u  \nonumber \\
&&-O^{-1}_{jL,j_2M}(f_M)^c[\langle \hat{Q}_{j_2}^{ca}(t^{-})\rangle
_u+i\frac \delta {\delta {{\cal U}}_{j_2}^{ca}(t^{-})}]  \nonumber \\
&&\times \{t_{j_1j_2}^{\bar{b}a}\langle TX_{j_1}^{\bar{b}}(t)\bar{\eta}%
\rangle _u+W_{j_1L_1,j_2a}^{}\langle Tc_{j_2L_2}^{\dagger }(t)\bar{\eta}%
\rangle _u\}.
\end{eqnarray}
The equation for the set of GFs $\langle TX_j^b(t)\bar{\eta}\rangle _u$ is
\begin{eqnarray}
&&[\delta _{ab}(i\partial _t-\Delta _{j\bar{a}})-A_{jb}^{a}(t)]\langle
TX_j^b(t)\bar{\eta}\rangle _u 
-A_{j{\bar{b}}}^{a}(t)\langle TX_j^{\bar{b}}(t)\bar{\eta}\rangle _u  
\nonumber \\
&=&i\langle \{X_j^a,\bar{\eta}\}\rangle _u\delta (t-t^{\prime })  \nonumber
\\
&&+[\langle \hat{Q}_j^{a\bar{b}}(t^{+})\rangle _u+i\frac \delta {\delta
{{\cal U}}_j^{a\bar{b}}(t^{+})}]t_{j_1j_2}^{\uparrow \bar{b}c}\langle
TX_{j_2}^c(t)\bar{\eta}\rangle _u  \nonumber \\
&&+[\langle \hat{Q}_j^{a\bar{b}}(t^{+})\rangle _u+i\frac \delta {\delta
{{\cal U}}_j^{a\bar{b}}(t^{+})}]w_{j\bar{b},j_2L_2}^{\uparrow }\langle
Tc_{j_2L_2}(t)\bar{\eta}\rangle _u  \nonumber \\
&&-[\langle \hat{Q}_j^{ca}(t^{-})\rangle _u+i\frac \delta {\delta {{\cal U}}%
_j^{ca}(t^{-})}]  \nonumber \\
&&\times \{t_{j_1j}^{\downarrow \bar{b}c}\langle TX_{j_1}^{\bar{b}}(t)\bar{%
\eta}\rangle _u+w_{j_1L_1,jc}^{\downarrow }\langle Tc_{j_1L_1}^{\dagger }(t)%
\bar{\eta}\rangle _u\}.
\end{eqnarray}
The equations for $\langle Tc_{j_1L_1}^{\dagger }(t)\bar{\eta}\rangle _u$
and $\langle TX_{j_1}^{\bar{b}}(t)\bar{\eta}\rangle _u$ can be obtained
from Hermitian conjugation of the corresponding equations of motion,
\begin{eqnarray}
&&[(-i)\partial _t
\langle Tc_{j_1L_1}^{\dagger }(t)\bar{\eta}\rangle _u
\delta _{j_1,j}\delta _{L_1,L}
-\langle Tc_{j_1L_1}^{\dagger }(t)\bar{\eta}\rangle
_u(HO^{-1})_{j_1L_1,jL}]-\langle TX_{j_1}^{\bar{b}}(t)\bar{\eta}\rangle _u
A^{X^{\dagger}c^{\dagger}}_{j_1\bar{%
b},jL}(t)  \nonumber \\
&-& \langle TX_{j_1}^{b}(t)\bar{\eta}\rangle _u
A^{Xc^{\dagger}}_{j_1b,jL}(t) 
=i\langle \{c_{jL}^{\dagger },\bar{\eta}\}\rangle _u
\delta (t-t^{\prime })+\langle TX_{j_2}^{\bar{b}}(t)\bar{\eta}%
\rangle _uw_{j_2\bar{b},jL}  \nonumber \\
&&+[\langle \hat{Q}_{j_1}^{a\bar{c}}(t^{-})\rangle _u+i\frac \delta {\delta
{{\cal U}}_{j_1}^{a\bar{c}}(t^{-})}]  \nonumber \\
&&\times \{\langle Tc_{j_2L_2}^{\dagger }(t)\bar{\eta}\rangle
_uW_{j_2L_2,j_1a}^{}+
\langle TX_{j_2}^{\bar{b}}(t)\bar{\eta}\rangle _u
t_{j_2j_1}^{\bar{b}a}\}(f_M^{\dagger })^{\bar{c}}(O^{-1})^*_{j_1M,jL}
\nonumber \\
&&-[\langle \hat{Q}_{j_2}^{ac}(t^{-})\rangle _u+i\frac \delta {\delta
{{\cal U}}_{j_2}^{ac}(t^{-})}]  \nonumber \\
&&\times \{W_{j_2\bar{a},j_1L_1}^{*}\langle Tc_{j_1L_1}^{}(t)\bar{\eta}%
\rangle _u+t_{j_2j_1}^{\bar{a}b}\langle TX_{j_1}^b(t)\bar{\eta}\rangle
_u\}(f_M^{\dagger })^{\bar{c}}(O^{-1})^*_{j_2M,jL}, 
\end{eqnarray}
and
\begin{eqnarray}
&&[\delta _{ab}(-i\partial _t-\Delta _{ja})-A_{j{\bar{b}}}^{\bar{a}}] 
\langle TX_j^{\bar{b}}(t)\bar{\eta}\rangle _u 
- A_{j b}^{\bar{a}} \langle TX_j^{b}(t)\bar{\eta}\rangle _u \nonumber \\
&=& -i\langle \{X_j^{\bar{a}},\bar{\eta}\}\rangle _u\delta (t-t^{\prime })
\nonumber \\
&&-[\langle \hat{Q}_j^{b\bar{a}}(t^{-})\rangle _u+i\frac \delta {\delta
{{\cal U}}_j^{\bar{b}a}(t^{-})}]  \nonumber \\
&&\times \{w_{j_2L_2,j_1b}^{\downarrow }\langle Tc_{j_2L_2}^{\dagger }(t)%
\bar{\eta}\rangle _u+t_{j_2j}^{\downarrow \bar{c}b}\langle TX_{j_2}^{\bar{c}%
}(t)\bar{\eta}\rangle _u\}  \nonumber \\
&&[\langle \hat{Q}_j^{ac}(t^{+})\rangle _u+i\frac \delta {\delta {{\cal U}}%
_j^{ac}(t^{+})}]  \nonumber \\
&&\times \{w_{jc,j_1L_1}^{\uparrow }\langle Tc_{j_1L_1}(t)\bar{\eta}\rangle
_u+t_{jj_1}^{\uparrow \bar{c}b}\langle TX_{j_1}^b(t)\bar{\eta}\rangle _u\}.
\end{eqnarray}
The products $(HO^{-1})$ and $(O^{-1}H)$ are the matrix products in the 
$cc$-subspace only, i.e.
\begin{equation}
(HO^{-1})_{jL,j'L'} \equiv H_{jL,j_1L_1}(O^{-1})_{j_1L_1,j'L'}.
\end{equation}
The labels $\pm $ near the time variable 
denote limits $t^{\pm }=t\pm 0^{+}$.

\subsubsection{Quasi-bosons}

Our recipe for calculation of the GFs is so far not complete; we have to
describe how to calculate the population numbers and their derivatives. The
latter, actually, describe the boson-like excitations in the system. 
Now we turn to the consideration of this part.

The ''gauge'' field coming from the source-term of the $S$-matrix is
\begin{equation}
A_j^{\xi \xi _2}(t)\equiv \varepsilon _{\xi _2}^{\xi \xi _1}{\mathcal U}%
_j^{\xi _1}(t),
\end{equation}
where  $\varepsilon _{\xi _2}^{\xi \xi _1} $, similar to Eqn.29, gives the 
coefficients of expansion $[Z^{\xi },Z^{\xi _1 }] 
= \epsilon ^{\xi,\xi_1}_{\xi _2}Z^{\xi _2}$.

Let us write down first the equation for $\langle Z_j^\xi (t)\rangle _u$.
For that we have to add to the average of (54) the contribution from the
source term. Then we find 
\begin{eqnarray}
\lbrack  &&\delta ^{\xi \xi _2}(i\partial _t-\Delta _{\bar{\xi}})-A_j^{\xi
\xi _2}(t)]\langle TZ_j^\xi (t)\rangle _u=  \nonumber \\
&&\sum_{n_1\lambda _1n_2\lambda _2}V_{n_1\lambda _1,n_2\lambda _2}^{j\xi
}\langle T\eta _{n_1\lambda _1}^{\dagger }(t)\eta _{n_2\lambda _2}(t)\rangle
_u.
\end{eqnarray}
This equation is valid also for the case of diagonal transition, $\langle
Z_j^{\xi =[\Gamma ,\Gamma ]}(t)\rangle _u=\langle h_j^\Gamma (t)\rangle _u$
. In this case $\Delta _{\bar{\xi}}=\Delta _{[\Gamma ,\Gamma ]}=E_\Gamma
-E_\Gamma =0$ and  at zero fields, ${\cal U}=0 $, we come to the case
of thermodynamical equilibrium , 
\begin{equation}
i\partial _t\langle Th_j^\Gamma (t)\rangle =\sum_{n_1\lambda _1n_2\lambda
_2}V_{n_1\lambda _1,n_2\lambda _2}^{\xi =[\Gamma ,\Gamma ]}\langle T\eta
_{n_1\lambda _1}^{\dagger }(t)\eta _{n_2\lambda _2}(t)\rangle .
\end{equation}
where the population numbers $N_\Gamma \equiv \langle Th_j^\Gamma (t)\rangle 
$ do not depend on time. Then, the right-hand side gives zero identically,
since summation in the right-hand side of (47) produces the difference of 
Hermitian adjoints.
This equation is useful for checking the approximation made but does not
give a recipe for finding $N_\Gamma $. Therefore, we have to go further and
to use connection between the GFs 
\begin{equation}
\langle TZ_j^\xi (t)Z_{j^{\prime }}^{\xi ^{\prime }}(t^{\prime })\rangle
_u\equiv K_{jj^{\prime }}^{\xi \xi ^{\prime }}(t,t^{\prime })
\end{equation}
and their end-factors 
\begin{equation}
P^{\xi \xi ^{\prime }}(t)\equiv \langle T[Z_j^\xi (t),Z_{j^{\prime }}^{\xi
^{\prime }}(t)]\rangle _u=\delta _{jj^{\prime }}\varepsilon _{\xi ^{\prime
\prime }}^{\xi \xi ^{\prime }}\langle TZ_j^{\xi ^{\prime \prime }}(t)\rangle
_u.
\end{equation}
Let us derive equations for the Bose-like GFs. They can be obtained in two
ways. One way is to use the equations of motion, Eq.(C18) and then to
connect the higher-order GF with the lower-order one via functional
derivative. Another way is to take the functional derivative $\delta /\delta 
{\mathcal U}_{j^{\prime }}^{\xi ^{\prime }}(t^{\prime })$ from both parts of
Eq.(46). These ways should lead to equivalent results. The equations of
motion give:
\begin{eqnarray}
\lbrack \delta _j^{\xi \xi _2}(i\partial _t-\Delta _{\bar{\xi}})-A_j^{\xi
\xi _2}(t)]K_{j\xi _2,j^{\prime }\xi ^{\prime }}(t,t^{\prime }) &=&i\delta
(t-t^{\prime })P_{jj^{\prime }}^{\xi \xi ^{\prime }}(t)+  \nonumber \\
&&V_{n_1\lambda _1,n_2\lambda _2}^\xi \langle T\eta _{n_1\lambda
_1}^{\dagger }(t)\eta _{n_2\lambda _2}^{}(t)Z_{j^{\prime }}^{\xi ^{\prime
}}(t^{\prime })\rangle _u
\end{eqnarray}


Let us consider now the following three-time GF $\langle T\eta _{n_2\lambda
_2}^{}(t_1)\eta _{n_1\lambda _1}^{\dagger }(t_2)Z_{j^{\prime }}^{\xi
^{\prime }}(t^{\prime })\rangle _u.$ On the one hand, the two-time GF in the
last term of Eq.50 can be expressed as the limit

\begin{equation}
\lim_{t_{2\rightarrow }t+0}\langle T\eta _{n_2\lambda _2}^{}(t)\eta
_{n_1\lambda _1}^{\dagger }(t_2)Z_{j^{\prime }}^{\xi ^{\prime }}(t^{\prime
})\rangle _u=\langle T\eta _{n_1\lambda _1}^{\dagger }(t)\eta _{n_2\lambda
_2}^{}(t)Z_{j^{\prime }}^{\xi ^{\prime }}(t^{\prime })\rangle _u
\end{equation}
of this function. On the other hand, we can again use Eq.39 and express it
in terms of the GF $G=\langle T\eta _{n_2\lambda _2}^{}(t)\eta _{n_1\lambda
_1}^{\dagger }(t_2)\rangle _u.$This gives the following equation for the
boson-like GF:

\begin{eqnarray}
(L^0(t,t_1)_{j\xi ,j_1\xi _1}^{-1} &&K_{j_1\xi _1,j^{\prime }\xi ^{\prime
}}(t_1,t^{\prime })=i\delta (t-t^{\prime })P_{jj^{\prime }}^{\xi \xi
^{\prime }}(t)+  \nonumber \\
&&\lim_{t_2\rightarrow t+0}V_{n_1\lambda _1,n_2\lambda _2}^\xi [\langle
TZ_{j^{\prime }}^{\xi ^{\prime }}(t^{\prime })\rangle _u+i\frac \delta
{\delta {\mathcal U}_{j^{\prime }}^{\xi ^{\prime }}(t^{\prime })}]\langle
T\eta _{n_2\lambda _2}^{}(t)\eta _{n_1\lambda _1}^{\dagger }(t_2)\rangle _u,
\end{eqnarray}

where ($L^0)^{-1}$ is defined by the equation:

\begin{equation}
\{[\delta _j^{\xi \xi _2}(i\partial _t-\Delta _{\bar{\xi}})-A_j^{\xi \xi
_2}(t)]\delta (t-t_2)\}L_{j\xi _2,j^{\prime }\xi ^{\prime }}^0(t_2,t^{\prime
})=i\delta (t-t^{\prime }).
\end{equation}

The procedure which remains to be discussed is how to calculate the
expectation values $\langle TZ_{3_b}(t_2)\rangle _u $, i.e., 
population numbers.
The natural way to do that is to use the multiplication rule for the 
Hubbard operators
$X_j^{p,q} \cdot X_j^{r,s}\equiv |j,p\rangle \langle j,q| \cdot
|j,r\rangle \langle j,s| = \delta _{q,r}X_j^{p,s}$.
In the limit $t' \rightarrow t+0$ the GFs $K(t,t')$ or $G(t,t')$ 
contain a product of two operators at the same moment of time and, 
therefore, we can use the rule of multiplication. 
In the case of retarded GFs an application of the spectral theorem
gives an equivalent system of equations in those of  approximations, 
in which the analytical extension of the vertex corrections
does not involve different values of vertexes from 
different time intervals:
\begin{equation}
N_j^\Gamma =[\lim_{t\rightarrow +0}\langle X_i^{\bar{a}%
}(t)X_j^a(0)\rangle ]_{i=j},  
\end{equation}
or from Bose-like GFs,
\begin{equation}
N_j^\Gamma =[\lim_{t\rightarrow +0}\langle Z_i^{\xi} (t)Z_j^{\xi}(0)\rangle ]_{i=j},
\end{equation}
where $\xi = [\Gamma,\Gamma_1].$

When the iterative 
perturbation theory at ${\cal U}(t) \neq 0$ is used none of expectation values 
of $\langle Z(t) \rangle $ is equal to zero. In the limit of thermodynamics 
(${\cal U}(t) = 0$ )
the non-diagonal numbers $\langle Z^{\xi} \rangle = 0$ while the diagonal 
ones $N_{\Gamma }$ and their derivatives 
$\frac{ \delta ^n N_{\Gamma }}{\delta {\cal U}_{\Gamma _1}
\delta {\cal U}_{\Gamma _2}... \delta {\cal U}_{\Gamma _n}}|_{{\cal U}=0}$ 
{\em are not}.
The derivatives of non-diagonal occupation numbers are 
charge and magnetic susceptibilities (see Eqs 94-95 below).
However, the correlators $\langle h^{\Gamma' }h^{\Gamma }\rangle $ also contribute 
to both of them.

\subsubsection{Comments on the system of equations}
Eqs. 40-43  and 52 are the basic equations of our theory which we will use 
for generating a series of perturbation theory.
The iterative terms and lowest approximations will be 
discussed in the next section.
Within this formulation of the diagram technique the 
difference between the
cases of orthogonal and non-orthogonal bases sets consists only in the 
number of
interactions involved and in the definitions of the matrix elements.
Contrary to the standard case of normal fermionic system 
the set of equations Eqs.40-43 for quasi-fermion transitions
is not yet a complete set even in terms of 
functional derivatives, as seen from Eqs.41,52 and will be seen also later. The
functional derivatives from expectation values of quasi-boson operators 
introduce also energy-dependent
contributions to the self-energy from
quasi-boson excitations and in addition, broadening of spectral
weights. The difference to the case of, for instance, standard Fermi-liquid theory 
is that
in the Fermi-liquid the boson excitations arise only as 
collective modes whereas 
in our case they exist even in the level of zero Hamiltonian .
At first sight the presence of the GFs $\langle
Tc_{j_2L_2}^{\dagger }(t)\bar{\eta}\rangle _u$ for $\eta = c^{\dagger}$ 
in the system of equations looks somewhat strange 
since we are considering the normal, not superconducting state. 
Actually, the zero GFs of this kind are indeed equal to zero in 
the normal state; this is simply a
manifestation of the fact that our external Bose-like field ${{\cal U}}$
contains components which describe the transitions $[\Gamma _n,\Gamma
_{n\pm 2}]$, \emph{i.e. }transfer of 2 electrons simultaneously.
The corresponding GF describes process of transformation of two conduction
electrons to $f$-electrons on the same site (or \emph{vice versa}) and
appears only as a perturbational contribution. Thus, at zero external field we have
$\langle c^{\dagger}c^{\dagger}\rangle =0$ whereas 
$\delta \langle c^{\dagger}c^{\dagger}\rangle /\delta U $ may have 
non-zero values. This, obviously, leads to the existence of some non-zero 
correlations of the 
kind $\langle c_1c_2(\tau )c_3^{\dagger }c_4^{\dagger }(\tau ')\rangle $, 
and as a consequence superconducting fluctuations above $T_c$ due to phonons. 
Some other mechanism which are described by the same function may also be 
influenced by the two-particle intra-atomic correlations (for example, 
it can make the solution with the $d$-wave symmetry of the order parameter more 
favorable compared to the one with $s$-symmetry\cite{plakida,izyumov_sc}.)

\section{Approximations}

Eqs. 40-43 and 52 can be used for iterations straightforwardly. 
However, since the subsystems of conduction electrons and 
quasi-atomic transitions  are
coupled,  it is more convenient  in practical calculations to 
make iterations
from a  more symmetrical form of these equations. Let us first rewrite the
equations in terms of the operators $\eta _1,\eta _1^{\dagger }$ 
with the\emph{\ 
}indexes $1\equiv (j_1,\lambda _1).$ In order to do this let us return for a moment to
the initial Hamiltonian of PAM. All its terms except the one ${\mathcal H}%
_0^f=\sum E_\Gamma ^0h_j^\Gamma $ can be expressed in terms of operators $%
\eta _1,\eta _1^{\dagger }$,

\begin{equation}
{\mathcal H}_{int}^{\prime }=\eta _2^{\dagger }V_{23}\eta _3\equiv
H_{23}c_2^{\dagger }c_3+W_{23}c_2^{\dagger }X_3+W_{23}^{*}X_2^{\dagger
}c_3+t_{23}X_2^{\dagger }X_3.
\end{equation}
The term ${ \mathcal H}_0^f$ does not cause difficulties since,
as seen from Eqns.C5,C6 and C18, in the equations of motion it 
generates a term which
also can be written in terms of $\eta $-operators. We denote this term as $%
\theta _{12}^\Gamma :\theta _{12}^\Gamma \eta _2\equiv [\eta _1^{},\sum
E_\Gamma ^0h_j^\Gamma ].$ Therefore, the equations for GFs, ${ \mathcal G}%
_{12}(t,t^{\prime })\equiv \langle T\eta _1(t)\eta _2^{\dagger }(t^{\prime
})\rangle _u$, can be expressed fully in terms of $\eta $-operators and
functional derivatives of GFs with respect to external fields,

\begin{eqnarray}
&(i\delta _{12}&\partial _t-A_{12}(t)-\Omega _{12}^0-\theta _{12}^\Gamma )%
{ \mathcal G}_{21^{\prime }}(t,t^{\prime })=i{ \mathcal P}_{11^{\prime
}}(t,t^{\prime })+  \nonumber \\
&&V_{23}\langle T\{\eta _1(t),\eta _2^{\dagger }(t)\}\eta _3(t)\eta
_{1^{\prime }}^{\dagger }(t^{\prime })\rangle - \\
&&V_{23}\langle T\eta _2^{\dagger }(t_1)\{\eta _1^{}(t),\eta _3(t)\}\eta
_{1^{\prime }}^{\dagger }(t^{\prime })\rangle ,
\end{eqnarray}
where $A_{jL,j^{\prime }L^{\prime }}=0.$ Since we have moved the non-$f$ part
of the LDA Hamiltonian from zero Hamiltonian to the matrix of interaction $V$%
, $\Omega ^0$ becomes a purely diagonal matrix:

\begin{equation}
\Omega _{12}^0=\Omega _1^0\delta _{12}:\;\Omega _{ja,j^{\prime }b}^0=\delta
_{jj^{\prime }}\delta _{ab}\Delta _{\bar{a}},\;\Omega _{jL,j^{\prime
}L^{\prime }}^0=\delta _{jj^{\prime }}\delta _{LL^{\prime }}\varepsilon _L^0.
\end{equation}
Here we have included a $\delta $-function into the definition of the
time-dependent population number,

\begin{eqnarray}
{\mathcal P}_{11^{\prime }}^{}(t_1,t_1^{\prime }) &\equiv &\delta
(t_1-t_1^{\prime })\langle \hat{Q}_{11^{\prime }}^{\eta \eta ^{\dagger
}}(t_1^{\prime })\rangle \equiv \\
&&\delta (t_1-t_1^{\prime })\langle \{\eta _1(t_1^{\prime }),\eta
_{1^{\prime }}^{\dagger }(t_1^{\prime })\}\rangle \equiv \delta
(t_1-t_1^{\prime }){ \mathcal P}_{11^{\prime }}^{}(t_1^{\prime }).
\end{eqnarray}

It is seen from Eqns.40-41 that in order to have a compact form of equations, we
have to form combinations of the operators $\langle Q\rangle
+i\frac \delta {\delta {\cal U}}$ with matrix elements, such that they
reproduce full combinations $\{\eta _1,\eta _2^{\dagger }\},\{\eta _1,\eta
_2^{}\}$ in the equations of motion. The GFs of higher order in terms of Eqns.40
and 41, which are expressed in terms of functional derivatives, can be written
in compact 
notations as follows: 

\begin{eqnarray}
V_{23}\langle T\{ &&\eta _1(t),\eta _2^{\dagger }(t)\}\eta _3(t)\eta
_{1^{\prime }}(t^{\prime })\rangle =  \nonumber \\
V_{23} &&[{ \mathcal P}_{12}(t^{+})+\hat{R}_{12}^l(t^{+})]{\mathcal G}%
_{31^{\prime }}(t,t^{\prime }),
\end{eqnarray}
\begin{eqnarray}
V_{23}\langle &&T\eta _2^{\dagger }(t_1)\{\eta _1(t_1),\eta _3(t_1)\}\eta
_{1^{\prime }}(t_1^{\prime })\rangle =  \nonumber \\
&&V_{23}[\bar{{ \mathcal P}}_{12}(t^{-})+\hat{R}_{12}^r(t^{-})]\langle T\eta
_2^{\dagger }(t)\eta _{1^{\prime }}^{\dagger }(t^{\prime })\rangle _u.
\end{eqnarray}
The operators $\hat{R}$ are here defined in such a way that
$\hat{R}_{12}^l(t^{+})=0$ and $\hat{R}_{12}^r(t^{-})=0$
for $c$-electrons, i.e., when $1=(j_1,L_1)$ and $2=(j_2,L_2)$. The cases
when one of these indexes$,1,2$, or both of them describe $f$-electron
transitions, are described with the help of the matrices $\alpha $ and $%
\beta $. They have to provide correct commutation relations and, therefore,
we define them by the equalities:

\begin{equation}
\{\eta _1,\eta _2^{\dagger }\}=\bar{\alpha} _{12,3_b}^{\eta \eta ^{\dagger
}}Z_{3_b}^{},\;\{\eta _1^{},\eta _3\}=\bar{\beta} _{12,3_b}^{\eta \eta
}Z_{3_b}^{},\;\{\eta _1^{\dagger },\eta _3^{\dagger }\}=\bar{\beta}
_{12,3_b}^{\eta ^{\dagger }\eta ^{\dagger }}Z_{3_b}^{}
\end{equation}
where $\xi =A,\bar{A},\Gamma $ , $\zeta =[\Gamma _n,\Gamma _{n+2}]$ and $%
\bar{\zeta}=[\Gamma _n,\Gamma _{n-2}]$. The complex index $3_b\equiv
(j_3,\xi _3)$ in the case of the $\bar{\alpha }$-matrix, then, $3_b\equiv
(j_3,\zeta _3)$ in the case of $\bar{\beta }_{12,3_b}^{\eta \eta }$-matrix , and,
at last, $3_b\equiv (j_3,\bar{\zeta}_3)$ for the $\beta _{12,3_b}^{\eta
^{\dagger }\eta ^{\dagger }}$-matrix. The subindex ''$b$'' points out that
the index runs over Bose-like variables. 
If, for example, $\eta =c,$ and $%
\eta ^{\dagger }=X^{\dagger }$, we have

\begin{equation}
\bar{\alpha }_{12,3_b}^{cX^{\dagger }}=O_{12}^{-1}\delta _{23}(f_M)^a\varepsilon
_\xi ^{ab_2},
\end{equation}
and so on. We will not repeat all these definitions since they are 
clearly seen from the expressions given in Eqs.40-43.
Then the operators $\hat{R}_{12}^l(t^{+})$ and $\hat{R}%
_{12}^r(t^{-})$ are:

\begin{equation}
\hat{R}_{12}^l(t^{+})\equiv \lim_{t_4\rightarrow t+0}\alpha _{12,3_b}^{\eta
\eta ^{\dagger }}\frac \delta {\delta {\cal U}_{j_3}^{\xi _3}(t_4)}\equiv
\lim_{t_4\rightarrow t+0}\alpha _{12,3_b}^{\eta \eta ^{\dagger }}\frac
\delta {\delta {\cal U}_{3_b}(t_4)},
\end{equation}

\begin{equation}
\hat{R}_{12}^r(t^{-})\equiv \lim_{t_4\rightarrow t-0}\beta _{12,3_b}^{\eta
\eta }\frac \delta {\delta {\cal U}_{j_3}^{\eta _3}(t_4)}\equiv
\lim_{t_4\rightarrow t-0}\beta _{12,3_b}^{\eta \eta }\frac \delta {\delta
{\cal U}_{3_b}(t_4)},
\end{equation}
where the relations between the matrices $\alpha $ and $\bar{\alpha }$, 
$\beta $ and $\bar{\beta }$ are as follows;
\[
\alpha _{12,3_b}^{\eta \eta ^{\dagger }} = (1 - \delta _{1,j_1L_1}
\delta _{2,j_2L_2})\bar{\alpha }_{12,3_b}^{\eta \eta ^{\dagger }}, \;\;\;\;\;\;
\beta _{12,3_b}^{\eta \eta } = (1 - \delta _{1,j_1L_1}
\delta _{2,j_2L_2})\bar{\beta }_{12,3_b}^{\eta \eta }.
\] 
Defining in this way $\alpha$- and $\beta$-matrices in the $\hat{R}$-operators 
we put in a formal language the requirement that
$\hat{R}_{12}=0$ when 
both ends, 1 and 2, are $c$-electron ones.
Further, as we remember, in Eqs.40-43 (see also here Eqns.78 and 81) the
term $\langle T\eta _2^{\dagger }\{\eta _1^{},\eta _3\}(t)\eta _{1^{\prime
}}^{\dagger }(t^{\prime })\rangle _u$ is expressed in terms of
functional derivatives of the GF $\langle T\eta _2^{\dagger }(t)\eta
_{1^{\prime }}^{\dagger }(t^{\prime })\rangle _u$. The matrix of this GF
consists of the same elements as the matrix $\langle T\eta _2^{}(t)\eta
_{1^{\prime }}^{\dagger }(t^{\prime })\rangle _u.$ Therefore, we can express
the former GF via the latter one as

\begin{equation}
\langle T\eta _2^{\dagger }(t)\eta _{1^{\prime }}^{\dagger }(t^{\prime
})\rangle _u=\tau _{23}{ \mathcal G}_{31^{\prime }}(t,t^{\prime }),
\end{equation}
where the matrix $\tau _{23}$ makes necessary permutations of the matrix
elements and, therefore, consists of zeroes and unities only. 
For example, if
we deal with a system which has $n$ conduction bands and $m$
intra atomic Fermi-type transitions, the matrix $\tau _{23}$ has all matrix
elements equal to zero except over- and sub-diagonals where all the 
matrix elements are equal to one,
\begin{eqnarray}
\tau _{(j_21,j_3n+m+1)} &=&\tau _{(j_22,j_3n+m+2)}=...=\tau
_{(j_2n+m,j_32(n+m))}=1,  \nonumber \\
\tau _{(j_2n+m+1,j_31)} &=&\tau _{(j_2n+m+2,j_32)}=...=\tau
_{(j_22(n+m),j_3n+m)}=1.
\end{eqnarray}
Note that the left $(t_1\rightarrow t_1^{+})$ and the right ( $%
t_1\rightarrow t_1^{-})$ limits are included into definitions of the
operators $R_{12}^l(t^{+})$ and $R_{12}^r(t^{-})$ correspondently.

Let us now introduce the auxiliary zero GF ${ \mathcal D}^{(0)}$ by the
following equation:

\begin{equation}
\int dt_2[\frac 1i(\delta _{12}\partial _{t_1}-A_{12}(t_1)-\Omega
_{12}^0)\delta (t_1-t_2)]{ \mathcal D}_{21^{\prime }}^{(0)}(t_2,t_1^{\prime
})=\delta _{11^{\prime }}\delta (t_1-t_1^{\prime }).
\end{equation}
As seen from (59), the matrix $\Omega _{12}^0$ has purely local, on-site,
characteristics. Therefore, it is natural to call the function ${ \mathcal D}%
_{21^{\prime }}^{(0)}$ ''locator''. However, mixing, hopping and overlap
connect different sites, which leads to dependence of the dressed locators $%
{ \mathcal D}(t,t^{\prime })$ on the variables of different sites; 
for this reason they will be named 
''pseudolocators''. The inverse matrix of ${\mathcal D}^{(0)}$, i.e. [$%
{\mathcal D}_{12}^{(0)}(t_1,t_2)]^{-1}\equiv {\mathcal D}_0^{-1}(1,2),$ is
given by the expression in the square brackets on the l.h.s of Eq.98.
Putting $\theta =V=0$ in Eq.(57-58) for the full GF we obtain the
equation for zero GF ${\mathcal G}_{21^{\prime }}^{(0)}(t_1,t_1^{\prime })$:

\begin{equation}
\int dt^{\prime \prime }{\mathcal D}_{0,12}^{-1}(t,t^{\prime \prime })%
{\mathcal G}_{21^{\prime }}^{(0)}(t^{\prime \prime },t_{}^{\prime })={\mathcal%
P}_{11^{\prime }}^{(0)}(t,t^{\prime }).
\end{equation}
We will use this equation below for iterations. In this notation the full
equation for the GF acquires the following form:

\begin{eqnarray}
&&\int dt_3{ \mathcal D}_{0,13}^{-1}(t,t_3){\mathcal G}%
_{31^{\prime }}((t_3,t^{\prime })  
-\{\frac 1i\theta _{13}^\Gamma +[{ \mathcal P}_{12}(t^{+})
+\hat{R}_{12}^l(t^{+})]V_{23} \nonumber \\
&&-\frac 1i V_{24}[\bar{{ \mathcal P}}_{14}(t^{-})+\hat{R}_{14}^r(t^{-})]\tau
_{23}\}{ \mathcal G}_{31^{\prime }}(t,t^{\prime }) =
{ \mathcal P}_{11^{\prime }}(t,t^{\prime }).
\end{eqnarray}
The same equation can be presented also in a way close to the standard form of
the Dyson equation:

\begin{equation}
\{ { \mathcal D}_{0,12}^{-1}(t,t_2)-S_{12}^{}(t,t_2)\}{\mathcal G}%
_{21^{\prime }}(t_2,t^{\prime })={ \mathcal P}_{11^{\prime }}(t,t^{\prime }).
\end{equation}
Here and below we omit the sign of integration over intermediate time (here $%
t_2$ ). 
Eq. 73 gives the definition of the self-operator matrix $S$.  However,
as  seen, this equation does not fully coincide with the standard
definition of self-energy. The difference between Eq.73 and the standard
Dyson equation consists of the presence of ${ \mathcal P}_{11^{\prime
}}(t,t^{\prime })$ in r.h.s., instead of ${ \mathcal \delta }_{11^{\prime
}}\delta (t,t^{\prime })$ and is caused by the non-fermion commutation
relations $\{\eta _1,\eta _2^{\dagger }\}\neq \delta _{12}.$ 
As will be seen later, 
these terms give rise to whole generation of new graphs. Part of graphs have 
the same structure that in the theory of fermions. The $S$-self-operator contains 
both type of graphs, of Fermi-liquid type, and the terms, generated by the right-hand 
side of Eqs. 72-73. In order to distinguish between them we use the term "self-operator" 
in definition for the full GF(23), 
while the term
 "self-energy" will be used in a more traditional way, for the set of graphs
which renormalize the denominators of the pseudolocators, and
will be introduced later. In
order to separate these two type of graphs
 we will search the solutions of (73) in the form
(71), i.e. we introduce the full pseudolocator 
${ \mathcal D}$ by the Dyson equation

\begin{equation}
\{ {\mathcal D}_{0,12}^{-1}(t,t_2)-S_{12}^{}(t,t_2)\}{\mathcal D}%
_{21^{\prime }}(t_2,t^{\prime })=\delta _{11^{\prime }}\delta
(t-t^{\prime }).
\end{equation}
Then the full GF is given by

\begin{equation}
{ \mathcal G}_{11^{\prime }}(t,t^{\prime })={\mathcal D}_{12}(t,t^{\prime
\prime }){ \mathcal P}_{21^{\prime }}(t^{\prime \prime },t^{\prime })={\mathcal %
D}_{12}(t,t^{\prime }){ \mathcal P}_{21^{\prime }}(t^{\prime }).
\end{equation}
(Note, that the population number ${ \mathcal P}$ in the equation above depends on
the right time $t^{\prime }$). 

The lowest approximation as well as the equation for the self-operator, which is
convenient for iterations, can be obtained from its explicit expression in
terms of functional derivatives. The equation for the self-energy follows 
from a comparison of the two forms of equations for
the GF, one, Eqn.73, in terms of self-energy, and another, in terms of 
functional derivatives Eqn.72, and we obtain,

\begin{eqnarray}
S_{11^{^{\prime \prime }}} &(t,t^{\prime \prime })=&\frac 1i\theta
_{11^{^{\prime \prime }}}^\Gamma \delta (t-t^{\prime \prime })+\{\frac 1i[%
{ \mathcal P}_{12}(t^{+})+\hat{R}_{12}^l(t^{+})]V_{23} \nonumber \\
&&-\frac 1iV_{24}[\bar{{ \mathcal P}}_{14}(t^{-})+\hat{R}_{14}^r((t^{-})]%
\tau _{23}\}{ \mathcal G}_{31^{\prime }}(t,t^{\prime })\}{\mathcal G}%
_{1^{^{\prime }}1^{^{\prime \prime }}}^{-1}((t^{\prime },t^{\prime \prime }).
\end{eqnarray}

\subsection{The Hubbard-I approximation. }

The Hubbard-I  approximation is obtained if we put
all the functional derivatives equal to zero, $\hat{R}{ \mathcal G}=0$, which results in

\begin{equation}
S^{(H)}_{11^{^{\prime \prime }}}(t,t^{\prime \prime })=
\Sigma^{(H)}_{11^{^{\prime \prime }}}(t,t^{\prime \prime })=
\frac 1i\delta
(t-t^{\prime \prime })\{\theta _{11^{\prime \prime }}^\Gamma +{ \mathcal P}%
_{12}(t^{+})V_{21^{\prime \prime }}-V_{24}\bar{{ \mathcal P}}_{14}(t^{-})\tau
_{21^{^{\prime \prime }}}\}.
\end{equation}
Since this term does not contain the factor $G^{-1}$, the
 $S$-self-operator here  can be interpreted also as self-energy (denoted as $\Sigma $).
If this expression will be used for a system in thermodynamical equilibrium
and far from a phase transition to a state with intermediate valence,
Hubbard's arguments can be applied, namely, that instead of fully
self consistent calculation of population numbers one can use just the
atomic-like solution for them, 
${ \mathcal P}_{12}(t_1^{+})\approx {\mathcal P}_{12}^{(0)}(t_1^{+})$. 
  Note that by
putting functional derivatives of the GF equal to zero, we automatically neglect
all the graphs, which renormalize the population numbers. Thus, this is the
lowest possible approximation. In the equilibrium state external fields $%
{ \mathcal U}=0$. Therefore, all non-diagonal population numbers ${\mathcal P}%
_{14}^{\eta \eta }$ =0. Then the corners $\langle Tcc\rangle ,\langle
TXX\rangle $ of the full matrix of GF give zeroes and it is sufficient to
consider only the sub-block with $\eta =c,X$ and $\eta ^{\dagger }=c^{\dagger
},X^{\dagger }$ . Then the expression

\begin{equation}
\Sigma _{13}^{(H)}(t_1,t_3)\mid_{{\cal U}=0}
=\frac 1i\delta (t_1^{+}-t_3)[\theta _{13}^\Gamma +{ \mathcal P}_{12}
(t_1^{+})V_{23}],
\end{equation}
gives the following system of equations in the supermatrix form,

\begin{eqnarray}
&&\left( 
\begin{array}{cc}
i[D_{cc}^0(t)]_{jL,j_1L_1}^{-1}-\Theta _{jL,j_2c}\tilde{W}_{j_2c,j_1L_1}^{*}
& -w_{jL,j_1b}-\Theta _{jL,j_2c}\tilde{t}_{j,j_1}^{cb} \\ 
-\tilde{w}_{ja,j_1L_1}^{\uparrow } & i[D_{XX}^0(t)]_{ja,j_1b}^{-1}-\tilde{t}%
_{j,j_1}^{\uparrow ab}
\end{array}
\right) 
\times  \nonumber \\
&& 
\left( 
\begin{array}{cc}
G_{j_1L_1,j^{\prime }L^{\prime }}^{cc}(t-t^{\prime }) & G_{j_1L_1,j^{\prime }%
\bar{a}^{\prime }}^{cX}(t-t^{\prime }) \\ 
G_{j_1b,j^{\prime }L^{\prime }}^{Xc}(t-t^{\prime }) & G_{j_1b,j^{\prime }%
\bar{a}^{\prime }}^{XX}(t-t^{\prime })
\end{array}
\right) 
=  \nonumber \\
&&i\delta (t-t^{\prime }) 
\left( 
\begin{array}{cc}
O_{jL,j^{\prime }L^{\prime }}^{-1} & \Theta _{jL,j^{\prime }b}P_{j^{\prime
}}^{b\bar{a}^{\prime }} \\ 
P_j^{a\bar{b}}\bar{\Theta}_{j\bar{b},j^{\prime }L^{\prime }} & \delta
_{jj^{\prime }}P_{j^{\prime }}^{a\bar{a}^{\prime }}.
\end{array}
\right) 
\end{eqnarray}
The Latin letters $D$ are used for $cc$- and $XX$-components of the 
pseudolocators, whereas their inverse pseudolocators here are the following
differential operators:

\begin{eqnarray}
i[D_{cc}^0(t)]_{jL,j^{\prime }L^{\prime }}^{-1} &\equiv &\delta _{jj^{\prime
}}\delta _{LL^{\prime }}i\partial _t-(O^{-1}H)_{jL,j^{\prime }L^{\prime }},
\\
i[D_{XX}^0(t)]_{ja,j^{\prime }b}^{-1} &\equiv &\delta _{jj^{\prime }}\delta
_{ab}(i\partial _t-\Delta _{\bar{a}}),
\end{eqnarray}
and the interactions which are renormalized by 
(convoluted with) the population
numbers,  are denoted by a tilde, 

\begin{equation}
\tilde{W}_{j_2c,j_1L_1}^{*}\equiv P_j^{c\bar{b}}W_{j_2\bar{b},j_1L_1}^{*},\;%
\tilde{t}_{j,j_1}^{cb}\equiv P_{j_2}^{c\bar{a}}t_{j_2j_1}^{\bar{a}b},\;%
\tilde{w}_{ja,j_1L_1}^{\uparrow }\equiv P_j^{a\bar{b}}w_{j\bar{b}%
,j_1L_1}^{\uparrow },\;\tilde{t}_{j,j_1}^{\uparrow ab}\equiv P_j^{a\bar{c}%
}t_{j,j_1}^{\uparrow \bar{c}b}.
\end{equation}

Finally, the $\Theta $-matrix is:

\begin{equation}
\Theta _{jL,j^{\prime }c}\equiv O_{jL,j^{\prime }M}^{-1}(f_M)^c,\;\bar{\Theta%
}_{j\bar{b},j^{\prime }L^{\prime }}\equiv (f_M^{\dagger })^{\bar{b}%
}O_{jM,j^{\prime }L^{\prime }}^{-1}.
\end{equation}
Thus, the new element, which appears in our Eqs.82
compared to normal LDA-based calculation, is that both effective
mixing and hopping matrix elements are decreased by the
population numbers of many-electron states of the ion, since usually they are less than
one. The latter is always the case if mixing of localized and delocalized 
electrons is not equal to zero.
It is worth noticing that any further corrections to
the population numbers, which do not change their locality 
and do not introduce an energy dependence, 
do not change the structure of these equations. The
approximation, where these population numbers are dressed, but still do not
depend on energy, we will call "Hubbard-I" approximation. 
The corresponding self-energy is shown in Fig.1. 


Let us emphasize what is achieved for a moment 
by this construction, {\em i.e.},
 which are the new features of our "Hubbard-I" equations, which differ 
them from the standard LDA ones.\\
{\em First}, we passed from the ground-state description of the system to the 
statistical one, therefore, some of thermodynamical properties can be 
calculated. \\
{\em Second}, we know now where the $f$-level is situated with respect to the
 Fermi energy
and other bands just by construction of the Hamiltonian. This is important 
for example, for calculation of contribution of mixing interaction into 
effective interaction of conduction and localized electrons. \\
{\em Third}, in spite of the allowance of mixing interaction between 
localized and delocalized electrons the atomic-like Hund's rules can 
be taken into account (already at the
level of constructing of the zero Hamiltonian ).\\
{\em Fourth}, the possible degeneracy of many-electron states involved is taken into
account by the form (25) of the population numbers.\\
{\em Fifth}, since the single-site contributions from the Coulomb and exchange
interactions are included by construction into the energies of 
many-electron states 
$E_\Gamma $, the calculation of an effective exchange interaction 
between the angular momenta (also, interaction between 
quadrupole momenta, ..., \emph{etc.,}) of
different atoms does not require special care, which is necessary within the
standard LDA model for heavy rare earths \cite{brooks}. \\
{\em Finally}, the interaction of quasilocalized electrons with itselves
 (self-interaction correction) is taken into
account by construction.

 It is obvious from a physical point of view, however, that the
approximation described above has very serious drawbacks. If some of $f$-ions has 
$n$   $f$-electrons in the ground state, the state with $n+1$ electrons 
belongs to the
excited ones and usually is separated by such a large energy gap from the
ground state, that the energy difference 
$\Delta_{\Gamma,\Gamma '}^{n+1,n} \equiv 
E_{\Gamma _{}}^{(n+1)}-E_{\Gamma _{}^{\prime }}^{(n)}$ exceeds 
the Fermi energy and lies in the energy
interval of the delocalized part of the
 spectrum of single-electron excitations.
Corresponding excitations unavoidably should become delocalized. 
Therefore, part of the spectral weight should be moved into 
this energy region and the
renormalization of the population numbers (combination of which describes in
this approximation also the spectral weights) and, also,
the energies of the transitions $\Delta_{\Gamma,\Gamma '}^{n+1,n}$, should contain the 
contribution from this delocalized part of spectrum. The Hubbard-I 
approximation misses it. These corrections arise 
only in  higher approximations. The other drawback of this
approximation is that the atomic-like $f$-levels remain 
non-renormalized by the interaction of the $f$-ion with surrounding medium.
This shift should arise due to admixture of the wave 
functions of neighboring
sites to the one of the reference site and interactions of $f$-electrons with
bosonic excitations (analogously to the Lamb shift in quantum
electrodynamics).
All these effects can be taken into account via the corrections to
the self-energy and we turn now to the formulation of this part of theory.

\subsection{ Vertexes and equation for self-operator in terms of vertex. }
In order to obtain the next
corrections we insert the GF, ${ \mathcal G }_{36}$, in the form of Eqn.75
into the expression for the self-operator, Eqn.76. 
The differential operators $\hat{R}$ acting on the
GF ${ \mathcal G}$ gives,
\begin{eqnarray}
&&\hat{R}_{12}(t_1){ \mathcal G}_{31^{\prime }}(t,t^{\prime })=\hat{R}%
_{12}(t_1)[{ \mathcal D}_{36}(t,t^{\prime }){\mathcal P}_{61^{\prime
}}(t^{\prime })]=  \nonumber \\
\lbrack  &&\hat{R}_{12}(t_1){ \mathcal D}_{36}(t,t^{\prime })]{\mathcal P}%
_{61^{\prime }}(t^{\prime })+{ \mathcal D}_{36}(t,t^{\prime })[\hat{R}%
_{12}(t_1){ \mathcal P}_{61^{\prime }}(t^{\prime })]=  \nonumber \\
&&-{ \mathcal D}_{39}(t,t_9)[\hat{R}_{12}(t_1){\mathcal D}_{98}^{-1}(t_9,t_8)]%
{ \mathcal G}_{81^{\prime }}(t_8,t^{\prime })+  \nonumber \\
&&{ \mathcal D}_{36}(t,t^{\prime })[\hat{R}_{12}(t_1){\mathcal P}_{61^{\prime
}}(t^{\prime })]
\end{eqnarray}
where the trick\cite{baym_kadanoff}, 
\begin{eqnarray}
&&\hat{R}_{12}(t_1)[{ \mathcal D}_{36}^{-1}(t_3,t_6){\mathcal D}%
_{67}(t_6,t_7)]=0\Rightarrow   \nonumber \\
\hat{R}_{12}(t_1){ \mathcal D}_{36}(t_3,t_6) &=&-{\mathcal D}_{39}(t_3,t_9)[%
\hat{R}_{12}(t_1){ \mathcal D}_{98}^{-1}(t_9,t_8)]{\mathcal D}_{86}(t_8,t_6),
\end{eqnarray}
has been used. Having in mind that the $\hat{R}$-operator is equal to zero
when both indexes run over $c$-electrons, we restore its definition via the
matrices $\alpha $ for the $\hat{R}^l$- and $\beta $ for the $\hat{R}^r$- 
operators and introduce the vertexes in a standard way, namely, as
functional derivative of the inverse GF with respect to external field,

\begin{eqnarray}
\hat{R}_{12}^l(t_1^{+}) &&{ \mathcal D}_{78}^{-1}(t_7,t_8)\equiv
\lim_{t_4\rightarrow t_1+0}\alpha _{12,4_b}^{\eta \eta ^{\dagger }}\frac
\delta {\delta {\cal U}_{j_4}^{\xi _4}(t_4)}{ \mathcal D}_{98}^{-1}(t_9,t_8) 
\nonumber \\
&\equiv &\lim_{t_4\rightarrow t_1+0}\alpha _{12,4_b}^{\eta \eta ^{\dagger
}}\Gamma _{98,4_b}^l(t_9,t_8;t_4)\equiv \alpha _{12,4_b}^{\eta \eta
^{\dagger }}\Gamma _{78,4_b}^l(t_7,t_8;t_1^{+}),\; \\
\hat{R}_{12}^r(t_1^{-}) &&{ \mathcal D}_{78}^{-1}(t_7,t_8)\equiv
\lim_{t_4\rightarrow t-0}\beta _{12,3_b}^{\eta \eta }\frac \delta {\delta
{\cal U}_{j_3}^{\zeta _3}(t_4)}{ \mathcal D}_{98}^{-1}(t_9,t_8) \nonumber \\
&\equiv &\lim_{t_4\rightarrow t_1-0}\beta _{12,4_b}^{\eta \eta }\Gamma
_{98,4_b}^r(t_9,t_8;t_4)\equiv \beta _{12,4_b}^{\eta \eta }\Gamma
_{98,4_b}^r(t_9,t_8;t_1^{-}).
\end{eqnarray}
It is worth noting that in our case the role of standard GF is played by the
pseudolocators, and in this respect the introduced vertexes are indeed the
standard ones. The new elements, which arise in the techniques for 
non-fermion operators, are the derivatives $\hat{R} { \mathcal P}$. They 
can be written via the correlation functions of Bose-like transitions (see 
Eq.95 below) and we will discuss them a little later.
 The other new element (compared to the standard case) is that%
\emph{\ }the vertexes which enter the equations in the '' left'' and ''right''
limits. The origin of  this difference is seen from the definitions of the
operators $\hat{R}_{12}^l(t_1^{+})$ and $\hat{R}_{12}^r(t_1^{-})$; the
vertex in the left limit arises in the usual technique too, while the one in the
right limit arises from transitions which involve creation or
destruction of two particles on one site simultaneously. This process is
described in the equations of motion by the anticommutator $\{\eta _1,\eta
_2\}$, which is equal to zero in the case of usual fermions, but is not
in the case of two-electron intra-atomic transitions. 

Using these expressions we obtain the following equation for the 
self-operator,

\begin{eqnarray}
S_{11^{\prime \prime }}(t,t^{\prime \prime }) &=&\Sigma _{11^{\prime
\prime }}^H(t,t^{\prime \prime })-\frac 1iV_{23}{ \mathcal D}%
_{39}(t,t_9)\alpha _{12,5_b}^{\eta \eta ^{+}}\Gamma _{91^{\prime \prime
},5_b}^{}(t_9,t^{\prime \prime };t^{+})+  \nonumber \\
&&\frac 1iV_{24}\tau _{23}{ \mathcal D}_{39}(t,t_9)\beta _{14,5_b}^{\eta \eta
}\Gamma _{91^{\prime \prime },5_b}^{}(t_9,t^{\prime \prime };t^{-})+ 
\nonumber \\
&&\left( 
\begin{array}{c}
\end{array}
\right. \frac 1iV_{23}{ \mathcal D}_{36}(t,t^{\prime })[\hat{R}_{12}^l(t^{+})%
{ \mathcal P}_{61^{\prime }}(t^{\prime })]-  \nonumber \\
&&\frac 1iV_{24}\tau _{23}{ \mathcal D}_{36}(t,t^{\prime })[\hat{R}%
_{14}^r(t^{-}){ \mathcal P}_{61^{\prime }}(t^{\prime })]\left) 
\begin{array}{c}
\end{array}
\right. { \mathcal G}_{1^{\prime }1^{\prime \prime }}^{-1}(t^{\prime
},t^{\prime \prime }).
\end{eqnarray}

Thus, structurally the self-operator can be separated into the sum of a few terms

\begin{equation}
S =S_H^{}+S_1^{}+S_P^{},\; \; S_H=\Sigma_H; \;\;
S_1=\Sigma_1; \;\;\Sigma _1^{}=\Sigma
_L^{}+\Sigma _R^{},
\end{equation}
where $\Sigma _L^{}$and $\Sigma _R^{}$ are the terms which contain left 
($\sim \alpha $) and
right ($\beta $) vertexes and are genius self-energies $\Sigma $ ({\em i.e.}, 
the graphs, which cannot be separated into two non-linked graphs but cutting one line 
of $D$-function), 
 while $S_P^{}$, the contribution, proportional to $G^{-1}$,
 contains derivatives of the population numbers and renormalizes end-factors.
 Let us also
consider the corrections to $\Sigma _{1.}^{}$
First we need the bare vertex. 
The definition of it arises if 
we replace the full inverse pseudolocator in Eqns.86-87 by the 
bare one, ${ \mathcal D}_{}^{-1}\rightarrow { \mathcal D}_0^{-1}.$ 
Since $\hat{R}^l(t){\mathcal D}%
_0^{-1}=\alpha (\delta { \mathcal D}_0^{-1}/\delta {\mathcal U})=-\alpha
(\delta A/\delta { \mathcal U})$, and a similar relation holds 
for $\hat{R}^r(t) { \mathcal D}$, the number of different bare 
vertexes is determined by the number
of different combinations of  external fields $A$, (Appendix D, Eqs.1-8). 
From the definition of the self-operator 
${ \mathcal D}_{}^{-1}={\mathcal D}_0^{-1}-S $ we see that the exact
total vertex 
being derivatives of the self-operator with respect to external fields,
can be rewritten in the form of a sum of the
bare vertex $\gamma $ and corrections, 
\begin{equation}
\Gamma _{98,5_b}^{}(t_9,t_8;t_5)=\gamma _{98,5_b}^{}(t_9,t_8;t_5)-\frac{%
\delta S_{98}(t_9,t_8)}{\delta { \mathcal U}_{5_b}(t_5)},
\end{equation}

\begin{equation}
\gamma _{98,5_b}^{}(t_9,t_8;t_5)=-\delta (t_9-t_8)\frac{\delta A_{98}(t_8)}{%
\delta { \mathcal U}_{5_b}(t_5)}.
\end{equation}
These relations are useful for obtaining subsequent
corrections to the vertex, substituting into Eqs.86,87
approximate expressions instead of the full self-operator. 
We will return to it a bit
later. Before that, 
let us rewrite with the help of Eqs. 76,88 the equation for the
full GF in a form which clarifies the physical 
meaning of different terms
in the self-operator (Eqn.89). Using the pseudolocator in the
Hubbard-I approximation,  ${ \mathcal D}_H^{-1}={\mathcal D}_0^{-1}-\Sigma _H^{}
$, we find,
\begin{eqnarray}
&&\{({ \mathcal D}_{}^H(t,t^{\prime \prime }))_{11^{\prime \prime }}^{-1}-\frac
1i{ \mathcal D}_{39}(t,t_9)[V_{23}\alpha _{12,5_b}^{\eta \eta ^{+}}\Gamma
_{91^{\prime \prime },5_b}^{}(t_9,t^{\prime \prime };t^{+})-  \nonumber \\
- &&\frac 1iV_{24}\tau _{23}\beta _{14,5_b}^{\eta \eta }\Gamma _{91^{\prime
\prime },5_b}^{}(t_9,t^{\prime \prime };t^{-})]\}{ \mathcal G}_{1^{\prime \prime
}1^{\prime }}(t^{\prime \prime },t^{\prime })=  \nonumber \\
&&{ \mathcal P}_{11^{\prime }}(t,t^{\prime })+\Delta {\mathcal P}_{11^{\prime
}}(t,t^{\prime }),
\end{eqnarray}
where the renormalization of the population numbers is given by the
following expression:
\begin{eqnarray}
\Delta {\mathcal P}_{11^{\prime }}(t,t^{\prime }) &\equiv &-\frac 1iV_{23}%
{ \mathcal D}_{36}(t,t^{\prime })[\hat{R}_{12}^l(t^{+}){\mathcal P}_{61^{\prime
}}(t^{\prime })]+  \nonumber \\
&&\frac 1iV_{24}\tau _{23}{ \mathcal D}_{36}(t,t^{\prime })[\hat{R}%
_{14}^r(t^{-}){ \mathcal P}_{61^{\prime }}(t^{\prime })],
\end{eqnarray}
Thus, we see that the self-operator can be represented as $S=\Sigma + \Delta P\cdot 
G^{-1}$ what allows to separate the graphs to the ones which renormalize the 
pseudolocator $D$ and are the contributions to the self-energy, and the ones which 
renormalize the end-factor $\Delta P$.
Let us turn now to the Bose-type of GFs. 
In order to obtain the 
equation for these GFs, we will
use again the short-index notations 1$_b\equiv (j_1\xi _1)$ and rewrite
the equation for the boson-like GF in terms of  vertex, Eqn.86 (we have to use
once more time the Kadanoff and Baym's trick-Eqn.85). Then we obtain,
\begin{eqnarray}
&&\lbrack  L^0(t_1,t_2)]_{1_b;2_b}^{-1}K_{2_b;1_b^{\prime }}(t_2,t_1^{\prime
})=\delta (t_1-t_1^{\prime })P_{1_b1_b^{\prime }}(t_1^{\prime })+  \nonumber
\\
&&\lim_{t_2\rightarrow t_1+0}\frac 1iV_{1;2}^{1_b}\langle TZ_{1_b^{\prime
}}(t^{\prime })\rangle _uG_{21}^{}(t_1,t_2)+  \nonumber \\
&&\lim_{t_2\rightarrow t_1+0}V_{1;2}^{1_b}[\frac 1iD_{24}(t_1,t_4)\Gamma
_{45;1_b^{\prime }}(t_4,t_5;t_1^{\prime })G_{51}(t_5,t_2)-  \nonumber \\
&&D_{23}^{}(t_1,t_2)\bar{\alpha}_{31^{\prime },3_b}^{\eta \eta ^{\dagger
}}[K_{3_b;1_b^{\prime }}(t_1^{\prime },t_2)-\langle TZ_{1_b^{\prime
}}(t_1^{\prime })\rangle _u\langle TZ_{3_b}(t_2)\rangle _u]. 
\end{eqnarray}
Here we have used the definition of $P_{31^{\prime }}(t_2)$, Eqs.60 and 61
and represented the derivative $\delta P/\delta U$  in the last term of this
equation in the form,

\begin{equation}
\bar{\alpha}_{31^{\prime },3_b}^{\eta \eta ^{\dagger }}\frac{\delta
P_{3_b}(t_2)}{\delta {\cal U}_{1_b^{\prime }}(t_1^{\prime })}=  
\bar{\alpha}_{31^{\prime },3_b}^{\eta \eta ^{\dagger }}\frac
1i[K_{3_b;1_b^{\prime }}(t_1^{\prime },t_2)-\langle TZ_{1_b^{\prime
}}(t_1^{\prime })\rangle _u\langle TZ_{3_b}(t_2)\rangle _u].
\end{equation}

This completes the derivation of the exact equation 
for the boson-like GFs.
Eqs.94,95 should be used also in Eqs.84,88 and 90
 for calculation of $\hat{R}^lP$ and $\hat{R}^rP$.

The way of deriving different approximations 
for the boson type of GFs is quite similar to the
one used above for the fermion-like excitations. First, of course, we have
to define the zero-order GF, putting $V=0$ in Eq.94, which gives
\begin{equation}
K_{2_b;1_b^{\prime }}^0(t_2,t_1^{\prime })=L_{2_b;3_b}^0(t_2,t_1^{\prime
})P_{3_b1_b^{\prime }}^0(t_1^{\prime }).
\end{equation}
Then an analogue of the Hubbard-I approximation for fermion-like 
transitions arises if we put in Eq.94  the vertex $\Gamma = 0$.  
As seen from the Hamiltonian, within the Hubbard-Anderson model,  the only 
way a renormalization of the boson-like GFs arises is due to 
mixing interaction and hopping, therefore, it requires knowledge of the electron GF. 
The latter creates fields which shift the energies of the bosonic transitions. 

\subsection{Mean field approximation and self-consistent equation for Hubbard $U$. }

As seen from Eq.92, the part of
the self-consistency which does not involve the 
effects of $S_P$, can be provided in the  usual way: 
a choice of certain approximation for the vertexes, $%
\Gamma \rightarrow \Gamma _{}^{(appr)}$, leads to a 
corresponding expression
for the self-energy via Eq.(88), while the GFs entering the $\Gamma
_{}^{(appr)}$ (if $\Gamma \neq \gamma $) contain the same $\Sigma $ and $%
\Gamma _{}^{(appr)}.$ 
According to this scheme the simplest approximation
which is beyond the Hubbard-1 approximation, should 
contain the vertex. It arises if we neglect the terms  
$\hat{R}\Sigma $ in the equations for $%
\Gamma $, i.e., $\Gamma \rightarrow \gamma .$ Since the $ff$-part of $%
D_0^{-1}$ is purely local, and the bare vertex does not depend on energy,
the diagonal parts of the loops

\begin{eqnarray}
\Sigma ^{loop}_{11^{\prime \prime }}(t,t^{\prime \prime }) 
&=&-\frac 1iV_{23}%
{ \mathcal D}_{39}(t,t_9)\alpha _{12,5_b}^{\eta \eta ^{+}}\gamma _{91^{\prime
\prime },5_b}^{}(t_9,t^{\prime \prime };t^{+})+  \nonumber \\
&&\frac 1iV_{24}\tau _{23}{ \mathcal D}_{39}(t,t_9)\beta _{14,5_b}^{\eta \eta
}\gamma _{91^{\prime \prime },5_b}^{}(t_9,t^{\prime \prime };t^{-})
\end{eqnarray}
renormalize the centers of $f$-bands.

Indeed, since we assumed that $f$-functions, belonging to different sites do
not overlap with each other, $O_{i,L=f;j,L^{\prime }=f}=0$, we see that the
energy $\Delta _{\bar{a}}$ of the transition $a=[\Gamma _n,\Gamma _{n+1}]$
is shifted by $\delta \Sigma ^{loop}$; if we neglect the non-diagonal
corrections, we find the following equation:
\begin{equation}
\Delta _{\bar{a}}^{*}=\Delta _{\bar{a}}+( \Sigma ^{loop})_{1=(i,\bar{a}%
),2=(i,\bar{a})}\equiv \Delta _{\bar{a}}+\delta \Delta _{\bar{a}}(\Delta _{%
\bar{a}}^{*}).
\end{equation}
This is self-consistent equation for transition energy; renormalization
comes from hopping and mixing interaction. If we take Hubbard's estimate for 
$E_{\Gamma _n},$ $E_{\Gamma _n}=\varepsilon _f^0n+\frac 12Un(n-1)$, then the
energies of the transitions $\Delta _2^0=\Delta _{\bar{a}=[\Gamma
_{n+1},\Gamma _n]}=E_{\Gamma _{n+1}}-E_{\Gamma _n}\approx U/2$; the lower
transition  $\Delta _1^0=\Delta _{\bar{a}=[\Gamma _n,\Gamma _{n-1}]}\approx
U/2$ too. Here we used the Hubbard's estimate\cite{hubb1} for $\varepsilon
_f^0$: the main contribution to this energy comes from attraction to
nucleous, which provides localization of certain number $n_0$ of $f$%
-electrons. Therefore, $dE_n/dn=0$ gives $\varepsilon _f^0=-(n_0-\frac 12)U.$
Correspondently, a fermion $ff$-GF has two poles. The lower one is
 located near energy of
lower transition $\Delta _1^0$ and the upper one $\Delta _2^0$ ; the Hubbard
bands are developed near these energies.  The shifts of $\Delta _i^0$ (i=1,2) 
due to hopping and mixing can be roughly estimated as 
\begin{equation}
\delta \Delta _i\approx \sum_p \tilde{t}_{hop}^i(p)f(E_p^i)
i+\sum_{p,\lambda}\frac{V_\lambda ^{*%
\bar{a}_i}(p)V_\lambda ^{a_i}(p)}{\Delta _i^{*}-\varepsilon _{\lambda p}}%
f(E_p^i),\;i=1,2.
\end{equation}
Here $\tilde{t}_{hop}^i(p)$ is an effective hopping, containing all chain 
($\sim $ Hubbard-I approximation) contributions. However, at large enough 
$\Delta$ main contribution comes from the second order 
($\sim t^2/\Delta \sim t^2/U $).\\
It is also  easy to see that the 
equation $\Delta^*=\Delta  + \delta \Delta (\Delta ^*)$
can have two solutions, one of which is located 
near Fermi energy, which is often called as "Kondo resonance"\cite{kondores}.
These two equations show that, contrary to the $s$-band Hubbard model, one
can expect that in reality the renormalizations of $\Delta _1^0$ and $\Delta
_2^0$ should be quite different, at least, for rare earth ions: on the one
hand, hopping and mixing are strongly suppressed in the region of energies $%
\omega \sim \Delta _1^0$ since this energy is very deep under Fermi surface
and even below the bottoms of conduction bands; on the other hand, in the
region of energies $\omega \sim \Delta _2^0$ hopping and mixing should be
much stronger, therefore, in the region of energies in vicinity of Fermi level
 the mechanism described by Eq.99 can be much
more effective. It is clear, however, that this mechanism is not able to
provide strong reduction of the bare Hubbard repulsion
 $\Delta _2^0-\Delta _1^0\simeq U$
from $20eV$ to $5\div 7eV$ (the latter is the value usually obtained within
LDA-based calculations). Therefore, other, stronger contributions, should be
generated by full Hamiltonian. Nevertheless, it worth to note that the
non-diagonal matrix elements of hopping, $t^{\lambda \mu }(f_\lambda
^{\dagger })^{\bar{a}}(f_\lambda )^b$ and effective hopping via mixing, $%
\sim [V_\lambda ^{*\bar{a}_i}(p)V_\lambda ^{b_i}(p)/(\omega -\varepsilon
_{\lambda p})](1-\delta _{ab})$, split the Eq.99 to the system of
equations.

At last, its $cf$-parts renormalize the mixing interaction. 
The self-energy which includes the Hubbard-I and  the loop 
correction Eqs.(97,98),  still does not depend on energy. 
Since this approximation is also beyond the  
Hubbard-I approximation, we will call it mean field approximation (MFA).
 Similar approximation for multiple-orbital Hubbard model 
but for the simpler case of orthogonal basis set has been 
considered earlier\cite{sand_joh}.
The fact, that the equations for fermion-like GFs always contain 
vertexes in combination with the matrices $\alpha $ and 
$\beta $ allows to use a shorter graphical notation:
the sums $\alpha \cdot \gamma $ and $\beta \cdot \gamma $ 
can be represented as four-leg vertexes, as shown in Fig.2. 
Then, in this 
graphical language the loop correction, Eq.97, acquires the 
form shown in Fig.3. 

\subsection{Further corrections to vertexes and self operator. }

In order to obtain the second-order corrections to vertexes and
self-energy we insert $\delta \Sigma =\delta \Sigma ^{loop}$,
Eq.97, 
into the definition of the vertexes Eqs.86,87,90,91. 
Since neither the interaction $V$, nor the
matrix $\tau $ depend on the external fields, only pseudolocators entering
the $\Sigma ^{(1)}$ give non-zero contribution when we calculate the
functional derivatives $\hat{R}\Sigma $. Therefore, we can construct the
higher-order corrections simply by formulation of the rule of
differentiation of the pseudolocator lines. 
This rule is displayed in Fig.4.
Applying it to the graphs of the $\Sigma ^{(1)}$ , shown in Fig.3, we obtain
the desired corrections of second order to vertexes (displayed in Fig.5). 
Then, inserting 
these corrections to the definition of the self-operator via the vertex, Eq.88,
we obtain the corresponding second-order corrections to the self-operator.
 The graphs for these corrections 
are shown in Fig.6. The corrections to the vertex in Fig.5
 describe contributions of electron-hole excitations to the kinematic 
interactions , which arises due to strong 
correlations and are caused by mixing and hopping. 
Note, that  there are also other, formal, 
 difference with the standard case:
instead of pure GF one of the  shoulders of the electron-hole loop contains the
product of the line of interaction $V$ (wiggle line) and the pseudolocator.

\subsection{The screening}

The next obvious step is to take into account the screening effects. 
The lowest
approximation is an analogue of the standard
random-phase approximation, used in the theory of normal metals. In order 
to avoid possible confusion, we remind the reader that 
here again it renormalizes not the Coulomb
interaction but only mixing and $ff$-hopping. 
One way to construct it would be to
solve the integral equation for vertex, generating the series of
''electron-hole'' loops (see Fig.4) and insert the solution of this
equation, i.e., corrected vertex, into self-energy for pseudolocator; 
then to derive similar corrections to the population numbers (r.h.s. of
Eq.92 for the GF). The equivalent, but shorter way, which automatically
renormalizes the interaction $V$ in all corrections, is to make an exact
transformation of the shift in the external field of Eqs. 40-43,52,92,94,
 introducing the
description in terms of effective fields, $\Phi=A+\Sigma ^{(loop)}$. The
quasifermion GFs depend on the external field ${\cal U}$ only via the
''gauge'' field $A_{12}(t)\equiv \nu _{12}^{6_b}{\cal U}_{6_b}(t)$, where 
the matrices $\nu $ are listed in Appendix D.
Therefore, we can express
equations for these GFs in terms of derivatives $\delta /\delta A^{eff},$
schematically, we have
\begin{equation}
V\frac \delta {\delta {\cal U}_{1_b}(t)}=(V\frac{\Phi _{12}(t_1)}{%
\delta {\cal U}_{1_b}(t)})\frac \delta {\Phi_{12}(t_1)}=[V(\nu
_{12}^{1_b}+\frac{\delta \Sigma ^{(loop)}_{12}(t_1)}{\delta {\cal U}_{1_b}(t)}%
)]\frac \delta {\Phi_{12}(t)}.
\end{equation}
Let us now perform this calculation. First we have to separate out 
the loop contribution (97) of the
self-energy  in the equation for GF (Eqn.92).
This one-loop correction contains zero vertex $\gamma $;
self-consistency requires that the full GF, which enters the loop, should
contain the same self-energy, which is taken into account in the equation
for GF. Taking into account that 
\begin{equation}
\gamma _{91^{\prime \prime },5_b}(t_9,t^{\prime \prime };t)=-\delta
(t_9-t^{\prime \prime })\delta (t_9-t)\nu _{91^{\prime \prime }}^{5_b},
\end{equation}
and including  the self-energy correction $\Sigma ^{(loop)}$ into the
effective field,

\begin{eqnarray}
\Phi_{11^{\prime \prime }}(t,t^{\prime \prime }) &\equiv &A_{11^{\prime
\prime }}(t,t^{\prime \prime })+\frac 1i\alpha _{12;5_b}^{\eta \eta
^{\dagger }}V_{23}D_{39}(tt^{+})\nu _{91^{\prime \prime }}^{5_b}\delta
(t^{+}-t^{\prime \prime })+  \nonumber \\
&&\frac 1i\beta _{12;5_b}^{\eta \eta }V_{24}\tau _{23}D_{39}(tt^{-})\nu
_{91^{\prime \prime }}^{5_b}\delta (t^{-}-t^{\prime \prime }),
\end{eqnarray}
we find, that the factor $Y_{11^{\prime \prime },2_b}(t,t^{\prime })\equiv
\delta \Phi_{11^{\prime \prime }}(t)/\delta {\cal U}_{2_b}(t^{\prime })$
which renormalizes the interaction, should satisfy the following equation,

\begin{eqnarray}
Y_{11^{\prime \prime },2_b}(t,t^{\prime }) &=&\nu _{91^{\prime \prime
}}^{5_b}[\delta _{91}\delta _{2_b5_b}\delta (t-t^{\prime
})-D_{3,10}(t,t_{10})\frac{\delta D_{10,11}^{-1}(t_{10},t_{11})}{\delta 
{\cal U}_{2_b}(t^{\prime })}\times   \nonumber \\
&&\{\alpha _{12;5_b}^{\eta \eta ^{\dagger
}}V_{23}D_{11,9}(t_{11},t^{+})\delta (t^{+}-t^{\prime \prime })+  \nonumber
\\
&&\beta _{14;5_b}^{\eta \eta }V_{24}\tau _{23}D_{11,9}(t_{11},t^{-})\delta
(t^{-}-t^{\prime \prime })\}].
\end{eqnarray}
Eq.103 contains full pseudolocator $D$.

\subsubsection{Fermi-liquid-like random phase approximation (RPA)}

If we substitute into  Eq.103 the bare vertex $\gamma =\delta
D_0^{-1}/\delta {\cal U}_{1_b}$ instead of the full vertex $\Gamma =\delta
D_{}^{-1}/\delta {\cal U}_{1_b}$, we obtain only the first correction to $Y.$ In
order to obtain a closed equation for $Y$, we have to take into account in $D$
the same correction to $\Sigma $, namely, $\Sigma ^{loop}.$ Since 

\begin{equation}
\delta D_{0l}^{-1}/\delta {\cal U}_{1_b}=-\delta \Phi/\delta {\cal U}%
_{1_b}=-Y,
\end{equation}
with $D_{0l}^{-1} \equiv D_{0l}^{-1}-\Sigma ^{(loop)}$ 
($\delta $-functions are  omitted) we obtain the desired equation,

\begin{eqnarray}
Y_{11^{\prime \prime },2_b}(t,t^{\prime }) &=&\nu _{91^{\prime \prime
}}^{5_b}[\delta _{91}\delta _{2_b5_b}\delta (t-t^{\prime
})-D_{3,10}(t,t_{10})Y_{10,11;2_b}(t_{10},t^{\prime })\times   \nonumber \\
&&\{\alpha _{12;5_b}^{\eta \eta ^{\dagger
}}V_{23}D_{11,9}(t_{10},t^{+})\delta (t^{+}-t^{\prime \prime })+  \nonumber
\\
&&\beta _{14;5_b}^{\eta \eta }V_{24}\tau _{23}D_{11,9}(t_{10},t^{-})\delta
(t^{-}-t^{\prime \prime })\}].
\end{eqnarray}
 Note that the complex structure of coupling of the Fermi-like excitations to 
the Bose-like ones makes the screening matrix dependent on three indices.

\subsubsection{Full equation for the GF in terms of RPA GF}
Now let us write down the self-energy part of the self-operator in the l.h.s. of Eq.112,
which can be named $\Sigma _D$ (the index $D$ here emphasizes that
it relates only to the pseudolocator 
$D$), in the 
form: $\Sigma _D=\Sigma ^H+\Sigma^{loop}+\Sigma ^{rest}.$ 
The effective field in these terms is 
$\Phi=A+\Sigma ^{loop}.$ Note that since $\Sigma ^H$ is not 
local, it cannot be included into effective field $\Phi$. Then, the vertex $\Gamma $, 
defined by Eqs. (86), (87) and (90) can be expressed in terms of 
$Y$  as follows,

\begin{equation}
\Gamma _{91^{\prime \prime },5_b}(t_9,t^{\prime \prime
};t_4)=-[Y_{91^{\prime \prime },5_b}(t_9,t_4)\delta (t^{\prime \prime }-t_9)+%
\frac{\delta \Sigma _{91^{\prime \prime }}^{rest}(t_9,t^{\prime \prime })}{%
\delta \Phi_{67}(t_6)}Y_{67,5_b}(t_6,t_4)].
\end{equation}
With the help of this expression for the vertex 
we can represent the equation for the full GF (92)
in a form which does not contain the non-screened interaction,

\begin{eqnarray}
&&\{D_{RPA,11^{\prime \prime }}^{-1}(t,t^{\prime \prime })-\frac
1iD_{3,9}(t,t_9)[\alpha _{12,5_b}^{\eta \eta ^{\dagger }}\frac{\delta \Sigma
_{91^{\prime \prime }}^{rest}(t_9,t^{\prime \prime })}{\delta
\Phi_{67}(t_6)}V_{23}Y_{67,5_b}(t_6,t^{+})-  \nonumber \\
&&\beta _{14,5_b}^{\eta \eta }\tau _{23}\frac{\delta \Sigma _{91^{\prime
\prime }}^{rest}(t_9,t^{\prime \prime })}{\delta \Phi_{67}(t_6)}%
V_{24}Y_{67,5_b}(t_6,t^{-})]\}G_{1^{\prime \prime },1^{\prime }}(t^{\prime
\prime },t^{\prime })=  \nonumber \\
&&P_{11^{\prime }}(1,1^{\prime })+\Delta P_{11^{\prime }}(1,1^{\prime }),
\end{eqnarray}
where the derivatives $\delta /\delta {\cal U}$ 
in the formula (93) for $\Delta P$ are also 
rewritten in the 
form $\delta /\delta {\cal U}$ $=Y\;\delta /\delta \Phi.$ 
At last, the equation for the pseudolocator $D_{RPA}$ becomes,

\begin{eqnarray}
&&\lbrack \delta (t,t^{\prime \prime })\delta _{12}i\partial _t-\Sigma
_{12}^H(t,t^{\prime \prime })+\frac 1iV_{23}\alpha _{12,5_b}^{\eta \eta
^{\dagger }}D_{3,9}^{MFA}(t,t^{\prime \prime })Y_{91^{\prime \prime
},5_b}(t_9,t^{+})-  \nonumber \\
&&\frac 1iV_{24}\tau _{23}\beta _{14,5_b}^{\eta \eta
}D_{3,9}^{MFA}(t,t^{\prime \prime })Y_{91^{\prime \prime
},5_b}(t_9,t^{-})]D_{RPA,24}(t^{\prime \prime },t^{\prime })=\delta
_{24}\delta (t^{\prime \prime }-t^{\prime }).
\end{eqnarray}
The role played by $Y$ in the strong-coupling theory is analogous 
to the one played by dielectric function in the weak-coupling theory 
of Coulomb systems: it describes the screening of the interaction $V_{12}$.

\section{Coulomb interaction}
 
The detailed consideration  of the Coulomb interaction is given in next paper,
here show the technical idea of description and emphasize the new 
elements arising within this approach.

Full Coulomb interaction have been written in $X$-representation as follows: 
\begin{equation}
H_{coul}= \nonumber \\
\sum v_{1234}(c_1^{\dagger }+(f_1^{\dagger })^{\bar{a}_1}X_1^{\bar{a%
}_1})(c_2^{\dagger }+(f_2^{\dagger })^{\bar{a}_2}X_2^{\bar{a}%
_2})(c_3+(f_3)^{a_3}X_3^{a_3})(c_4+(f_4)^{a_4}X_4^{a_4}).
\end{equation}

The single-site terms $vf^{\dagger }f^{\dagger }ff$ have been transformed
into $\sum E_\Gamma h_j^\Gamma ;$ the terms which contain three single-site $%
f$-operator and one operator of conduction electron are included to mixing
interaction $Vc^{\dagger }X+h.c.;$ the terms of the kind $v(f^{\dagger
}f^{\dagger }f)_jf$ $_{j^{\prime }}(1-\delta _{jj^{\prime }})$ are included
into Hubbard's hopping Hamiltonian. Note that in case if some selection
rules happen to be valid for the single-particle matrix elements of hopping
and mixing interaction, the \emph{effective} matrix elements have different
selection rules due to presence of Coulomb terms in it. All the other terms
should be considered. Remind that due to large strength of on-site
interaction we are not allowed to decouple the operators belonging to the
same site. Thus, if these $f$-operators are not neighbors in the
Hamiltonian, we have to move (using rules for anti commutations) them to each
other and then either to transform the $f$-product into Hubbard operator(s)
or to transform each $f$-operator into $X$-operator and multiply them. This
procedure also helps to clarify a physical meaning of corresponding terms.
The  number of different terms in the Hamiltonian is large and we 
are not able to give here all formulas. Besides, a degree of 
importance of different terms is material- and scenario- dependent.
Thus, we will consider here few terms in order to show how the
developed above perturbation theory is to be modified in order to include 
Coulomb interaction between different sites (the description of the
 Coulomb intersite interaction within slave-boson technique has been 
considered earlier in Ref.3).

Let us start with the terms, which describe simultaneous ''hopping'' of two
electrons from one site to another: 
\begin{equation}
H_1^C=\sum v_{j1j2j^{\prime }3j^{\prime }4}(f_1^{\dagger }f_2^{\dagger })^{%
\bar{\zeta}}(f_3f_4)^{\zeta ^{\prime }}Z_j^{\bar{\zeta}}Z_{j^{\prime
}}^\zeta \equiv
 \sum C_{jj'}^{\bar{\zeta} \zeta '}Z_j^{\bar{\zeta}}Z_{j^{\prime}}^\zeta.
\end{equation}
One can expect that this term can be neglected in the lowest
approximations since it involves large-energy transition (Hubbard U in
denominator): although $[X^a,Z^{\eta}]$ gives $X^{\bar{b}}$ in equation 
of motion remaining operator $Z^{\eta}$ requires $Z^{\bar{\eta}}$
in order to obtain non-zero result; the GF
 $\langle T Z^{\eta} Z^{\bar{\eta}} \rangle $ gives makes 
the contribution small.
The reason why these processes have to be considered is that there are 
high-order processes (like assisted spin-flips) which contain in denominator 
small difference of these large energies.

The terms 
\begin{eqnarray}
&&H_1^C=\sum [v_{j\mu _1j^{\prime }\mu _2j^{\prime }\mu _3j\mu _4}(f_{_{\mu
_1}}^{\dagger }f_{_{\mu _4}})^\xi (f_{_{\mu _2}}^{\dagger }f_{_{\mu
_3}})^{\xi ^{\prime }}-  \nonumber \\
&&v_{j\mu _1j^{\prime }\mu _2j\mu _3j^{\prime }\mu _4}(f_{_{\mu
_1}}^{\dagger }f_{_{\mu _3}})^\xi (f_{_{\mu _2}}^{\dagger }f_{_{\mu
_4}})^{\xi ^{\prime }}]Z_j^\xi Z_{j^{\prime }}^{\xi ^{\prime }}\equiv \sum
C_{jj^{\prime }}^{\xi \xi ^{\prime }}Z_j^\xi Z_{j^{\prime }}^{\xi ^{\prime
}}.
\end{eqnarray}
do not change number of $f$-electrons in ions. Part of these terms describe
Coulomb screening of ions, part can be transformed into spin-spin,
orbital-orbital interactions \cite{irkhiny} and so on. We will not consider 
particular terms from the sum over $\xi ,\xi ^{\prime }$ here, however, note
that on one hand, instead of one Heisenberg term $-I_{jj^{\prime }}\mathbf{J}%
_j\mathbf{J}_{j^{\prime }}$ the expansion 
\begin{equation}
H_{exch}=-\sum I_{\alpha \beta \gamma }(\mathbf{J}_1\mathbf{R})^\alpha (%
\mathbf{J}_2\mathbf{R})^\beta (\mathbf{J}_1\mathbf{J}_2)^\gamma /R^{\alpha
+\beta }
\end{equation}
arises\cite{irkhiny} (with $\alpha +\gamma \leq 2l_1+1,\beta +\gamma \leq $ $%
2l_2+1,\alpha +\beta $ being odd and vector $\mathbf{R}$ connecting ions);
on the other hand, the different-site-exchange integrals $v_{j\mu
_1j^{\prime }\mu _2j\mu _3j^{\prime }\mu _4}$ are small due to small overlap
of $f$-functions belonging to different sites, 
what makes the exchange-type of contributions from mixing and hopping 
visible on the background of positive direct exchange integrals and 
make non-ferromagnetic ordering (e.g., antiferromagnetism) possible. Now let us 
write down the terms, which are added 
into the equations of motion for the functions 
$\langle Tc_{n1}\eta ^{\dagger }\rangle _u$:
\begin{eqnarray}
&&\lim_{t^{\prime \prime }\rightarrow t^{-}}O_{n1,j\mu }^{-1}f_\mu
^a\varepsilon _b^{a\xi }C_{jj^{\prime }}^{\xi \xi ^{\prime }}[\langle
Z_{j^{\prime }}^{\xi ^{\prime }}(t^{\prime \prime })\rangle +i\frac \delta
{\delta {\cal U}_{j^{\prime }}^{\xi ^{\prime }}(t^{\prime \prime
})}]\langle TX_j^b(t)\eta ^{\dagger }(t^{\prime })\rangle _u+  \nonumber \\
&&\lim_{t^{\prime \prime }\rightarrow t^{+}}O_{n1,j^{\prime }\mu }^{-1}f_\mu
^c\varepsilon _{b^{\prime }}^{c\xi }C_{jj^{\prime }}^{\xi \xi ^{\prime
}}[\langle Z_j^\xi (t^{\prime \prime })\rangle +i\frac \delta {\delta 
{\cal U}_j^\xi (t^{\prime \prime })}]\langle TX_{j^{\prime }}^{b^{\prime
}}(t)\eta ^{\dagger }(t^{\prime })\rangle _u.
\end{eqnarray}
They contribute to the self-operator $\Sigma_{cX}G_{X\eta }$.
The structure of this correction is similar to the term which comes in Eq.78
from hopping. Obviously, the Hartree corrections which is obtained by
putting $\delta \langle Tc\eta ^{\dagger }\rangle _u/\delta {\cal U}=0$,
are the biggest terms, and of these terms the biggest one is the diagonal
term, with $\xi =[\Gamma ,\Gamma ]$.
Next terms, generated by  
$\delta \langle Tc\eta ^{\dagger }\rangle _u/\delta {\cal U}$, 
in lowest approximation give  the loop-corrections to the dielectric function and
depend on the scenario under study. 

Let us obtain now the contribution from
this interaction to the equation for $\langle TX^a\eta ^{\dagger }\rangle
_u. $ Calculating corresponding commutator, we find quite similar result,
namely, the term which should be added to $\Sigma _{XX}G_{X\eta }$, has the
form: 
\begin{eqnarray}
&&\lim_{t^{\prime \prime }\rightarrow t^{-}}\varepsilon _b^{a\xi
}C_{jj^{\prime }}^{\xi \xi ^{\prime }}\delta _{nj}[\langle Z_{j^{\prime
}}^{\xi ^{\prime }}(t^{\prime \prime })\rangle +i\frac \delta {\delta 
{\cal U}_{j^{\prime }}^{\xi ^{\prime }}(t^{\prime \prime })}]\langle
TX_j^b(t)\eta ^{\dagger }(t^{\prime })\rangle _u+  \nonumber \\
&&\lim_{t^{\prime \prime }\rightarrow t^{+}}\varepsilon _{b^{\prime }}^{a\xi
^{\prime }}C_{jj^{\prime }}^{\xi \xi ^{\prime }}\delta _{nj^{\prime
}}[\langle Z_j^\xi (t^{\prime \prime })\rangle +i\frac \delta {\delta 
{\cal U}_j^\xi (t^{\prime \prime })}]\langle TX_{j^{\prime }}^{b^{\prime
}}(t)\eta ^{\dagger }(t^{\prime })\rangle _u.
\end{eqnarray}
This term has the structure similar to the one $\sim t_{jj^{\prime
}}^{\uparrow \bar{b}c}$ in Eq.40. In short notations we can write the two
ones above as 
\begin{eqnarray}
&&\lim_{t^{\prime \prime }\rightarrow t^{-}}\varepsilon
_2^{11_b}C_{1_b2_b}[\langle Z_{2_b}(t^{\prime \prime })\rangle +i\frac
\delta {\delta {\cal U}_{2_b}(t^{\prime \prime })}]\langle T\eta
_2(t)\eta ^{\dagger }(t^{\prime })\rangle _u+  \nonumber \\
&&\lim_{t^{\prime \prime }\rightarrow t^{+}}\varepsilon
_2^{12_b}C_{1_b2_b}[\langle Z_{1_b}(t^{\prime \prime })\rangle +i\frac
\delta {\delta {\cal U}_{1_b}(t^{\prime \prime })}]\langle T\eta
_2(t)\eta ^{\dagger }(t^{\prime })\rangle _u,
\end{eqnarray}
where notations are obvious from comparison with explicit form. Thus, these
terms of Hamiltonian do not generate essentially new elements in the digram
technique. The Hartree term $\varepsilon C\langle Z\rangle _u$ (the terms at $%
\varepsilon C[$ $\delta /\delta {\cal U}]\langle TX^a\eta ^{\dagger
}\rangle _u=0$) describes the static screening of charges of neighboring
ions, $(Z-\rho _{core})-\rho _f$. The
non-spherical part of it is responsible also for the crystal field splitting
of the states $|\Gamma \rangle $. Usually this splitting is less important
for the Fermi-like excitations than
for Bose-like, where it can be seen directly in neutron scattering
experiments. Sometimes, however, a scattering of 
conduction electrons on Bose-like excitations in crystal electric field
can strongly renormalize effective electronic masses
\cite{white_fulde,fulde_jensen}. 
 The terms $\sim $ $\varepsilon C[$ $\delta /\delta {\cal U}%
]\langle TX^a\eta ^{\dagger }\rangle _u$ describe effects of exchange and
the dynamical screening from those $f$-bands which cross Fermi energy. The $%
f $-density in the close vicinity of Fermi level $\varepsilon _F$ in normal
materials is usually small and, therefore, these terms can be neglected.
These contributions may play an important role in physics of 
 the intermediate-valence (IV) 
compounds\cite{kuz'min_sandalov} and, also, 
of rare earth metals\cite{urban}. In both latter cases
the $f$-level can be situated just slightly above $\varepsilon _F$.
It can be important also in $%
d-d$-electron scattering in $Fe,Co,Ni$ metals and their compounds.

Let us pass to a next term, $v_{1234}a_1^{\dagger }a_2^{\dagger }a_3a_4$. 
 Again we have to calculate the 
corrections which arise in equations of motion for GFs.
 Of course, all terms with two $f$-operators and two $c$-operators
 should be taken into account. 
Since they give similar contributions, for briefness we consider only one
of them, 
\beq
v_{nL,j\mu j\mu ^{\prime }n^{\prime }L^{\prime
}}(c_{nL}^{\dagger }c_{n^{\prime }L^{\prime }})(f_{j\mu }^{\dagger }f_{j\mu
^{\prime }})^\xi Z_j^\xi \equiv C_{nLn^{\prime }L^{\prime }}^{j\xi
}(c_{nL}^{\dagger }c_{n^{\prime }L^{\prime }})Z_j^\xi. 
\eeq
Summation over
transitions $\xi $ includes both, the Coulomb interaction of the
density-density type, and exchange interactions. 
 Commutation of $c$-operator with this term generates in 
the equation for the GFs
$\langle Tc_{iL}(t)\eta ^{\dagger }(t^{\prime })\rangle_u$ the GFs:
 $\langle
Tc_{n_2L_2}(t)Z_j^\xi (t)\eta ^{\dagger }(t^{\prime })\rangle _u$ and $%
\langle Tc_{n_1L_1}^{\dagger }(t)c_{n_2L_2}(t)X_j^b(t)\eta ^{\dagger
}(t^{\prime })\rangle _u$. One can express them
 via functional derivatives of GFs as follows: 
\begin{eqnarray}
&&\lim_{t^{\prime \prime }\rightarrow
t^{-}}(O^{-1})_{iL,n_1L_1}C_{n_1L_1n_2L_2}^{j\xi }[\langle TZ_j^\xi
(t^{\prime \prime })\rangle _u+i\frac \delta {\delta {\cal U}_j^\xi
(t^{\prime \prime })}]\langle Tc_{n_2L_2}(t)\eta ^{\dagger }(t^{\prime
})\rangle _u+  \nonumber \\
&&\lim_{t^{\prime \prime }\rightarrow t^{+}}(O^{-1})_{iL,j\mu }f_\mu
^a\varepsilon _b^{a\xi }C_{n_1L_1n_2L_2}^{j\xi }[\langle
Tc_{n_1L_1}^{\dagger }(t)c_{n_2L_2}(t)\rangle _u+  \nonumber \\
&&i\frac \delta {\delta F_{n_1L_1n_2L_2}(t^{\prime \prime })}]\langle
TX_j^b(t)\eta ^{\dagger }(t^{\prime })\rangle _u.
\end{eqnarray}
First line gives contribution to $S_{cc}G_{c\eta };  $the Hartree term, $%
\sim O^{-1}C\langle TZ\rangle $, describes a simple fact, that the
conduction electron ''sees'' screened by $f$-electrons ion's charge, $%
(Z-\rho _{core})-\rho _f$, as well as with average angular moment if the
phase under consideration has a long-range magnetic order. At zero external
fields $\langle TZ\rangle \rightarrow \langle h^\Gamma \rangle \equiv
N_\Gamma $ and $\langle T\mathbf{J}_j\rangle \rightarrow \langle
J_j^z\rangle .$ Second line in this formula contains new element, $\delta
/\delta F$, describing creation of the conduction-band electron-hole pair
(contribution to $S_{cX}G_{X\eta }$).

Let us turn now to the GFs $\langle TX_i^a(t)\eta ^{\dagger }(t^{\prime
})\rangle _u.$ From equations of motion we find that the correction to
right-hand side, 
\begin{equation}
C_{n_1L_1n_2L_2}^{j\xi }\times (\delta G_{IV})_{n_1L_1n_2L_2}^{j\xi ,i}
\end{equation}
involves quite complex GFs: 
\begin{eqnarray}
(\delta G_{IV})_{n_1L_1n_2L_2}^{j\xi ,i} &=&(O^{-1})_{i\mu ,n_1L_1}(f_\mu
^{\dagger })^{\bar{b}}\varepsilon _b^{a\xi }\langle TZ_i^{\bar{\xi}^{\prime
}}(t)c_{n_2L_2}(t)Z_j^\xi (t)\eta ^{\dagger }(t^{\prime })\rangle _u- 
\nonumber \\
&&(O^{-1})_{n_2L_2,i\mu }f_\mu ^b\varepsilon _\eta ^{ba}\langle
Tc_{n_1L_1}^{\dagger }(t)Z_i^\eta (t)Z_j^\xi (t)\eta ^{\dagger }(t^{\prime
})\rangle _u+  \nonumber \\
&&\delta _{ij}\varepsilon _b^{a\xi }\langle Tc_{n_1L_1}^{\dagger
}(t)c_{n_2L_2}(t)X_i^b(t)\eta ^{\dagger }(t^{\prime })\rangle _u.
\end{eqnarray}
The GFs in first two lines contain per \emph{two} $Z$-operators, therefore,
in our equations appears the new element,  derivatives of \emph{second}
order: 
\begin{eqnarray}
&\langle T&Z_i^{\bar{\xi}^{\prime }}(t)c_{n_2L_2}(t)Z_j^\xi (t)\eta
^{\dagger }(t^{\prime })\rangle _u=  \nonumber \\
&&\lim_{t^{\prime \prime }\rightarrow t^{+}}[\langle TZ_i^{\xi ^{\prime
}}(t^{\prime \prime })\rangle _u+i\frac \delta {\delta {\cal U}_i^{\xi
^{\prime }}(t^{\prime \prime })}]\langle Tc_{n_2L_2}(t)Z_j^\xi (t)\eta
^{\dagger }(t^{\prime })\rangle _u=  \nonumber \\
&&\lim_{t^{\prime \prime }\rightarrow t^{+}}[\langle TZ_i^{\xi ^{\prime
}}(t^{\prime \prime })\rangle _u+i\frac \delta {\delta {\cal U}_i^{\xi
^{\prime }}(t^{\prime \prime })}]\times   \nonumber \\
&&\{\lim_{t_3\rightarrow t^{-}}[\langle TZ_j^\xi (t_3)\rangle _u+i\frac
\delta {\delta {\cal U}_j^\xi (t_3)}]\langle Tc_{n_2L_2}(t)\eta ^{\dagger
}(t^{\prime })\rangle _u\}.
\end{eqnarray}
Thus, we've obtained the correction to $\Sigma _{Xc}G_{c\eta }.$ In the same
fashion we find that 
\begin{eqnarray}
&\langle T&c_{n_1L_1}^{\dagger }(t)Z_i^\eta (t)Z_j^\xi (t)\eta ^{\dagger
}(t^{\prime })\rangle _u=  \nonumber \\
&&\lim_{t^{\prime \prime }\rightarrow t^{-}}[\langle TZ_i^\eta (t^{\prime
\prime })\rangle _u+i\frac \delta {\delta {\cal U}_i^\eta (t^{\prime
\prime })}]\langle Tc_{n_1L_1}^{\dagger }(t)Z_j^\xi (t)\eta ^{\dagger
}(t^{\prime })\rangle _u=  \nonumber \\
&&\lim_{t^{\prime \prime }\rightarrow t^{-}}[\langle TZ_i^\eta (t^{\prime
\prime })\rangle _u+i\frac \delta {\delta {\cal U}_i^\eta (t^{\prime
\prime })}]\times   \nonumber \\
&&\{\lim_{t_3\rightarrow t^{-}}[\langle TZ_j^\xi (t_3)\rangle _u+i\frac
\delta {\delta {\cal U}_j^\xi (t_3)}]\langle Tc_{n_1L_1}^{\dagger
}(t)\eta ^{\dagger }(t^{\prime })\rangle _u\}.
\end{eqnarray}
Note, that this term does not give corrections of mean-field type, of first
order, $\sim v_{n_1L_1n_2L_2}^{j\xi }$. At last, the function $\langle
T(c^{\dagger }cX)\eta ^{\dagger }\rangle _u$ can be written down as 
\begin{eqnarray*}
\langle  &&Tc_{n_1L_1}^{\dagger }(t)c_{n_2L_2}(t)X_i^b(t)\eta ^{\dagger
}(t^{\prime })\rangle _u= \\
&&\lim_{t^{\prime \prime }\rightarrow t^{+}}[\langle Tc_{n_1L_1}^{\dagger
}(t^{\prime \prime })c_{n_2L_2}(t^{\prime \prime })\rangle _u+i\frac \delta
{\delta {\cal F}_{n_1L_1n_2L_2}(t^{\prime \prime })}]\langle X_i^b(t)\eta
^{\dagger }(t^{\prime })\rangle _u
\end{eqnarray*}
and gives contribution to $\Sigma _{XX}G_{X\eta }.$ 

  The mean-field contribution from the last term, $\delta \Delta
_b^{Coul}\equiv v_{n_1L_1,n_2L_2}^{i\xi }\langle Tc_{n_1L_1}^{\dagger
}c_{n_2L_2}\rangle _u\varepsilon _b^{a\xi },$ should be added to the
equation for effective energies of transitions $\Delta _1^{*}$ and $\Delta
_2^{*}$ (see Eqs.98,99):
\begin{equation}
\Delta _i^{*}=\Delta _i^0+\Delta _i^{mixing}+\Delta _i^{hopping}+\Delta
_i^{Coul},\;i=1,2.
\end{equation}
As we discussed above, the correction $\Delta _1^{mixing}+\Delta _1^{hopping}
$ to the deep level $\Delta _1^0$ are small due to strong suppression of
mixing and hopping in this region of energy. The corrections from $%
vc^{\dagger }cZ$ are not so sensitive to the position of the $\Delta _i^0$%
-level and depends mainly in the density of $c$-charge or, since the
contribution is local, just on number of $c$-electrons per $f$-cite. In
order to visualize (see discussion in Ref.\cite{mcmahan}) how this
correction works, let us write it for a moment in the form $Vn_cn_f$. Then $%
E_{\Gamma _n}=\varepsilon _0^fn_f+\frac 12Un_f(n_f-1)+Vn_cn_f$ and we find
that $\Delta _{\Gamma _{n+1}\Gamma _n}=\varepsilon _0^f+Un_f+Vn_c,\;\Delta
_{\Gamma _n\Gamma _{n-1}}=\varepsilon _0^f+U(n_f-1)+Vn_c$. Thus, the
positions of both levels are essentially shifted by the Coulomb interaction
with conduction electrons towards Fermi energy. But  the effective Hubbard $%
U^{*}$ remains unchanged, $U^{*}=\Delta _{\Gamma _{n+1}\Gamma _n}-\Delta
_{\Gamma _n\Gamma _{n-1}}=U$, until number of condition electrons $n_c$
remains unchanged.  Therefore, main mechanism of a decrease of Hubbard $U$
is expected from the processes of dynamical screening or, in other words,
from partial delocalization of wave functions of $f$-electrons. The
calculations, based on the density-functional theory take into account these
processes very efficiently (even slightly overestimate) giving, thus, much
smaller values of  $U^{*}$ than the bare value of $U$.
 Accurate evaluation of screening effects
requires special analysis and we postpone it for a future work. 

\section{Discussion and Conclusions}

The theory presented here is intended
mainly for those systems which contain a
group of correlated electrons in the {\em strong-coupling} regime.
The correlations have local, intra-atomic nature and
therefore it is reasonable to construct a perturbation theory for
the description of these electrons starting from the atomic limit.
The technical difficulties in our approach come from
the necessity to perform renormalization of fermion-fermion
interactions in
the {\em site} representation, which unavoidably leads to non-fermionic
commutation relations of new on-site collective variables. This,
in turn,
makes it impossible to use standard Wick's  theorem for constructing
a diagram technique. Besides, the site-centered single-electron 
wave functions, used for the formation of many-electron ones,
are not orthogonal to functions of other sites.
In the case of weak Coulomb interaction one does not require
the use of a non-orthogonal basis. However, for many
band structure methods this may still be desirable,
for technical reasons.
In this work we have shown 
that Schwinger's method\cite{schwinger}
of sources,  which has become a 
standard method in solid state physics due to the
work by Baym and Kadanoff\cite{baym_kadanoff}, can be used for
constructing a regular perturbation theory for systems with
strongly correlated electrons and we have shown that this leads to important advantages.
On one hand,
this approach allows one  to overcome the difficulties existing in the original
formulation of the diagram technique from the atomic 
limit\cite{zaitsev,izyumov1} and to generalize
the technique for $d$- and $f$-electron systems. On the other hand,
the approach happens to be fruitful also in overcoming
the difficulties which come from the necessity to use
a non-orthogonal basis set. 
We hope also that, it gives a convenient tool to derive effective
low-energy equations and to perform an analysis of corrections, arising
beyond the standard Kohn-Sham equations.
It is worth to note that since the theory is formulated in
real space, it can
be applied not only to the case of a regular lattice,
but also to the case of
impurity problems, or non-regular lattices. Of course, for the latter case an
additional configurational averaging should be made.
The perturbation theory developed here has a very close similarity to the 
ones for the boson and fermion systems. 
In particularly we have demonstrated this by the derivation of equations 
for the GFs: 
taking vertex in zero, first, and RPA approximations 
correspondingly, and we obtained the equations in 
 the Hubbard-I, mean field, and RPA approximations. 

Some important questions, however, remain to be solved. 
It is a well-known fact, that a direct calculations of
Hubbard's parameters
for the  Coulomb repulsion, for instance via the
Slater integrals, give too large
values.
Within our theory the intra-atomic parameters enter the zero Hamiltonian,
while the mixing and hopping 
become perturbations. 
The energies
of intra atomic transitions of fermion type, which are
determined by the bare value of Hubbard $U$ in the
atomic limit and in the Hubbard-I approximation,
 are renormalized by mixing and hopping in our theory
already in the next, mean-field approximation.
These renormalized parameters
still determine the positions of the centers of the correlated bands and 
the structure of the equations for the renormalized energies of 
transitions is almost coinciding with the one within well-known non-crossing 
approximation. 
It is interesting to note that mathematically
the mechanism is analogous to the Lamb shift in electrodynamics, but
here it is provided by conduction electrons.
The energy difference between the centers
of these bands can be
interpreted as renormalized Hubbard $U$ values.
In a  parallel work\cite{urb_model}
we have shown that, on one hand, the corrections,
generated by the mixing interaction, can
decrease or increase the value of effective Hubbard $U$ depending on
the particular solution chosen. On other hand,
this mechanism gives small corrections and cannot provide the 
expected reduction 
of Hubbard $U$.

In a series of works\cite{eschrig1,swane,erikssonPu} a separation of 
$f$-electrons 
into two subsystems, localized and delocalized, have been considered within
the DFT approach on a phenomenological grounds. Application of the technique, 
developed in present work to this issue\cite{fLDA_SEC,lundin} 
gives solid grounds to this
 assumption and provides also a better insight. Particularly, it is shown 
that the role of the localized $f$-wave functions is played by the combination (25) 
of the transitions $[\Gamma_{n-1},\Gamma_n ]$ with an effective  single-electron 
wave function,
\beq
\Psi_f^{(\Gamma_{n-1},\Gamma_n )}({\bf r}) \equiv \Psi_f^{lower}({\bf r}) \equiv
\int d({\bf r_1})d({\bf r_2})...d({\bf r_{n-1}})
\Psi^*_{\Gamma_{n-1}}(({\bf r_1,r_2,...,r_{n-1}})
\Psi_{\Gamma_{n}}(({\bf r_1,r_2,...,r_{n-1},r}),
\eeq
corresponding to the lower-energy transition $\Delta_1$,
while the delocalized $f$-wave functions correspond to combinations of 
wave functions
\beq
\Psi_f^{(\Gamma_{n},\Gamma_{n+1} )}({\bf r}) \equiv \Psi_f^{upper}({\bf r}) \equiv
\int d({\bf r_1})d({\bf r_2})...d({\bf r_{n}})
\Psi^*_{\Gamma_{n}}(({\bf r_1,r_2,...,r_{n}})
\Psi_{\Gamma_{n+1}}(({\bf r_1,r_2,...,r_{n},r}),
\eeq
The question how much
the intra-atomic Coulomb $f-f$-interactions are renormalized by
those corrections from the
$(spd-f)$-interactions, which are not included in the standard LDA scheme
is still open.
One can expect that these interactions should provide a
strong reduction of the effective $U^*$-s. 
 This question is especially important for a calculation
of the self-interaction correction (SIC), which plays a decisive role
in the problem of volume-collapse transitions in rare earth
metals\cite{mcmahan} and the theory of Mott insulators. We expect that 
the expression for this correction, which arises in
the equations of motion for the GFs when the self-energy is
approximated by the difference between the static part of RPA-screened
Coulomb interaction and the matrix element of the LDA potential,
should be accurate. No problem
with double counting and orthogonalization arises within such an
approach.

The method suggested here
is a special type of perturbation theory, which
does not contain any approximations in its general formulation.
The Hubbard-I approximation (HIA) and mean field approximation
from atomic limit (AL-MFA) can be made with reasonably moderate
efforts.
It is relevant to emphasize here the similarities and differences
between our AL-MFA
from the  popular 
method of dynamic mean field (DMF)\cite{kotliar}.
The AL-MFA differs from the HIA
by taking into account the renormalization of energies of intra-atomic
transitions by mixing interaction and hopping. Therefore,
besides self-consistent equations for many-electron population numbers,
which arise in HIA, the system of equations for self-consistent
centers of gravity of the {\em Hubbard sub bands} is present in
the calculation in
AL-MFA\cite{urb_model}. A renormalization of this type
is accounted for also in the DMF theory. However, it is not clear at the moment,
to what extent they are different. Eq.(122) for the renormalized 
energies of intra atomic Fermi-type transitions in the  AL-MFA
is non-linear and sometimes contains more than one solution;  the number of
 poles in the fermionic GF (or, number of Hubbard sub bands) is
equal to the number of intra atomic
fermionic transitions only near atomic limit. The DMF method is analogous
to the coherent potential approximation and, therefore, can give some
other poles. To be specific, the DMF gives a central peak in the density
of states of fermionic excitations for the paramagnetic state of
the $s$-band Hubbard model. The AL-MFA equation for renormalized energy 
of transition also may have second solution in vicinity of Fermi 
energy\cite{fuldebook}, what can provide the central peak in the density of 
electron states. However, in many cases the upper transition $\Delta_2$
being slightly above Fermi energy hybridizes with non-$f$ bands
and gives birth to some tails under Fermi energy. These peaks are seen in 
electron spectroscopy\cite{gschneider} and are
 close enough to Fermi energy in order to mask weak Kondo peak. The details
of the mechanism of formation
of these additional bands in the region of moderate coupling constants
in terms of Hubbard-band excitations remain to be elucidated.
There is a difference between DMF and our approach in the treatment
of spatial correlations. Namely, in applications of the DMF approach 
only a local, momentum-independent, self-energy has been considered, whereas
in our theory first $(k, \omega)$-dependent corrections in the expansion
of the self-energy arises in the RPA (due to kinematic interactions).

At last, we note that it is easily seen from the close similarity
between
our construction of perturbation theory and the one by Baym-Kadanoff 
\cite{baym_kadanoff}
that the
description of interacting delocalized
and localized electrons in
non-equilibrium states needs only slight
generalization of our theory. Example of calculation of effects of non-orthogonality 
on the tunneling processes via a region with strongly interacting electrons 
is given in Ref. 48.
 An application of the technique to 
spin systems will be given elsewhere.

\section{Acknowledgments}
The authors are grateful to the Swedish Natural Science Council (NFR) 
and G\"oran Gustafsson Foundation
for financial support. Part of this work has been performed 
in the Grup d'Electromagnetisme, 
Universitat Aut\'onoma de Barcelona. I.S. thanks 
Prof. F.L\'opez-Aguilar and members of the group 
for hospitality and Ministerio
de Educaci\'on y Ciencia of Spain for support (grant SAB95-0225) during 
his stay in Bellaterra. I.S. is also greatful to 
Prof. A.K. McMahan for valuable 
discussion of mechanisms of screening of Hubbard $U$.

\appendix
\renewcommand{\thesection}{\Alph{section}}
\section{ Parameters of the Hubbard-Anderson model}
Let us now discuss 
the parameters of the model.  The energies 
$E_{\Gamma}^0$ according to definition, are $E_{\Gamma}^0=
\langle \Gamma |H_{ion}|\Gamma \rangle $. For the case of Russel-Saunders
coupling, i.e. for rare earths, corresponding expressions are given by 
Eqs. (7.18),(7.19) in Ref.\cite{irkhiny}. 
Main contributions to these energies come from pure Coulomb interaction,
\begin{equation}
\delta E_{\Gamma, c}= \frac{1}{2}n(n-1)F^{(0)}_{l=3} - 
\langle \Gamma |v^{LDA}|\Gamma \rangle.
\end{equation}
Typically, the bare $F^{(0)}$ are very large, $ \sim 20eV $,
 what lead Hubbard\cite{hubb1} to the 
conclusion that only the states with $n=n_0, n_0 \pm 1$ from the parabolic 
dependence of $E_{\Gamma}(n)$ can give contribution to the Hamiltonian. Here 
$n_0$ is dictated by the valence of the ion. As has been explained 
by McMahan\cite{mcmahan}, the interaction with non-$f$ electrons 
strongly decreases the $U_{model} =(\Delta_{\Gamma_{n+1},\Gamma_{n}}-
(\Delta_{\Gamma_{n},\Gamma_{n-1}})$, where
$\Delta_{\Gamma,\Gamma_1} \equiv
(E_{\Gamma}- E_{\Gamma_{1}})$. 
It worth to emphasize 
the difference between our and Hubbard's schemes.  Hubbard and, also, 
Irkhin, diagonalized all on-site electrons, treating all of them 
in equal footing; this means that they have accepted the assumption that all on-site 
electrons are in strong-coupling regime. We assume that the conduction 
electrons (which are above tops of Coulomb barriers, 
can be described as a weakly-coupled subsystem.

Let us now discuss the mixing interaction. 
The matrix element $W$ in Eq.20 consists of three terms: $%
W_{mix}=W^{(1)}+W^{(2)}+W^{(3)}.$ The term $W^{(1)}\sim Hc^{+}f+H.c.$
 describes the single-electron LDA-type of mixing, which is present in 
Eq.(8). This  matrix element is:
\beq
W_{jL,j^{\prime }a^{\prime }}^{(1)}=\sum_{m_l^{\prime }}H_{jL,j^{\prime
}m_l^{\prime }\sigma }(f_{m_l^{\prime }\sigma })^{a^{\prime }}\equiv
\sum_{m_l^{\prime }}H_{jL,j^{\prime }m_l^{\prime }\sigma }\langle \Gamma
_n|f_{m_l^{\prime }\sigma }|\Gamma _{n+1}^{^{\prime }}\rangle
\eeq
The other contribution to $H_{mix}$ arises from those terms of 
$(H-H_{LDA})$ which have the following operator structure;
$ c^{+}X^a$ and $X^{\bar{a}}c.$ Obviously, the terms
of the Coulomb interaction $\delta
H_{ee}^{Coul}=C_{1234}^{fffc}f_1^{+}f_2^{+}f_3c_4+C_{1243}^{ffcf}f_1^{+}f_2^{+}c_4f_3+C_{4231}^{cfff}c_4^{+}f_3^{+}f_2f_1+C_{3421}^{fcff}f_3^{+}c_4^{+}f_2f_1,
$ which contain the $f-$operators related to the same site, should not be
decoupled. We rewrite these products of $f$-operators in terms of $X$%
-operators,
\beq
(f_1^{+}f_2^{+}f_3)=\sum_{\bar{a}}(f_1^{+}f_2^{+}f_3)^{\bar{a}}X^{\bar{a}%
}, \;\;\;
(f_3^{+}f_2^{}f_1)=\sum_{\bar{a}}(f_3^{+}f_2^{}f_1)^aX^a.
\eeq
In this place a possibility exists to consider a simplified models, 
omitting some terms of genius Coulomb interaction, but keeping 
corresponding terms in LDA.
The Hartree part of the LDA potential contains these terms in the decoupled
form. If we perform the decoupling of these terms in the Hartree-Fock
fashion, there appears also exchange terms. However, the exchange interaction
is treated within LSDA theory by the term $\sim (\alpha \rho
^{1/3}(r))_{jL,jL^{\prime }}a_{jL}^{+}a_{jL^{\prime }}$ in the potential.
Therefore, it is reasonable to regroup the terms in order to represent 
 every term of the perturbation Hamiltonian in the form of a  
deviation of the non-approximated terms from the ones used in the  
LDA approach, as has been discussed in Sec.II. 
In order to avoid double counting we should not
touch other terms of the Coulomb interaction if we do not consider
its LDA-counterpart, 
if we do not take into account some
of the Coulomb terms $(H-H_{LDA})_{ee}^{Coul}$, 
corresponding terms enter all equations
in the LDA form.
 Particularly, we treat in this way the terms of
the form $\sim Cc^{+}c^{+}cf$ and similar ones. Then we come to the
following form for the remaining terms of the mixing matrix elements.
For the density-density part we have,
\begin{eqnarray}
W_{jL,j^{\prime }a^{\prime }}^{(2)}
&=&\sum_{\{m_l\},\sigma }
C_{jL,j^{\prime }m_{l1}\sigma ^{\prime };
j^{\prime }m_{l2}\sigma ^{\prime };j^{\prime }m_{l3}\sigma }^{cfff} \nonumber
\\ &\times &
[(f_{m_{l1}\sigma ^{\prime }}^{\dagger}f_{m_{l2}\sigma ^{\prime }}
f_{m_{l3}\sigma })^{a^{\prime }}
-(f_{m_{l3}\sigma ^{\prime }})^{a^{\prime }}
\langle f_{m_{l1}\sigma ^{\prime }}^{\dagger}f_{m_{l2}\sigma%
^{\prime }}\rangle ^{LDA}] \nonumber \\
&\simeq &\sum_{\{m_l\},\sigma }
C_{jL,j^{\prime }m_{l1}\sigma ^{\prime };j^{\prime }m_{l2}\sigma ^{\prime };
j^{\prime }m_{l3}\sigma }^{cfff} \nonumber \\
&\times & [(f_{m_{l1}\sigma ^{\prime }}^{\dagger}
f_{m_{l2}\sigma ^{\prime }}
f_{m_{l3}\sigma })^{a^{\prime }}
-(f_{m_{l3}\sigma })^{a^{\prime }}
\delta _{m_{l1},m_{l2}}
n_{j^{\prime }m_{l1}\sigma ^{\prime }}^{LDA}],
\end{eqnarray}
and for the exchange part,
\begin{eqnarray}
W_{jL,j^{\prime }a^{\prime }}^{(3)} &=&\sum_{\{m_l\},\sigma ^{\prime
}}C_{jL,j^{\prime }m_{l1}\sigma ^{\prime };j^{\prime }m_{l2}\sigma ^{\prime
};j^{\prime }m_{l3}\sigma }^{cfff}\delta _{m_{l2},m_{l3}}\delta _{\sigma
\sigma ^{\prime }}(f_{m_{l1}\sigma ^{\prime }}^{})^{a^{\prime }}n_{j^{\prime
}m_{l1}\sigma ^{\prime }}^{LDA} \nonumber \\
&&-\sum_{m_l,\sigma }\left(v_{xc}(r)\right) _{jL,m_l\sigma
}(f_{m_{l1}\sigma }^{})^{a^{\prime }}.
\end{eqnarray}
As seen the term $\sim W_{jL,j^{\prime }a^{\prime }}^{(2)}\equiv 0$ for the
empty orbitals. It is this matrix element that 
is responsible for the different
mixing of conduction electrons with the 
intra atomic transitions $f^{n-1}\rightarrow
f^n$ and $f^n\rightarrow f^{n+1}.$ Obviously, this is a pure correlation
effect: {\em decoupling} within the weak-coupling perturbation theory 
will {\em make it vanish}.

The prohibition to decouple the single-site products of $f$-operators leads
to Coulomb renormalization of the LDA-hopping too:
\begin{equation}
t_{jj^{\prime }}^{\bar{a}b}=t_{1jj^{\prime }}^{\bar{a}b}+t_{2jj^{\prime }}^{%
\bar{a}b},
\end{equation}
\begin{equation}
t_{1jj^{\prime }}^{\bar{a}b}=\sum_{\lambda \nu }[t_{jj^{\prime }}^{\lambda
,\nu }-(v_{LDA})_{jj^{\prime }}^{\lambda \nu }](f_\lambda ^{\dagger })^{\bar{%
a}}(f_\nu )^b,
\end{equation}
\begin{eqnarray}
t_{2jj^{\prime }}^{\bar{a}b} &=&[v_{j\mu _1,j\mu _2,j\mu _3,j^{\prime }\mu
_4}(f_{\mu _1}^{\dagger }f_{\mu _2}^{\dagger }f_{\mu _3})^{\bar{a}}(f_{\mu
_4})^b+v_{j\mu _1,j^{\prime }\mu _2,j^{\prime }\mu _3,j^{\prime }\mu
_4}(f_{\mu _1}^{\dagger })^{\bar{a}}(f_{\mu _2}^{\dagger }f_{\mu _3}f_{\mu
_4})^b-  \nonumber \\
&&v_{j\mu _1,j^{\prime }\mu _2,j\mu _3,j\mu _4}(f_{\mu _2}^{\dagger })^{\bar{%
a}}(f_{\mu _1}^{\dagger }f_{\mu _3}f_{\mu _4})^b-v_{j\mu _1,j\mu
_2,j^{\prime }\mu _3,j\mu _4}(f_{\mu _1}^{\dagger }f_{\mu _2}^{\dagger
}f_{\mu _4})^{\bar{a}}(f_{\mu _3})^b.
\end{eqnarray}
Note   the $t_2$-contribution  from Coulomb interaction works similar to
corresponding contributions to mixing: it makes hopping very different for,
say, 
transitions $[\Gamma _{n-1},\Gamma _n]$ and  $[\Gamma _n,\Gamma _{n+1}]$.

\section{ Example: Three-electron states and commutation relations }

Let us consider a simple example of three electron states and corresponding
commutation relations for the Hubbard operators, constructed on these states
and belonging to the same site. A general three-electron state $|\Gamma
\rangle $ can be written as follows:
\begin{equation}
|\Gamma \rangle =\sum C_{\nu _1\nu _2\nu _3}^\Gamma \theta _{\nu _1\nu
_2}\theta _{\nu _2\nu _3}f_{\nu _1}^{\dagger }f_{\nu _2}^{\dagger }f_{\nu
_3}^{\dagger }|0\rangle ,\;\langle \Gamma ^{\prime }|=\sum C_{\mu _1\mu
_2\mu _3}^{*\Gamma ^{\prime }}\theta _{\mu _1\mu _2}\theta _{\mu _2\mu
_3}\langle 0|f_{\mu _3}f_{\mu _2}f_{\mu _1}.
\end{equation}
Here  $\theta$-functions are needed in order to establish a sign convention, 
i.e. in which order have to be taken the $f$-operators in the many-electron 
wave function; $\theta_{\nu_1,\nu_2}=1$ if $\nu_1>\nu_2$ and zero otherwise.
Then, according to the definition of the $X$-operators, $X^{\Gamma \Gamma
^{\prime }}=|\Gamma \rangle \langle \Gamma ^{\prime }|,$ the multiplication
rule is:
\begin{equation}
X^{\Gamma \Gamma _1}X^{\Gamma _2\Gamma ^{\prime }}=\langle \Gamma _1|\Gamma
_2\rangle X^{\Gamma \Gamma ^{\prime }}=\nu_{\Gamma _1\Gamma _2}X^{\Gamma
\Gamma ^{\prime }}.
\end{equation}
From (B.1) we have  
\begin{equation}
\langle \Gamma _1|\Gamma _2\rangle =\sum_{\{\nu ,\mu \}}C_{\mu _1\mu _2\mu
_3}^{*\Gamma _1}C_{\nu _1\nu _2\nu _3}^{\Gamma _2}\theta _{\mu _3\mu
_2}\theta _{\mu _2\mu _1}\theta _{\nu _1\nu _2}\theta _{\nu _2\nu _3}S_{\mu
_3\mu _2\mu _1,\nu _1\nu _2\nu _3}.
\end{equation}
Then, from the anticommutation relation
\begin{equation}
\{f_{n\mu },f_{m\nu }^{\dagger }\}={\mathcal O}_{n\mu ,m\nu }^{-1},
\end{equation}
For the same site $n=m$ and we see that 
\begin{equation}
\nu_{\mu _3\mu _2\mu _1,\nu _1\nu _2\nu _3}=\det \left| 
\begin{array}{ccc}
{\mathcal O}_{\mu _1\nu _1}^{-1} & {\mathcal O}_{\mu _1\nu _2}^{-1} & {\mathcal %
O}_{\mu _1\nu _3}^{-1} \\ 
{\mathcal O}_{\mu _2\nu _1}^{-1} & {\mathcal O}_{\mu _2\nu _2}^{-1} & {\mathcal %
O}_{\mu _2\nu _3}^{-1} \\ 
{\mathcal O}_{\mu _3\nu _1}^{-1} & {\mathcal O}_{\mu _3\nu _2}^{-1} & {\mathcal %
O}_{\mu _3\nu _3}^{-1}
\end{array}
\right| 
\end{equation}
and, therefore, the commutation relation is given by
\begin{equation}
\{X^{\Gamma \Gamma _1},X^{\Gamma _2\Gamma ^{\prime }}\}=\nu_{\Gamma _1\Gamma
_2}X^{\Gamma \Gamma ^{\prime }}+\nu_{\Gamma ^{\prime }\Gamma }X^{\Gamma
_2\Gamma _1}.
\end{equation}
The equation of motion for the operator $X_i^{\Gamma \Gamma _1}$ contains 
the commutator 
\begin{equation}
\sum_{n,\Gamma }\varepsilon _\Gamma [X_i^{\Gamma _1\Gamma _2},h_n^{\Gamma
\Gamma }]=\sum_{n,\Gamma }[\nu_{i\Gamma _2,n\Gamma }\varepsilon _\Gamma
X_{in}^{\Gamma _1\Gamma }-\nu_{n\Gamma ,i\Gamma _1}\varepsilon _\Gamma
X_{in}^{\Gamma \Gamma _2}],
\end{equation}
where 
\begin{equation}
X_{in}^{\Gamma _1\Gamma }\equiv |i,\Gamma _1\rangle \langle n,\Gamma |.
\end{equation}
According to our assumption the direct overlap integral on wave functions of 
$f$-electrons belonging to different sites is negligible, therefore, main
contribution comes from the terms with $i=n$, since $\nu_{i\Gamma _2,n\Gamma }$
involves determinant of the inverse of overlap matrix on $f-f$-matrix
elements and each of them is proportional at least to second order of
overlap between wave functions of band electrons and $f$-electrons. Without
this approximation one has to solve problem of interacting via conduction
electrons clusters. In this work we use the approximation 
\begin{equation}
\sum_{n,\Gamma }\varepsilon _\Gamma [X_i^{\Gamma _1\Gamma _2},h_n^{\Gamma
\Gamma }]\approx \sum_\Gamma [\nu_{i\Gamma _2,i\Gamma }\varepsilon _\Gamma
X_i^{\Gamma _1\Gamma }-\nu_{i\Gamma ,i\Gamma _1}\varepsilon _\Gamma
X_i^{\Gamma \Gamma _2}]\approx \Delta _{\Gamma _2\Gamma _1}X^{\Gamma _1\Gamma
_2}
\end{equation}
with 
\begin{equation}
\Delta _{\Gamma _2\Gamma _1}\equiv \nu_{i\Gamma _2,i\Gamma _2}\varepsilon
_{\Gamma _2}-\nu_{i\Gamma _1,i\Gamma _1}\varepsilon _{\Gamma _1.}
\end{equation}
\section{ Equations of motion for operators}
\subsection{ Equations of motion for quasi-fermion operators}
In order to construct equations for the Green's functions we need the
equations of motion $i\partial _t\hat{F}=[\hat{F},\hat{H}-\mu \hat{N}]$ for
(quasi)fermions. As seen from Eq.(28), the particle number operator 
has the same operator structure as the Hamiltonian (26-27). Therefore, 
in order
to obtain $\tilde{H}=\hat{H}-\mu \hat{N},$ we should make the simply
the following shifts:
\beq
E_\Gamma ^0\rightarrow \tilde{E}_\Gamma ^0=E_\Gamma ^0-\mu
\sum_M(n_{MM})_\Gamma  \\,
\eeq
\beq
H_{jL,j^{\prime }L^{\prime }}\rightarrow \tilde{H}_{jL,j^{\prime }L^{\prime
}}=H_{jL,j^{\prime }L^{\prime }}-\mu O_{jL,j^{\prime }L^{\prime }},
\eeq
\beq
t_{jj^{\prime }}^{\bar{a}a^{\prime }}\rightarrow \tilde{t}_{jj^{\prime }}^{%
\bar{a}a^{\prime }}=t_{jj^{\prime }}^{\bar{a}a^{\prime }}-\mu
\sum_{M_1M_2}(1-\delta _{jj^{\prime }})(f_{M_1}^{\dagger})^{\bar{a}%
}O_{jM_1,j^{\prime }M_2}(f_{M_2}^{})^{a^{\prime }},
\eeq
and
\beq
W_{jL,j^{\prime }a}\rightarrow \tilde{W}_{jL,j^{\prime }a}=W_{jL,j^{\prime
}a}-\mu \sum_MO_{jL,j^{\prime }M}(f_{M_{}}^{})^{a^{}},
\eeq
in the Hamiltonian $H$. Below we assume that such a shift has been
 made, and will accordingly drop the tilde. 
Then the equations of motion for the $c$- and $X$-operators 
can be written as:
\beq
i\partial
_tc_{jL}=\sum_{j_2L_2}(O^{-1}H)_{jL,j_2L_2}c_{j_2L_2}+(M_c)_{jL}^{},
\eeq
\beq
i\partial _tX_j^a=\Delta _{\bar{a}}X_j^a+(M_X)_j^a
\eeq
where  for $a=[\Gamma_{n-1},\Gamma_n ]$ the energy of the transition 
$\Delta _{\bar{a}} = E_{\Gamma_n} \nu_{\Gamma_n,\Gamma_n} - 
E_{\Gamma_{n-1}} \nu_{\Gamma_{n-1},\Gamma_{n-1}}$. The coefficient $\nu =1$ 
when the $f$-orbitals are orthogonal to other and $\nu <1 $ when they are not.
The operator $M$ describes deviations from the solution at zero
mixing interaction in the particular potential chosen (say, LDA).
The $M$-operator for the $c$-subsystem consists of linear and non-linear parts,
\beq
(M_c)_{jL}=(M_c^{lin})_{jL}+[(M_c^{\uparrow })_{jL}^{}-(M_c^{\downarrow
})_{jL}^{}],
\eeq
where the linear one is given by
\beq
(M_c^{lin})_{jL}=\sum_{j_2L_2}w_{jL,j_2L_2}X_{j_2}^{a_2},
\eeq
and
\beq
w_{jL,j_2a_2}\equiv (O^{-1}W)_{jL,j_2a_2}+\sum_{\Gamma
,M,a}(O^{-1})_{jL,j_2M}(f_{M_{}}^{})^{a^{}}\varepsilon _{a_2}^{a\Gamma
}E_{j_2\Gamma }^0.
\eeq
The  non-linear terms are given by the 
operators $(M_c^{\uparrow })_{jL},(M_c)^{\downarrow }_{j,L}$ which can
be written in the form
\beq
(M_c^{\uparrow })_{jL}^{}=\sum
(O^{-1})_{jL,j_1M}(f_{M_{}}^{})^{c^{}} 
\hat{P}_{j_1}^{c \bar{a}}[t_{j_1j_2}^{\bar{a}b}
X_{j_2}^b+W_{j_1a,j_2L_2}^{*}
c_{j_2L_2}^{}]
\eeq
and
\beq
(M_c^{\downarrow })_{jL}^{}=\sum
(O^{-1})_{jL,j_2M}(f_{M_{}}^{})^{c^{}}[t_{j_1j_2}^{\bar{b}a}X_{j_1}^{\bar{b}}%
 +W_{j_1L_1,j_2a}^{}c_{j_1L_1}^{\dagger}]
\hat{P}_{j_2}^{ca}.
\eeq
Here
$\hat{P}^{a,\bar{b}} \equiv \{X^a,X^{\bar{b}}\} \equiv 
\epsilon{a,\bar{b}}_{\xi} Z^{\xi}, \,\,\,
\hat{P}^{a,b} \equiv \{X^a,X^{b}\} \equiv
\epsilon{a,b}_{\eta} Z^{\eta} $.
The expression for the $M$-operator of the 
$f$-subsystem can also be written in the form
\beq
(M_X)_j=[(M_X^{\uparrow })_j^a-(M_X^{\downarrow })_j^a]
\eeq
with
\begin{eqnarray}
(M_X^{\uparrow })_j^a &=&\sum t_{jj_2}^{\uparrow \bar{b}c}\hat{P}_{j_{}}^{a%
\bar{b}}X_{j_2}^c+\sum w_{j_1b,j_2L_2}^{\uparrow }\hat{P}_{j_{}}^{a\bar{b}%
}c_{j_2L_2}^{} \\
(M_X^{\downarrow })_j^a &=&\sum t_{j_1j_{}}^{\downarrow \bar{b}c}X_{j_1}^{%
\bar{b}}\hat{P}_{j_{}}^{ca}\ +\sum w_{j_1L_1,jc}^{\downarrow }
c_{j_1L_1}^{\dagger}\hat{P}_j^{ca}.  \nonumber
\end{eqnarray}
Here the matrix elements are given by the following combinations
\begin{eqnarray}
w_{j_1b,j_2L_2}^{\uparrow } &\equiv
&W_{j_1b,j_2L_2}^{*}+\sum_M(f_{M_{}}^{\dagger})^{\bar{b}}
(O^{-1*}H)_{jM,j_2L_2}^{},
\\
t_{jj_2}^{\uparrow \bar{b}c} &\equiv &t_{jj_2}^{\bar{b}c}+%
\sum_M(f_{M_{}}^{\dagger})^{\bar{b}}
(O^{-1*}W)_{ jM,j_2c } ^{} \\
  w^{\downarrow } &=(w^{\uparrow })^{\dagger}, \;\;\;
t^{\downarrow }=(t^{\uparrow })^{\dagger}.
\end{eqnarray}
Some features of the correlated electron system can be
seen already from these equations. 

{\em First,} we see
that instead of equations for $f$-operators, as in case of non-correlated
electrons, we have to write equations for their complicate combinations
$X^a$. This  leads to an unpleasant consequence: 
the number of variables and,
correspondently, equations, increases drastically:
 every
$f$-operator is split into {\em many }Hubbard sub-amplitudes. 
The number of transitions involved depends on the valence of the
 $f$-ion.

{\em Second }, we find that due to non-orthogonality the $f$-level
also contributes to the
mixing interaction: the term in Eq.45 describing linear mixing 
would disappear for the orthogonal basis. 

{\em Third,} the non-orthogonality transfers non-linearity into the
 conduction electron subsystem. This correctly reflects 
the underlying physics; the
$c$-electron spends part of its life  as an $f$-electron, therefore,
the  mixing and hopping which it experiences should be
 correlated to the same extent.
It is this reasoning which forced us to perform operations in this 
particular order.
First, we take into account
correlations, and then perform orthogonalization. The other
motivation is of a technical nature; this construction allows
 us to avoid consideration of a
combinatorics of tails for vertexes with one $c$-tail and many $%
f$-tails. On the other hand if we now decide to write the equations 
in terms of $f$-electrons ({\em i.e.}, fermion $f$-operators), some 
features of the many-tail vertexes are clarified by these equations. 
Particularly, the many-electron GFs, which are not allowed to be 
decoupled, can be obtained simply by transformation of all Hubbard's 
operators back to the $f$-operator form.

 {\em Fourth,} the equation for the  $X$-operator does not contain
linear mixing.

{\em Fifth.} The right-hand side of the  equation has some 
similarity to the
Boltzmann equation; both contributions, from mixing and from hopping,
consist of two terms, "incoming" and "outgoing"; we denote them by the
superscripts $\uparrow $ and $\downarrow $, correspondingly, in order to
emphasize that the first describe intra-atomic transitions "up",
from the states with $n$ electrons to the states with $n+1$ electrons;
the  $\downarrow $-terms describe the inverse processes. As we have seen above
(see, e.g., discussion after Eq.110)
later, the contributions from the latter terms always contain a 
small parameter.

\subsection{Equations of motion for quasi-boson operators}

Let us introduce the vector-operator $\eta $, 
which will describe all fermion or fermion-like transitions in our system.
For example, for the system with $n$ conduction bands 
and $m$ intra-atomic fermi-like transitions it is given by: 

\begin{equation}
\eta _{j\lambda }^{\dagger }=(c_{jL_1}^{\dagger },...,c_{jL_n}^{\dagger
},X_j^{\bar{b}_1},...,X_j^{\bar{b}%
_m},c_{jL_1}^{},...,c_{jL_n}^{},X_j^{b_1},...,X_j^{b_m}).
\end{equation}

As is seen from this notation, 
for $\eta =c$ or $c^{\dagger }$ the index $\lambda $ 
takes a value $L$ while for $\eta
=X$ or $X^{\dagger }$, and $\lambda $ can either be $b$ or 
$b^{\dagger }$. 
The Hamiltonian without external fields generates the following equation of
motion for the operators $Z^\xi $ ;

\begin{eqnarray}
i\partial _tZ_j^\xi  &=&\sum\limits_\Gamma \bar{\delta}_{\xi ,\Gamma }\Delta
_{\bar{\xi}}Z_j^\xi +V_{n_1L_1,jb}^\xi c_{n_1L_1}^{\dagger }X_j^b+V_{j\bar{b}%
,n_1L_1}^\xi X_j^{\bar{b}}c_{n_2L_2}^{}  \nonumber \\
&&+V_{n_1\bar{a}_1,n_2a_2}^\xi [\delta _{n_1,j}\delta _{\bar{a}_1,\bar{b}%
}+\delta _{n_2,j}\delta _{a_2,b}]X_{n_1}^{\bar{a}_1}X_{n_2}^{a_2};
\nonumber
\\
&\equiv &\sum\limits_\Gamma \Delta _{\bar{\xi}}Z_j^\xi +V_{n_1\lambda
_1,n_2\lambda _2}^\xi \eta _{n_1\lambda _1}^{\dagger }\eta _{n_2\lambda
_2}^{}.
\end{eqnarray}

Let us remind the reader that for the diagonal operators (i.e., for the transition $%
\xi =[\Gamma ,\Gamma ]$) the first term in the right-hand side of Eqn.C18
vanishes, since  for this transition  $\Delta _{\bar{\xi}=[\Gamma
,\Gamma ]}=\nu_{\Gamma \Gamma }E_\Gamma 
-\nu_{\Gamma \Gamma }E_\Gamma =0$.  Furthermore, we have defined 
 the constants of the interaction in such a way that
 $V_{n_1\lambda _1,n_2\lambda _2}^\xi =0$
when ($n_1\lambda _1)=(n_1L_1)$ and $(n_2\lambda _2)=(n_2L_2)$
simultaneously, i.e. when both indices relate to $c$-electrons.
In all other cases they  denote
the following combinations of the matrix elements : 

\begin{eqnarray}
V_{j\bar{b},n_2L_2}^\xi  &\equiv &-O_{jM,n_1L_1}^{-1}(f_M^{\dagger })^{\bar{c%
}}\varepsilon _{\bar{b}}^{\bar{c},\xi
}h_{n_1L_1,n_2L_2}^{}+W_{ja_1,n_2L_2}^{*}\varepsilon _{\bar{b}}^{\xi \bar{a}%
_1}, \\
V_{n_1L_1,jb}^\xi  &\equiv &-O_{n_2L_2,jM}^{-1}(f_M^{})^c\varepsilon
_b^{c,\xi }h_{n_1L_1,n_2L_2}^{}+W_{n_1L_1,ja_2}^{}\varepsilon _b^{\xi a_2},
\\
V_{n_1\bar{a}_{1,}jb}^\xi  &\equiv &-O_{n_2L_2,jM}^{-1}(f_M^{})^c\varepsilon
_b^{c,\xi }W_{n_1\bar{a}_1,n_2L_2}^{*}, \\
V_{n_1\bar{a}_1,jb,}^\xi  &\equiv
&-W_{jL_1,n_2a_2}^{}O_{jM,n_1L_1}^{-1}(f_M^{\dagger })^{\bar{c}}\varepsilon
_{\bar{b}}^{\bar{c},\xi }.
\end{eqnarray}

\section{
Auxiliary fields $A_{j\lambda ,j^{\prime }\lambda ^{\prime }}^{}(t)$ }
The auxiliary fields 
$A_{j\lambda ,j^{\prime }\lambda ^{\prime }}(t)$
have the following non-zero elements: \\
For $\;\eta _\lambda  =c $ we have
\begin{equation}
A_{jL,j^{\prime }a}^{(cX)}(t)=
\sum_{j^{\prime } Mb,\xi=\eta,A,\Gamma}
\varepsilon _a^{b\xi }{\cal U}_{j^{\prime }\xi }^{}(t)
(f_M^{})^b(O^{-1})_{jL,j^{\prime }M}
\end{equation}
and 
\begin{equation}
A_{jL,j^{\prime }a}^{(cX^{\dagger})}(t)=
\sum_{j^{\prime } Mb,\bar{\eta}}
\varepsilon _{\bar{a}}^{b\bar{\eta} }{\cal U}_{j^{\prime }\bar{\eta} }^{}(t)
(f_M^{})^b(O^{-1})_{jL,j^{\prime }M}.
\end{equation}
For $\eta _\lambda  =X $ we have 
\begin{equation}
A_{ja,j^{\prime }b}^{(XX)}(t)=
\;\sum_{\xi=\eta,A,\Gamma}
 \varepsilon _b^{a\xi }{\cal U}_{j\xi }^{}(t)\delta_{jj'} \equiv
\delta_{jj'} A_{jb}^{a}
\end{equation}
and
\begin{equation}
A_{ja,j^{\prime }\bar{b}}^{(XX^{\dagger})}(t)=
\;\sum_{\bar{\eta}}
 \varepsilon _{\bar{b}}^{a\bar{\eta} }
{\cal U}_{j\bar{\eta} }^{}(t)\delta_{jj'} \equiv
\delta_{jj'} A_{j\bar{b}}^{a}.
\end{equation}
For $\;\eta _\lambda =c^{\dagger } $ we have 
\begin{equation}
A_{jL,j^{\prime }\bar{b}}^{(c^{\dagger} X^{\dagger })}(t)=
-\sum_{\xi=A,\bar{\eta},\Gamma Ma}\varepsilon _{\bar{b}}^{\bar{a} \bar{\xi}}
{\cal U}_{j^{\prime }\bar{\xi}}^{}(t)(f_M^{})^{\bar{a} }
(O^{-1*})_{j^{\prime }M,jL}
\end{equation}
and
\begin{equation}   
 A_{jL,j^{\prime }\bar{b}}^{(c^{\dagger} X)}(t)=
-\sum_{\bar{\eta} Ma}\varepsilon _{\bar{b}}^{\bar{a} \bar{\xi}}
{\cal U}_{j^{\prime }\bar{\xi}}^{}(t)(f_M^{})^{\bar{a} }
(O^{-1*})_{j^{\prime }M,jL}.
\end{equation}
For $\;\eta _\lambda = X^{\dagger } $ we have
\begin{equation}
A_{j\bar{a },j^{\prime }\bar{b}}^{(X^{\dagger} X^{\dagger })}(t)=
-\delta _{jj^{\prime }}
\sum_{\xi=\bar{\eta},A,\Gamma}
 \varepsilon _{\bar{b}}^{\bar{a}\xi }{\cal U}_{j^{\prime }\xi }^{}(t) 
\equiv \delta_{jj'} A_{j\bar{b}}^{\bar{a} }
\end{equation}
and
\begin{equation}
A_{j\bar{a },j^{\prime }b}^{(X^{\dagger}X)}(t)=
-\delta _{jj^{\prime }}
\sum_\eta \varepsilon _{b}^{\bar{a}\eta }{\cal U}_{j^{\prime }\eta }^{}(t) 
\equiv \delta_{jj'} A_{jb}^{\bar{a} }.
\end{equation}

\begin{figure}
\caption{Corrections to the self-energy in the
approximation Hubbard-I (Eq. 78).
The dashed wiggle line
denotes the part of mixing $\theta ^{\Gamma }$, the triangle with
inscribed $\alpha $ denotes
the $\alpha $-matrix, given by Eq.64, the mixing interaction is
denoted by the wiggle line (see Eq.84).
 The expectation value
$\langle Z_{3_b} \rangle $
is denoted
by circle-cross on the top of the $\alpha $-triangle. Summation over
bosonic
index $1_b$ is implied in this top what transforms the product
$\alpha _{12,3_b}\langle Z \rangle $ into $P_{12}$.}
\end{figure}

\begin{figure}
\caption{ Simplification of the graphical notations in the
case of equations for fermions. In the vertex containing $\alpha
$-matrix
the line of mixing interaction {\em goes out} of the vertex,
while in the one
containing $\beta $-matrix it {\em enters} the vertex;
$\beta $-matrix appears always together with the $\tau $-matrix
(see Eqs.64, 72 and 88), which inverts direction of the
line of mixing. This vertex describes
the process involving two-electron
$(\Gamma_{n-2} \rightarrow \Gamma _n)$-transition. }
\end{figure}

\begin{figure}
\caption{$R^l$-rule of taking functional
derivatives of the pseudolocator line; changing $\alpha $ by $\beta $
gives $R^r$-rule (see Eqs.85 and 86,87).}
\end{figure}

\begin{figure}
\caption{Mean field corrections to the
self-energy of pseudolocator.
 Note that both vertexes have three tails of pseudolocators and
one of interaction; {\em left} vertex has the {\em
 outgoing} interaction tail, while {\em right} one has
{\em incoming }interaction tail. The point from which the dashed line of
$\eta $-pseudolocator and the wavy line of the $\eta-\eta $-interaction
$V_{12}$ go out denotes the $\tau $-matrix.}
\end{figure}

\begin{figure}
\caption{The second-order corrections to the
vertexes from kinematic interactions. }
\end{figure}

\begin{figure}
\caption{Corrections to the self-energy of the
$\eta $-pseudolocator generated by the second-order corrections to
vertex.}
\end{figure}

\end{document}